\documentclass[review]{elsarticle}

\usepackage{lineno,hyperref}
\modulolinenumbers[5]

\journal{Physics Reports}

\usepackage{amsmath,amssymb}              
\usepackage{slashed}
\usepackage{graphicx}
\usepackage{bm}
\usepackage[top=1.0in,bottom=1.0in,left=1.0in,right=1.0in]{geometry}

\newcommand{\be}{\begin{equation}}
\newcommand{\ee}{\end{equation}}
\newcommand{\bea}{\begin{eqnarray}}
\newcommand{\eea}{\end{eqnarray}}
\newcommand{\ba}{\begin{array}}
\newcommand{\ea}{\end{array}}

\begin{document}

\begin{frontmatter}

	\title{How to observe the QCD instanton/sphaleron processes at hadron colliders?}
	\author{Edward Shuryak and Ismail Zahed}
	\address{ Center of Nuclear Theory,
		Department of Physics and Astronomy,  \\ Stony Brook University,
		Stony Brook, NY 11794, USA}

	\begin{abstract}
 The instanton/sphaleron processes involve gauge fields with changing topology,  including  a nonzero variation of the Chern-Simons number $\Delta N_{CS}=\pm 1$. In QCD such processes lead to the production of $2N_f \Delta N_{CS}$ units of the axial charge, proportional to the number of light quark flavors $N_f=3$, for $u,d,s$.
 In the  electroweak theory such processes lead to the  production of 12 fermions, with $\Delta B=\Delta L=\pm 3$ units of baryon and lepton number. While this is all  known for a long time, and is one of the pillars of the nonperturbative theory of the QCD vacuum, in this paper we discuss what we call a ``reclined tunneling",
 in which external forces are added to the tunneling processes and create certain gluonic objects
 with positive invariant mass.  The idea to observe these objects experimentally at hadronic colliders have
 been proposed before, but insofar without success.
  Motivated by the recent CERN workshop on the topic, we review these ideas. We put forward our own suggestions, in particular to utilize a double-diffractive (Pomeron-Pomeron) collisions to this goal,  which we believe maximizes the entrance factor and  	minimizes the backgrounds.
 We consider clusters of small ($M=3-10\, {\rm GeV}$), medium ($M=10-30\, {\rm GeV}$) and high $M\sim 100 \, {\rm GeV}$ invariant masses, subsequently. Among the proposed signals are 
		specific flavor combination of channels,
	originating from well-defined 6-, 8- and 10-quark-antiquark operators, as well as correlation of quark chiralities to be potentially detected via $\Lambda$ hyperon decays.
	\end{abstract}
	

\end{frontmatter}

	
	
	\section{Introduction}
	\subsection{ The instanton/sphaleron processes}
	
In going to the mountains, one needs a reliable map and preferably from a  variety of sources.
Building a tunnel  is expensive, so one needs to think a lot. Should it be horizontal
or reclined? Where should the entrance be, and where should the exit be?
As we will see shortly, all such questions also appear when we think of optimal paths
producing topologically nontrivial final states in a collision. 

Some of these questions have  already been answered in theory decades ago, 
but  some still require further analyses and calculations.  There is a vast
 number of applications of instanton-induced effects
 in vacuum and hadronic structure. The instanton-sphaleron processes, producing certain topologically nontrivial gluonic clusters, are 
 only seen in low-mass domain, like $\eta_c$ decays.
 This paper is about possible ways to see such objects
 at hadron colliders, with variable masses.

 One can split the expected cross section into three distinct parts:  (i) tunneling action; (ii) the semiclassical prefactor;  (iii) the entrance factor; and, last but not least,
 (iv) branching ratios into particular exclusive channels, or sequentially
\begin{equation} 
\sigma \sim \bigg[{{\rm entrance \over factor}}\bigg]\bigg[{{\rm semiclassical \over prefactor}}\bigg]\bigg[ e^{- S_{cl}}\bigg]\bigg[ B({\rm final\, state})\bigg]
\end{equation}
As we  show below, the classical part is well under control. The semiclassical prefactor is
not yet calculated for gauge theory beyond one loop, but it was done in  relevant toy models, so it can  be
evaluated if needed. The entrance factor is really the difficult part still in deliberation.
	
The produced unstable magnetic objects, generically known as ``sphalerons", explode
into outgoing gluon and quark waves, which eventually ``hadronize" into mesons and baryons
going to the detectors.
 This initial part of this  process is described by a solution of  the  classical
YM equation, see section \ref{sec_explode}. Quark pair production is described by the solution of the Dirac equation in the exploding background. For all light quark flavors one has the same solution,
which results in an effective	
	6-quark operator of the flavor composition
	 $(\bar{u} u)(\bar{d} d)(\bar{s} s)$ and with specific chiral properties.
Its projection onto the  distribution amplitudes of pertinent mesons,  provides the decay probabilities
to exclusive final states, e.g. for three mesons those are $KK\pi,\pi\pi\eta,\pi\pi\eta'$.
 	 
A further discussion of heavy  ($c,b$) quark
production is done in section \ref{sec_c_and_b}.
We will estimate the sphaleron scales at which
such production becomes possible, and then
discuss 8- and 10-fermion operators and  their
final states such as  $KKDD$ and $BBD_sDK$.
	 
	
\subsection{Double diffraction and Pomerons, glueballs and clusters}	\label{sec_PP}
	 
In multiple papers the instanton/sphaleron process was calculated with the assumption
that it is initiated at a partonic level, gluon-gluon or quark-antiquark collisions. 	 
In this case the secondaries originated from sphalerons are imbedded in
complicated  $pp$ collisions, together with many other jets and secondaries produced.
The separation of the signal from backgrounds is in this case very difficult. 
	 
Perhaps the main idea first expressed in \cite{Shuryak:2003xz},
is that one may avoid large backgrounds associated with min.bias $pp$ collisions, by using
double diffractive events (also known as  Pomeron-Pomeron  abbreviated as {\bf PP}, distinct from $pp$).  We start with a brief summary of the experimental situation,
before elaborating on  various theoretical models.

The two  early experiments of interest are WA102~\cite{Kirk:1998et} and WA8~\cite{Brandt:2002qr}, both of which were carried decades ago   at CERN .

WA102 was driven by an idea that  the {\bf PP} processes provide
a ``glueball filter", allowing to separate glueball-type hadrons from other mesons.
It provided important data on the  distribution
over the azimuthal angle $\phi$, between the transverse momentum transfers to  both surviving protons. The collaboration  identified a certain number of scalar, pseudoscalar and tensor hadronic resonances in the  invariant mass distribution. As a subsequent discussion has shown, the collaboration has indeed seen not only a scalar glueball  candidate, but also  ta  tensor $2^+$ glueball at a mass of $2.2\, {\rm GeV}$.  Both have angular distributions in $\phi$
completely different  from those for ordinary mesons.

WA8  was focused on larger invariant masses. 
Unfortunately, its central detector was just a calorimeter, so no exclusive channels were studied. 
This collaboration  reported observation of some isotropically decaying ``clusters", with a mass peaked at 
$M\approx 3\, {\rm GeV}$, which we will discuss further in sections \ref{sec_Hooft} and \ref{sec_sph_in_PP}. They also 
reported non-isotropic background events with masses $M=5-20\, GeV$.
	
	The  revival of the diffractive field at the LHC has began relatively recently. The Atlas Forward Proton (AFP) detector was installed and ran, but
	it is focused on producing  systems with very large mass in the range of hundreds of GeVs. In our understanding, ATLAS ALFA forward detectors were designed to measure forward elastic scatterings.
On the other hand, the CMS-TOTEM collaborations, working together, have 
	 addressed soft {\bf PP} reactions, and
	recently reported soft exclusive $\pi^+\pi^-$ data~\cite{2002.06959}. They focused on small invariant mass region $M(\pi^+\pi^-)< 2 \,{\rm GeV}$ and have confirmed the production of (at least) four scalar and tensor mesons, all seen previously by WA102.
	From a wider perspective, apart from
	 focusing on glueball spectroscopy, these experiments have provided important insights into the structure of the Pomeron-Pomeron-hadron vertices, and thus the 
	structure of the Pomeron itself. 
	
	From the theory side, during the last decade there were basically two major 
	developments:
	
	(i) The Pomeron is defined as a virtual state, described by continuation of a glueball Regge trajectory from positive to small negative $t$ domain, where it is observed in scattering.
	Note that the nearest physical state to it is the tensor $2^{++}$ glueball.
	In holographic models of QCD  this glueball is identified with the excitation of bulk gravitons~\cite{Polchinski:2002jw,Brower:2006ea}.
	This, and other consideration, relate Pomeron exchanges with Reggeized 
	graviton exchanges.  These ideas have  explained well the WA102 data production of pseudoscalar \cite{Anderson:2014jia} 
	and tensor \cite{Iatrakis:2016rvj} glueballs.   In this latter paper the {\bf PP}-$2^+$ vertex used was a 3-tensor coupling deduced from the Einstein-Hilbert action of general relativity. 
	Earlier suggestions that the Pomeron phenomenology needs some  {\em polarization tensor} 
	were made by the Heidelberg group, see review in \cite{Ewerz:2016onn}.

	(ii) From the 1970's the perturbative theory of the Pomeron was based first on two-gluon exchanges, and then developed into inclusion of  ladder diagrams with gluonic rungs.
	This theory culminated in the famed  papers \cite{Kuraev:1976ge}
	and is known as the BFKL Pomeron.  It is successful in describing diffraction with
	large $|t| >  few\, GeV$.
	Yet at small $|t|$ (or large impact parameter $b\sim 1/\sqrt{|t|}$)
	there is no reason to use perturbative QCD. Alternative theory for this case,
	based on ``stringy instanton" solution to Nambu-Goto string exchange
	was developed in \cite{Basar:2012ra}, and is known as the BKYZ Pomeron. 
	 Although the two theories start from very different descriptions, with different interpretations of the parameters, the scattering amplitude
	 is nearly the same. The latter BKYZ version, unlike BFKL, tells when the transition
	 in impact parameter should happen, as strings undergo Hagedorn transition.

	 \section{The topological landscape} 
	 The Hilbert space of all possible gauge potentials $A^\mu_a(x)$ is the manifold over which we need
	 to integrate the Feynman path integral. 
	 Of course, we would not discuss infinite-dimensional maps, and focus
	 on two main variables, for all
	  static ( 3-dimensional and  purely magnetic) configurations of the lowest energy, consistent with 
	  the walue of those coordinates.	 
	 One of the coordinates  is the topological Chern-Simons number
	 \begin{equation} 
	 N_{CS}\equiv { \epsilon^{\alpha\beta\gamma} \over 16\pi^2}\int d^3x  \left( A^a_\alpha \partial_\beta A^a_\gamma +{1\over 3}\epsilon^{abc}A^a_\alpha A^b_\beta A^c_\gamma \right)
	 \label{eqn_Ncs}
	 \end{equation}
which is integer for pure gauge configurations, at which the energy is zero. 
Those points will be referred to as "valley minima".
The other coordinate is the mean square radius 
	 of the magnetic  field strength squared
	 \begin{equation}
	 	\rho^2\equiv { \int d^3 x \vec x^2 \vec B^2
	 		\over \int d^3 x  \vec B^2 }
 		\label{eqn_rho}
	 \end{equation}
	 Without fixing it, there is no minimal energy, as simple rescaling can change it.

\begin{figure}[h!]
	\begin{center}
		\includegraphics[width=8cm]{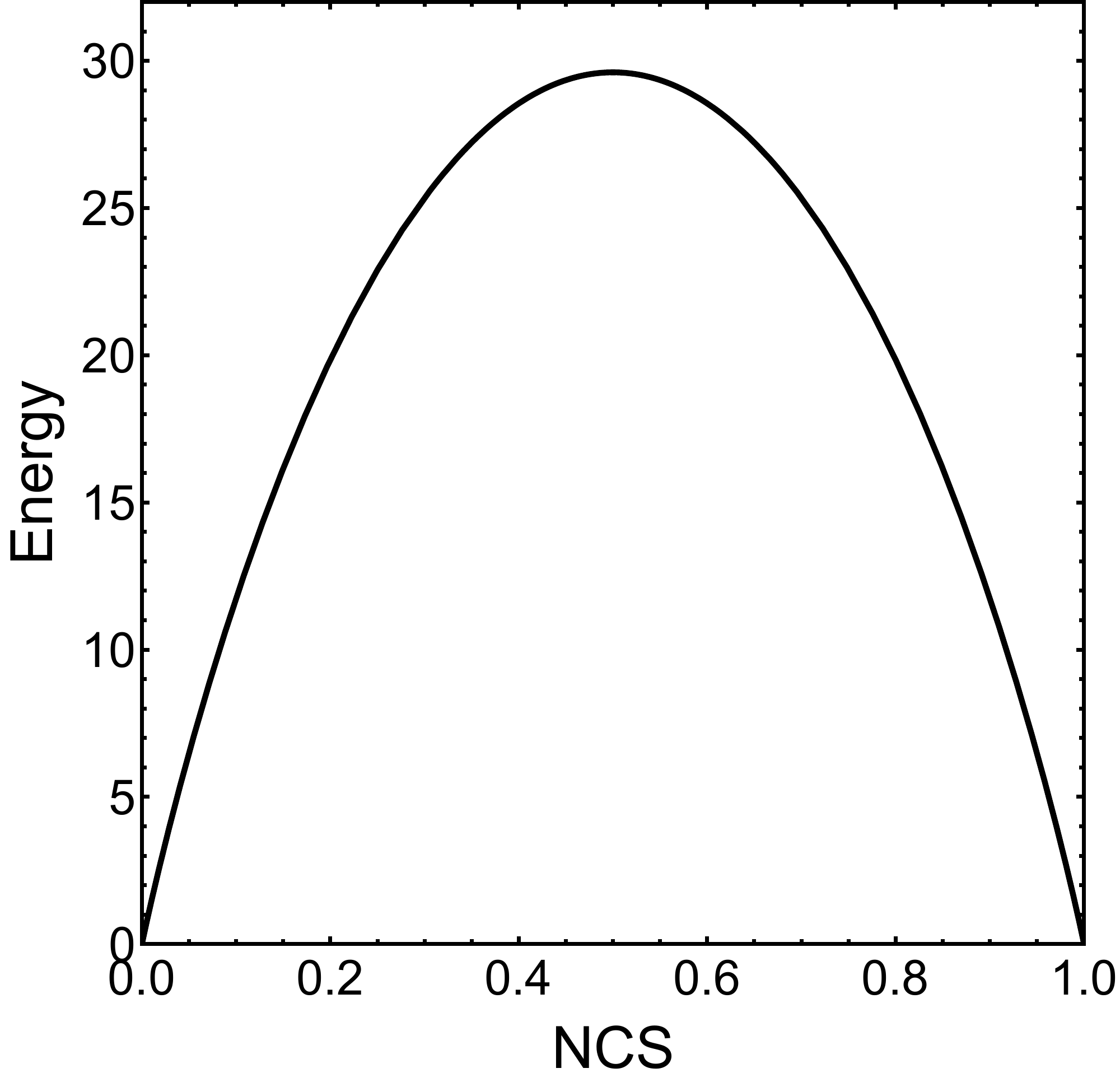}
		\caption{ 
			The potential energy $U_{\rm min}\bigg(N_{CS} ,\rho\bigg)$ (in units of $1/g^2\rho$)
			versus the Chern-Simons number ${N}_{CS}$  for the 
			``sphaleron path"  between $N_{CS}=0$ and $N_{CS}=1$.
		} \label{fig_sphaleron_path}
	\end{center}
\end{figure}

	 
	 By the ``topological landscape" we mean the 2-dimensional profile of
the minimal energy $U_{\rm min}(N_{CS},\rho)$ of gauge field configurations with those two coordinates fixed. For pure gauge theory, such minimal energy configurations themselves, are known  
 as the ``sphaleron path": as changing  $N_{CS}$ from one integer to the next we are leading from one topological valley to another, keeping  minimal energy at any point.	  
	Those were derived  by
	 Carter, Ostrovsky and one of us~\cite{Ostrovsky:2002cg} by two different methods.
The one related with the instanton/sphaleron process
will be discussed in the next section. The second method is minimization with	 two
	 Lagrange multipliers times two conditions, (\ref{eqn_Ncs}) and (\ref{eqn_rho}).	 
	 The minimal energy along the path was obtained  in a parametric form
	 \begin{eqnarray}
	 	U_{\rm min}(\kappa, \rho)&=&(1-\kappa^2)^2{3\pi^2\over g^2\rho} \\ \nonumber
	 	{N}_{CS}(\kappa)&=&\frac 14 {\rm sign}(\kappa)(1-|\kappa|)^2(2+|\kappa|) 	
	 \end{eqnarray}	
The result shows a profile of the ``topological mountain" of the gauge theory, see Fig.~\ref{fig_sphaleron_path}, also known as ``the sphaleron path". Its maximum, at
  $\kappa=0$, has $N_{CS}=\frac 12$ and its energy is known as the {\em sphaleron mass} 
 	 
	 \begin{eqnarray} 
	 M_{sph}=U_{\rm min}\bigg(\frac 12 ,\rho\bigg)={3\pi^2 \over g^2 \rho}
	 \end{eqnarray}

	 When the momentum scale $1/\rho$ is high,
the gauge fields are very strong and one can 
neglect the  ``vacuum structure" effects and keep only the
	   classical Yang-Mills equation. Of course, this equation is  scale-invariant, and therefore  $U_{\rm min}\sim 1/\rho$. However, if $1/\rho$
	   is small enough, one can no longer neglect vacuum effects.
	 
	 In the electroweak theory these ``vacuum effects" are described by the Higgs field $\phi$. The scale is set by 
its VEV $v=\langle \phi \rangle \approx 246\, {\rm GeV}$.   Ignoring first the Weinberg angle and using variational methods, Klinkhamer and Manton~\cite{Klinkhamer:1984di}
have  included  the  Higgs field and solved the YM+scalar equations.
	They observed that the Higgs field must vanish at the origin $r=0$ to keep the topology of the gauge field intact, and 
	 therefore the sphaleron is a  ``semi-empty bag" inside the Higgs vacuum. This adds
	 some energy: e.g.
	the leading term at large $\rho$ proportional to the bag volume and the highest power of the Higgs VEV $$U_{\rm Higgs}\sim \rho^3 v^4 $$
The $1/\rho$ gauge term plus Higgs terms with  positive powers of the size leads to a minimum
	 at a certain $\rho^*$, which fixes the configuration completely. The mass of the
	 electroweak sphaleron is found to be about 9 TeV. Thus, going from one topological valley of the electroweak theory to another, one needs
	 at least that much energy. All of that was clear since 1984.
	 
   Proceeding to our main object of interest,
  the sphaleron path 
	 in QCD, we note that  the QCD vacuum also
	 has a rather complicated vacuum structure, with nontrivial VEVs
	 of various operators (known as ``vacuum condensates"). 	 
	 There are lattice data and models of the 
	 QCD vacuum structure, that provides some understanding of the topological landscape at large $\rho$.  It will be discussed below, and here we just state that they also
	 force the  minimal energy $U_{\rm min}(1/2,\rho)$ to start growing
	 at large $\rho$, also producing some optimal size $\rho^*$.

	 \section{The tunneling paths}
	 
	 The first ``tunneling path" was the original BPST
	 instanton discovered in 1975. It is the solution of the Euclidean
	 YM equation. In terms of the  landscape illustrated in  Fig.~\ref{fig_sphaleron_path}, it is a horizontal path, at energy zero, leading directly from the
	 bottom of one valley to the bottom of the next one.  The tunneling amplitude along it is 
	 $\sim {\rm exp}(-8\pi^2/g^2)$, given by the action of the instanton.
	 
	 Many of the discussions on how to improve the tunneling path,
including the final state gluon radiation, were the subject of complicated technical papers in the late 1980's. 
We would not describe them, but just state that 
Khoze and Ringwald 
\cite{Khoze:1991sa},  as well as Verbaarschot and Shuryak \cite{Shuryak:1991pn}, have suggested that 4-dimensional  configurations describing
the $optimal$	instanton-antiinstanton  $I\bar{I}$ also provide
 the {\em optimal path for sphaleron production}. 
 Since then, most  calculations follow this idea.

\begin{figure}
	\centering
	\includegraphics[width=0.7\linewidth]{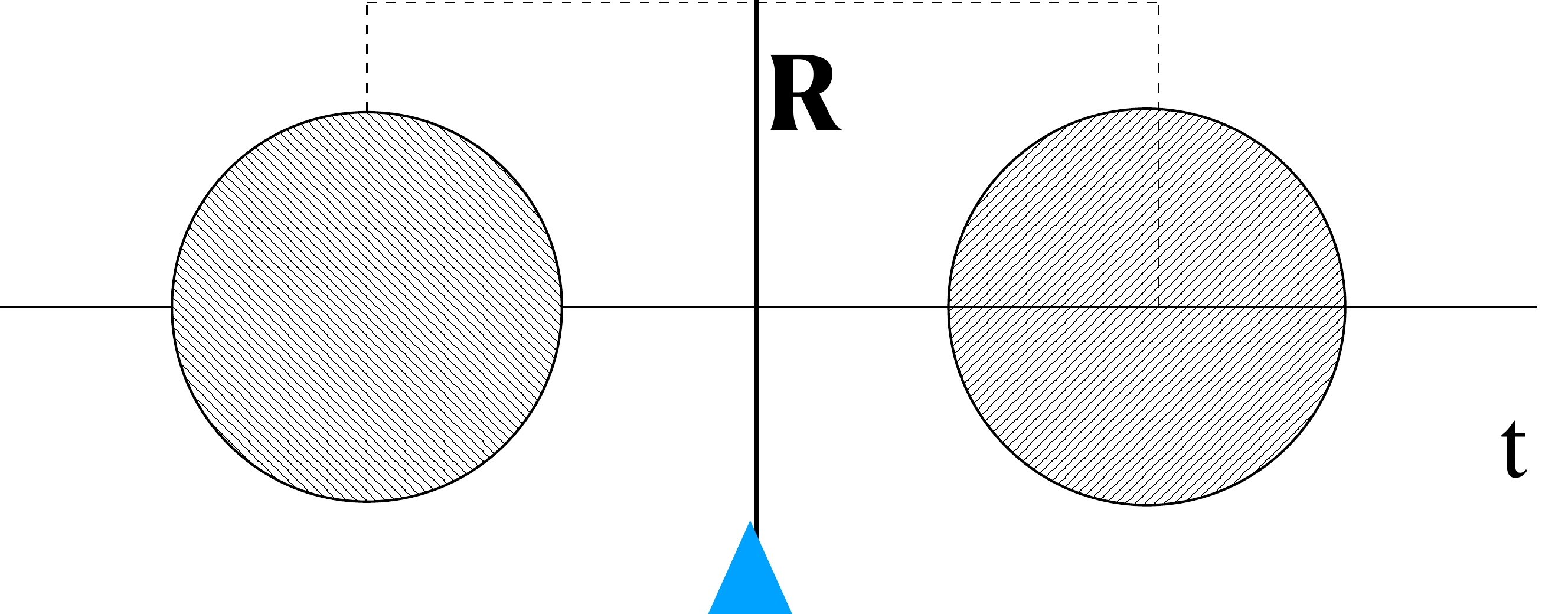}
	\includegraphics[width=0.6\linewidth]{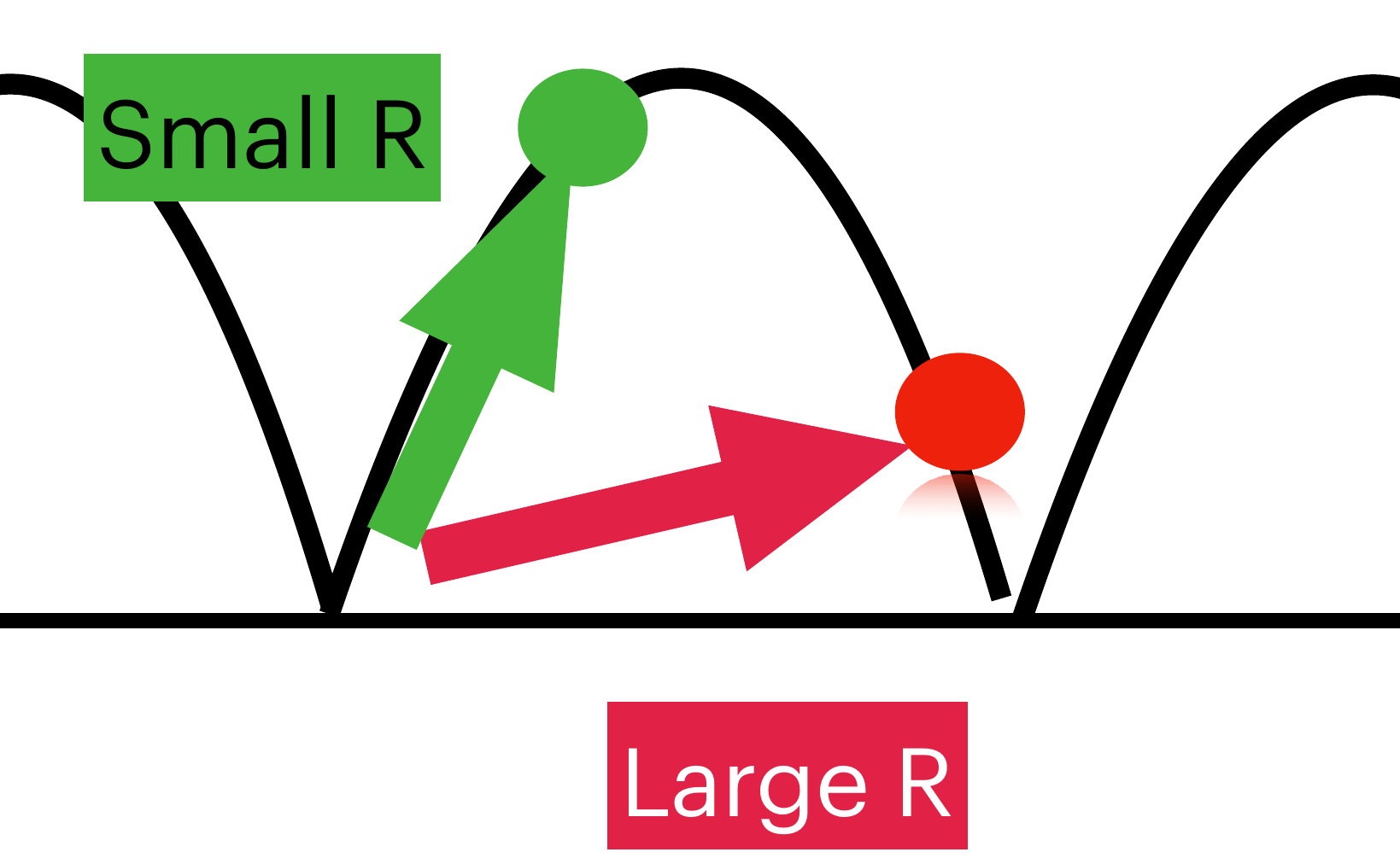}
	\caption{The upper plot shows a schematic picture of the  instanton-antiinstanton configuration. The horizontal axis is (Euclidean) time $t$, and 
	$R$ is the distance between their centers. The blue triangle indicates $t=0$, a  3-dimensional  hyper-surface in which the produced magnetic object resides.
The lower plot with the red arrow, refers to the ``reclined tunnel" corresponding to large $R$. The green arrow
on the left indicates tunneling for small $R$, with a Chern-Simons number of the produced object  $N_{CS}<1/2$.
For this last  case,  the classical explosion returns the system to the original valley.  	
}
	\label{fig:iibar}
\end{figure}

Before going to the details, let us explain why it is the case, following important insights from~\cite{Ostrovsky:2002cg,Shuryak:2000df,Nowak:2000de}. The 
$I\bar{I}$ configuration is schematically shown in the upper plot of Fig.~\ref{fig:iibar}. The shaded
circles indicate regions where the instanton and antiinstanton fields are strong. In one of them, the
fields are (close to) self-dual $\vec E=\vec B$, and  in the other antiself-dual $\vec E=-\vec B$. If both have the same sizes and color orientations, symmetry of the plot suggest that the electric field $\vec E$ changes sign at $t\rightarrow -t$, and is therefore zero at $t=0$. The 3-dimensional  plane in the middle of the plot must contain a pure magnetic configuration, corresponding
to semiclassical ``turning points" where the momentum vanishes. At such points the paths go from a tunneling process  in Euclidean time to  a ``real life" process in Minkowski
space.

So, the $t=0$ plane is nothing else but the ``unitarity cut", familiar from perturbative Feynman diagrams. The object on it
$is$ the object produced. the two halfs of the $I\bar{I}$
4-dimensional  configurations are the amplitude and the conjugated amplitude. They  describe the probability to produce this object. Any discussion of the multi-gluon production, with the
complicated interferences between them, are not needed at all: what is 
produced is this classical magnetic object.
	 
Now, why is it that this tunneling path is better
than the original instanton (or $R\rightarrow \infty$
limit)? It may be confusing at first glance, since
the $I\bar{I}$ configurations are $not$ solutions of the YM equations. They are  tunnels $inclided$ $upward$
as indicated  by the arrows
in the lower plot of Fig.~\ref{fig:iibar}.
Indeed, to go uphill an external force is needed: but
on those paths the action is reduced! (As we will see, roughly by a factor 2, or many decades in the rates.)

 Instead of ending at the bottom of the valley, these paths end
at some points on the slopes (indicated by red and green balls). After the corresponding magnetic objects are born, they roll down (explode) classically. Their action is real, $|exp(iS)|=1$, 
and therefore their decays have probablity one. 
Whatever decays they have, it does not affect the cross section. 

As it  is clear from the lower plot of Fig.~\ref{fig:iibar}, for small $R$ the instanton and antiinstanton nearly annihilate each other and their combined action
can be reduced all the way to zero. If the product has $N_{CS}<1/2$, it will not roll to the second valley, but return to the original one. The
anomaly relation then would indicate  no fermionic transitions. In summary: there exist 
 {\em topologically trivial} and {\em topologically nontrivial} tunnelling paths! One cannot economize more than (roughly) $half$ of the instanton+antiinstanton action.

	 The issue of instanton-antiinstanton ($I\bar{I}$) interactions has its own long history. When instanton and aniinstanton are well separated, the simple ``sum ansatz" gives
	 twice the action, but what if the distance between them is comparable to their size? One can
	 invent many arbitrary interpolations.
	 
	 The idea   in~\cite{Balitsky:1986qn}
	 was to ``follow the force" using $\partial S /\partial x(\tau)$. In this way one gets a set
	 of configurations, which are action minima  for infinitely many perturbations, except along one
	 direction --  the set itself.
	One of us has independently generated this set
	 numerically~\cite{Shuryak:1987tr} for the double well~\footnote{We did not know then that 
	 	in mathematics our ``sreamlines" were 
	 	known in complex analysis as ``Lefschitz thimbles", special lines connecting saddles in the complex plane.} . 
		Yung~\cite{Yung:1987zp}  proposed the  ``Yung ansatz" solving the streamline equation at large $R$.
	 
	 For gauge theory instantons the problem looked more complicated. First of all, even a sum ansatz
	 could not be used, as special cancellations near the centers (in singular gauge) were spoiled  as the 
	 field strength gets singular. For this purpose a  ``ratio ansatz" was developed to cure this. Furthermore,  it looked that  the
	 interaction should    depend on at least 3 variables, $\rho_I, \rho_A, R$,
	 even for identical color orientations.  
	 Verbaarschot~\cite{Verbaarschot:1991sq}  however, noticed
	 that since the classical YM theory has conformal symmetry, the answer should depend on 
	 their single 
	 conformal-invariant combination~\footnote{Many years later, it was realized that this is
	 	a geodesic distance between two points in AdS$_5$ space, if $\rho$ is the extra coordinate.} 
	 \begin{equation}
	 	{ R^2+(\rho_1^2-\rho_2^2)^2 \over \rho_1\rho_2}
	 \end{equation}
	 Using an appropriate conformal map, the antiinstanton was set inside the instanton, and the
	 problem was mapped onto the double-well potential.  As a general surprise, Verbaarschot's configurations
	 happen to be described rather accurately by Yung ansatz, not only at large $R$ as was originally claimed, but in fact
	 all the way to $R\rightarrow 0$ \footnote{The formula itself remained complicated, and nobody
	 	-- Yung included -- suspected it to be a pure gauge, with zero field strength!}. 

To complete the story, let us mention that in~\cite{Ostrovsky:2002cg} it was shown that  the  ``streamline" $I\bar{I}$ configurations, well approximated by Yung ansatz, do indeed describe 
$U_{\rm min}(N_{CS},\rho)$ in complete agreement with  the
constrained minimization already shown in  Fig~\ref{fig_sphaleron_path}. The  parameter $\kappa$ is played by the relative separation $R$.

\begin{figure}[h!]
\begin{center}
 \includegraphics[width=10cm]{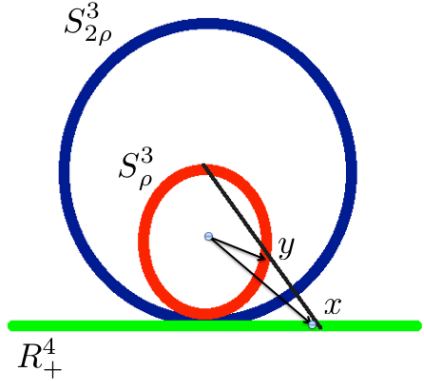}
  \caption{The inversion of  the 3-sphere $S_\rho^3$ of radius $\rho$ onto the upper part of Minkowski space with $t\geq 0$ or $R_+^4$,  through the 3-sphere $S^3_{2\rho}$
  of radius $2\rho$.}
 \label{fig_inverse}
 \end{center}
\end{figure}

\section{Classically exploding sphalerons} \label{sec_explode}
Both static and time-dependent exploding solutions for the pure-gauge sphaleron have been originally discussed 
by Carter,Ostrovsky and Shuryak (COS) ~\cite{Ostrovsky:2002cg}. A simpler derivation, to be used below, has been subsequently  given 
in~\cite{Shuryak:2002qz}.
The technique relies on an  {\em off-center conformal transformation} of the $O(4)$ symmetric Euclidean instanton
solution, which is analytically continued to  Minkowski space-time. However, the chief focus of that work~\cite{Shuryak:2002qz} was   the 
 description of the fermion production. 

The original $O(4)$-symmetric solution is given by the following ansatz

\begin{equation}  
\label{FO4}g A_\mu^a=\eta_{a\mu\nu} \partial_\nu F(y), \,\,\,\, F(y)=2\int_0^{\xi(y)} d\xi'   f(\xi')     
\end{equation}
with $\xi=  {\rm log}(y^2/\rho^2)$ and $\eta$ the 't Hooft symbol. 
Upon substitution of the gauge fields in  the gauge Lagrangian 
one finds  the effective  action for $f(\xi)$ 
\begin{equation} S_{\rm eff}=   \int d\xi \left[{\dot{f}^2\over 2}+2f^2(1-f)^2 \right]
\end{equation}   
corresponding to the motion of a particle in a double-well potential. 
In the Euclidean formulation, as written, the effective potential is inverted
\begin{equation} V_E=-2f^2(1-f)^2 \end{equation}
and the corresponding solution  is the well known BPST instanton, 
a path connecting the two maxima of $V_E$, at $f=0, f=1$. 
Any other solution of the  equation of motion
following from $S_{\rm eff}$
obviously generalizes to a solution of the Yang-Mills equations for $A_\mu^a(x)$ as well.
The sphaleron itself is the static solution at the top of the potential between the minima $f=-1/2$.

The next step is to perform an off-center conformal transformation as illustrated in Fig.~\ref{fig_inverse}
\begin{equation}  
\label{OFFCENTER}
(x+a)_\mu={2 \rho^2 \over (y+a)^2} (y+a)_\mu
\end{equation}
with $a_\mu=(0,0, 0, \rho) $.  It changes the original spherically symmetric 
solution 
to a solution of Yang-Mills equation depending on new coordinates
$x_\mu$, with a separate dependences on the time
$x_4$ and the 3-dimensional radius $r=\sqrt{x_1^2+x_2^2+x_3^2}$. 

The last step  is the analytic continuation to Minkowski time $t$, via $x_4\rightarrow i t$. 
The original parameter $\xi$ in terms of 
these   Minkowskian coordinates, which we still call  $x_\mu$, has the form
\begin{equation} \xi ={1\over 2}  {\rm log}\bigg({y^2\over \rho^2}\bigg)={1\over 2} {\rm log}\left( {(t+i\rho) ^2-r^2 \over (t-i\rho) ^2-r^2 } \right) \end{equation}
which is purely imaginary. To avoid carrying the extra $i$, we use the real combination
\begin{equation} \xi_E \rightarrow -i \xi_M = {\rm arctan}\left( { 2 \rho t \over t^2-r^2-\rho^2 } \right)   \label{arctan}  \end{equation} 
and in what follows we will drop the suffix $E$.
Switching from imaginary to real $\xi$,  corresponds to switching from the Euclidean to
Minkowski spacetime solution. It changes the sign of the acceleration, or the sign of the effective potential $V_M=-V_E$,
to that of the normal double-well problem.

The solution of the equation of motion is  given in ~\cite{Shuryak:2002qz}~\footnote{There was a misprint in the index of this expression in the original paper.} 
\begin{equation} 
\label{SOLUTION}
f(\xi)={1 \over 2} \left[ 1- \sqrt{1+\sqrt{2\epsilon}}\, {\rm dn} \left(  \sqrt{1+\sqrt{2\epsilon}} (\xi-K), {1 \over \sqrt{m}} \right) \right] 
\end{equation}
where  ${\rm dn}(z,k) $ is one of the elliptic Jacobi functions, $2\epsilon=E/E_s,2m=1+1/\sqrt{2\epsilon}$,
$E=V(f_{\rm in})$ is the conserved energy of the mechanical system normalized to that of the sphaleron $E_s=V(f=1/2)=1/8$. 
(\ref{SOLUTION})reduces to the  SO(4) solution derived by Luscher~\cite{Luscher:1977cw}, and the hypertorus solution obtained by Schechter~\cite{Schechter:1977qg}.
Since  starting from exactly
the maximum takes a divergent rolling time, we will start  from the nearby  turning point
with 
\begin{equation} f(0)=f_{\rm in}={1\over 2} - \tau, \,\,\,\,\,\, f'(0)=0 \end{equation}
where a small displacement $\tau$ ensures that  the  ``rolling downhill" from the maximum takes a finite time and
that the half-period $K$ -- given by an elliptic integral -- in the expression is not divergent. 
In the plots below we will use $\kappa=0.01$, but the results  dependent on its value very weakly.

The solution above describes a particle tumbling periodically between two turning points, 
and so the expression above defines a periodic function for all $\xi$. However, as 
it is clear from (\ref{arctan}), for our particular application the only relevant domain is $\xi \in [-\pi/2,\pi/2]$. Using the first 3 nonzero terms of its Taylor expansion
\begin{equation} f\approx 0.4929 - 0.00706 \,\xi^2 - 0.0011\,\xi^4 - 0.000078\,\xi^6 \end{equation}
we find a parametrization with an accuracy of $10^{-5}$, obviously invisible in the plot and
more than enough for our considerations. 

The gauge potential has the form \cite{Shuryak:2002qz} 
\begin{equation} \label{eqn_sph_field}
	gA_4^a=-f(\xi) { 8 t\rho x_a \over [(t-i\rho)^2-r^2]  [(t+i\rho)^2-r^2]  }  \end{equation}
$$ gA^a_i=4\rho f(\xi) { \delta_{ai}(t^2-r^2+\rho^2)+2\rho \epsilon_{aij} x_j +2 x_i x_a \over [(t-i\rho)^2-r^2]  [(t+i\rho)^2-r^2]  } $$
which is manifestly real.  
From those potentials we
generate rather lengthy expressions for the electric and magnetic fields,  and eventually for
CP-violating operators, using Mathematica. 

Note that the sphaleron solution corresponds to $t=0$ or static,  which is a  pure magnetic solution with $gA_4^a=0$. The magnetic field squared
is spherically symmetric and simple
\begin{equation} \vec{B}^2={96 \rho^4 \over (\rho^2+r^2)^4 } \end{equation}
\begin{figure}[h!]
	\centering
	\includegraphics[width=0.4\linewidth]{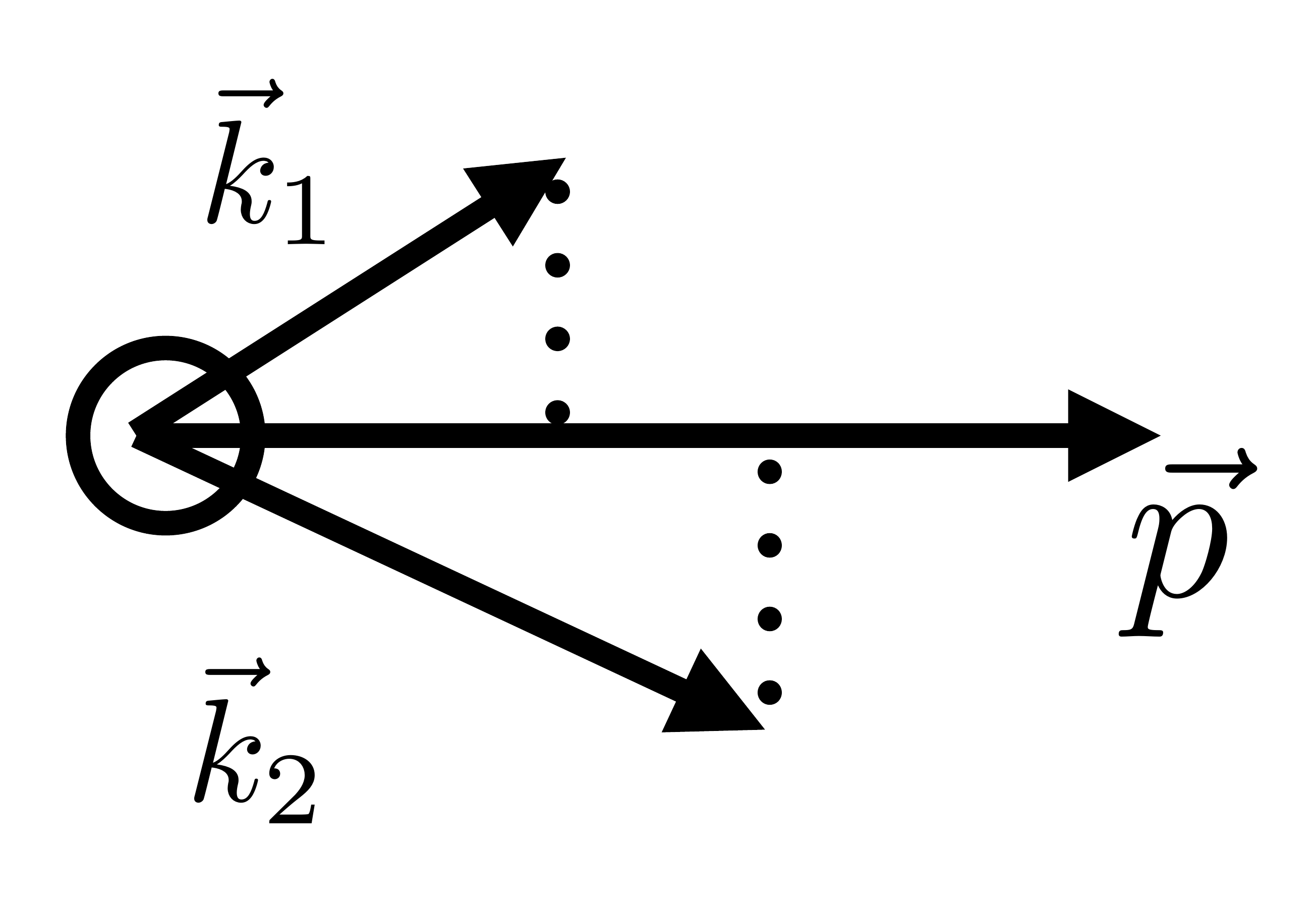}
	\includegraphics[width=0.4\linewidth]{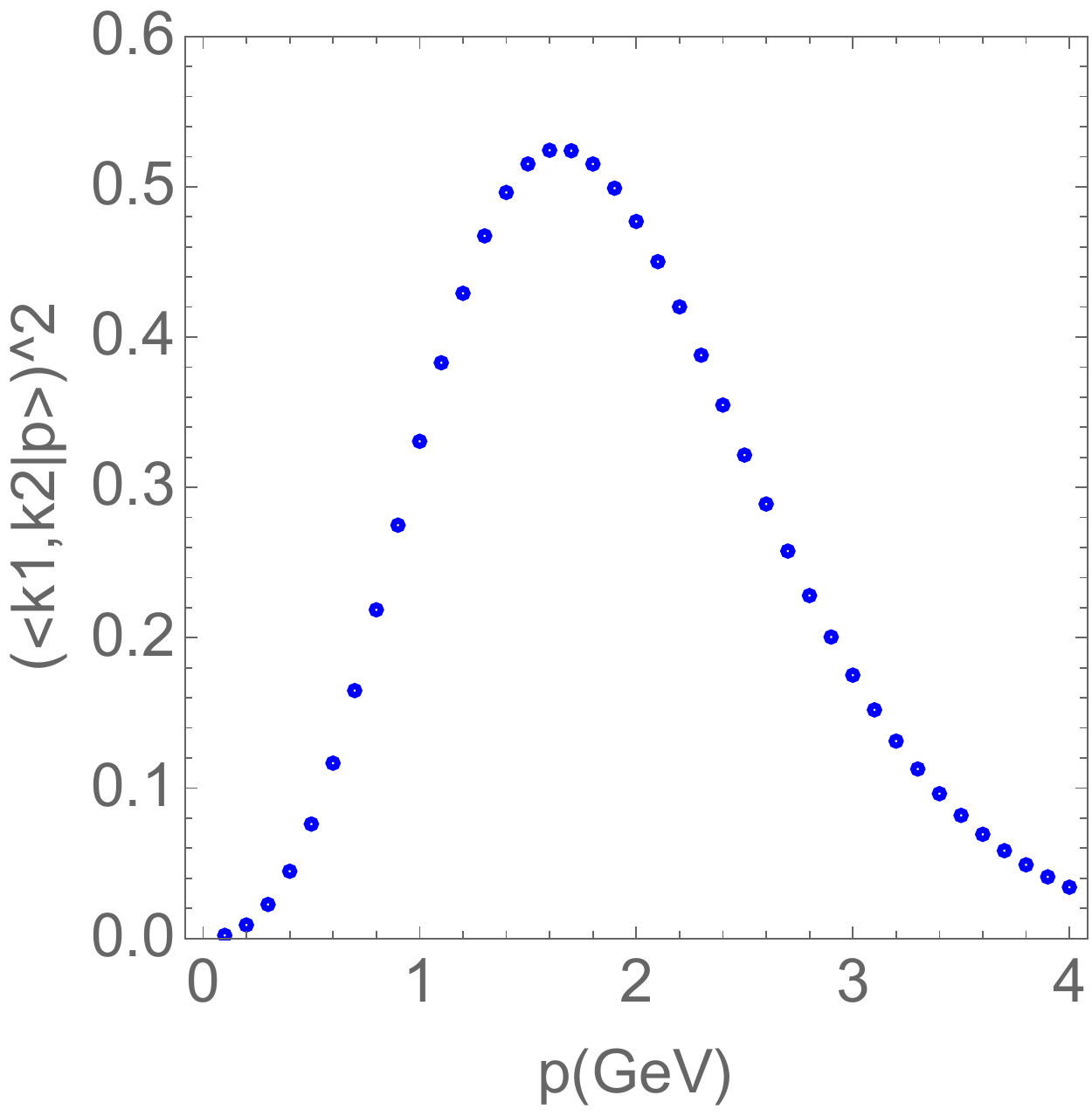}
	\caption{Upper plot: kinematics for two emerging  quarks with momenta $\vec k_{1,2}$ from an exploding sphaleron,  fusing into a meson of momentum $\vec p$; 
	Lower plot: squared projection of outgoing quark waves into a mesonic distribution amplitude with momentum $p \, ({\rm GeV})$.}
	\label{fig:projsquared}
\end{figure}

The ensuing  explosion starting from the sphaleron at $t=0$, we describe using  the YM equations. This is justified only if the sphaleron size is sufficiently small
$\rho \ll \rho_{\rm max}$ compared to the maximal size
defined by the vacuum effects.
 If $\rho v\sim 1$ (where $v$ is  the Higgs VEV in the electroweak theory,  or the dual Higgs VEV in QCD), the equations get
modified by additional terms with scalar fields. 

It is known for a long time, that the Adler-Bell-Jackiw 
chiral anomaly relates the divergence
of the axial current to the topological charge $Q$.
Also divergence of the Chern-Simons topological current is proportional to it. As a result, their combination is conserved, and changes in the axial charge and Chern-Simons number are locked
\begin{equation} \Delta Q_5 = (2N_f) \Delta N_{CS} \end{equation}
where $N_f=3$ is the number of light quark flavors.	

In agreement with these general arguments, we  have constructed in~\cite{Shuryak:2002qz}
the  analytical solutions of the Dirac equation in the field of exploding sphalerons. As a function of time, one can see how  quarks are accelerated from the lower Dirac sea to the physical 
continuum of positive energies.
The spectrum of outgoing quarks has a simple form
\begin{equation} \label{eqn_psi(k)}
	{dN \over dk}= 4\pi k^2| \psi(k)|^2=\rho (2 k \rho)^2 e^{-2k\rho}
\end{equation}	
which is  similar to a Planckian spectrum with an effective temperature $T_{\rm eff}=1/(2\rho)$. For the maximum value set by the vacuum instanton size 
distribution $\rho_{\rm max}\approx 0.3\, {\rm fm} \approx 1/(600\, {\rm MeV})$, the effective temperature is $300\, {\rm MeV}$. The mean momentum is 
$$ \langle k \rangle ={3 \over 2 \rho}\sim 900\, {\rm MeV}$$


A simple quantum estimate of the production amplitude, follows from the projection
of the emerging  quarks onto the distribution amplitudes (DAs) of the outgoing mesons
at any momenta as illustrated in Fig.\ref{fig:projsquared} (upper).  More specifically, we have

\begin{equation}
\langle \vec k_1, \vec k_2 | \vec p \rangle \sim 	\int d k_{1\perp}^2 d k_{1l} d k_{1\perp}^2 d k_{1l} \psi(\vec k_1) \psi(\vec k_2) f(p_\perp^2)
\varphi(x=k_{1l}/P)\delta(k_{1\perp}^2-p_\perp^2) \delta(k_{2\perp}^2-p_\perp^2)\delta(k_{1l}+k_2l-P)
\end{equation}
where $\psi(k)$ is the outgoing quark wave (\ref{eqn_psi(k)}), and the functions $\varphi(x)f(p_\perp^2)$ 
are the DAs of the corresponding
mesons with  longitudinal fraction ($x$) and transverse ($p_\perp$) momenta. For simplicity, we take  a Gaussian $f(p_\perp)\sim {\rm exp}(-{\rm const}*p_\perp^2)$ and
a flat $\varphi(x)=1$ which approximate well say a pion. The squared projection
for a sphaleron of size $\rho=0.3\, {\rm fm} =1/(0.6\, {\rm GeV})$
is shown in the  lower plot, as a function of the outgoing meson momentum $p_\perp$. 

(Note that in this estimate we  
ignored all the other particles produced.
In reality the total mass of the cluster puts its own kinematical 
restrictions. For example, for three-meson decay modes to be discussed, the tail at large momenta
is cutoff above $M/3$.)

\section{  The instanton size distribution  in the QCD vacuum} \label{sec_size_distribution}
\begin{figure}[t]
\begin{center}
\includegraphics[width=8cm]{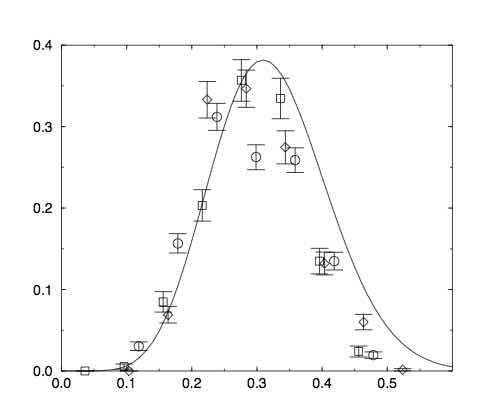}
\includegraphics[width=8cm]{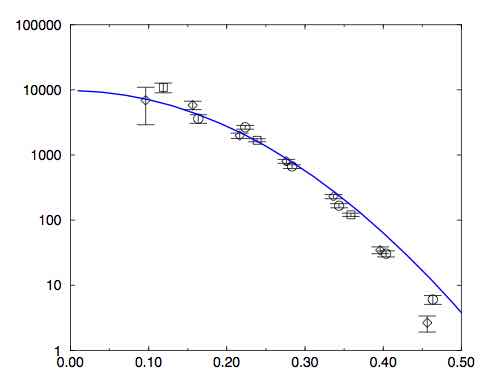}
\caption{(left) The instanton size $\rho$ [fm]
distribution  $dn/(d\rho d^4z)$. (right) The combination $\rho^{-6}dn/d\rho d^4z$, in which the main one-loop behavior drops
 out for $N_c = 3,N_f = 0$. 
 The points are from the lattice work 
 for this theory, with $\beta$=5.85 (diamonds), 6.0 (squares) and 6.1 (circles). Their comparison should demonstrate that results are lattice-independent. The line corresponds to the proposed expression, see text.
 }
\label{fig_inst_sizes}
\end{center}
\end{figure}

By now, the subject of  instantons in the QCD vacuum is  well established and broad, and clearly goes 
beyond the scope of this review. For us, the only relevant issue 
is the
	  instanton size distribution $dn/d\rho$ in the vacuum. 
	  It has been evaluated  in various models and on the lattice. For definiteness we use the lattice results from
	    \cite{Hasenfratz:1999ng}. The average size 
was found to be $\langle \rho \rangle\approx 0 .30 \pm 0.01\, {\rm fm}$,  a bit smaller than in the ILM.
The mean distance was found instead to be $0.61\pm 0.02 \, fm$. 
The data on the instanton size distribution 
  are shown in Fig.\ref{fig_inst_sizes}.
(The figure is taken from~\cite{Shuryak:1999fe} and  the lattice data from
Hasenfratz et al~\cite{Hasenfratz:1999ng}). The left plot shows the 
size distribution itself. Recall that the semiclassical theory predicts it to be
$dn/d\rho\sim \rho^{b-5}$ at small sizes, with $b=11N_c/3=11$ for pure gauge $N_c=3$  theory. 
The right plot -- in which this power is taken out --is constant at small $\rho$, which agrees with
the semiclassical prediction.

The other feature is a peak at $\rho\approx 0.3 \, {\rm fm}$ -- the value first proposed phenomenologically
in \cite{Shuryak:1981ff}, decades before the lattice data. 
The peak is due to a suppression at large sizes.
 Trying to understand its origin, one may factor out all known effects. The right plot shows that 
 after this is done, a
rather simple suppression factor $\sim {\rm exp}(-{\rm const}*\rho^2)$ describes it well, for about 3 decades.
What is the physical origin of this suppression?

There are two answers to that question, which are perhaps ``Poisson dual" to each other~\cite{Ramamurti:2018evz}.
 The first is that it is due to the mutual repulsion between  an instanton and the rest
 of the instanton-antiinstanton ensemble. (It is described in the mean field approximation and numerically, see the review~\cite{Schafer:1996wv}).
 
Another one, proposed in~\cite{Shuryak:1999fe},   is that the coefficient is proportional to
the dual magnetic condensate, that of Bose-condensed monopoles. It has been further argued there that
it can be related to the string tension $\sigma$, so that the suppression factor should be 

\be 
{dn \over d\rho}= \bigg({dn \over d\rho}\bigg)_{\rm semiclassical} exp[-2\pi \sigma \rho^2]
\ee
where the Higgs VEV is traded for the string tension $\sigma$ via the dual Higgs model of confinement.
If this idea is correct, this suppression factor should be missing at $T>T_c$, in which the dual magnetic condensate
is absent. However, in this regime, quantum/thermal fluctuations generate at high $T$ a similar factor~\cite{Pisarski:1980md} 
\be  {dn \over d\rho}= \bigg({dn \over d\rho}\bigg)_{T=0} exp[- (2N_c+{N_f \over 3}) (\pi \rho T)^2]\ee
 related to the scattering of the quarks and gluons of the quark-gluon-plasma (QGP) on the instanton~\cite{Shuryak:1994ay}. 
Empirically, the suppression factor at all temperatures looks Gaussian in $\rho$, interpolating between those limiting 
regimes.

\section{Quark flavor signatures of instanton/sphaleron process at small  $M=3-10\, {\rm GeV}$}

\subsection{The instanton-induced  glueball and $\eta_c$ decays}
As emphasized earlier, the instantons   are virtual paths at zero
energy, playing a significant role in  the theory of the QCD vacuum, for a review see e.g.
\cite{Schafer:1996wv}. Their existence has been unambiguously  revealed by the 
cooled lattice simulations carried by many groups,  e.g. Leinweber~\cite{Leinweber:1999cw}.

The subject of this work, the instanton/sphaleron process, requires
external forces able to provide the amount of energy needed for the sphaleron production.
Furthermore,  one should look for reactions involving gluons in  combinations 
to which 
 instantons are naturally coupled. Those are in scalar combinations or $G_{\mu\nu}^2$
and pseudoscalar combinations or $G_{\alpha\beta}G_{\gamma\delta}\epsilon^{\alpha\beta\gamma\delta}$.
In order of increasing mass, they are: \\
(i) decays of a scalar glueball ($M\approx 1.6\, {\rm GeV}$);\\
(ii) decays of a pseudoscalar glueball ($M\approx 2.5\, {\rm GeV}$);\\
(iii) decays of pseudoscalar quarkonia\footnote{We are not aware of any study of $\eta_b$ decays
in this resepect.}
 ($M_{\eta_c}\approx 3\, {\rm GeV}$)\\ 
 
 The main idea of this work is that gluon-gluon or (especially) Pomeron-Pomeron collisions
 also couple to such operators by fusion, and through them one can possibly investigate
 instanton/sphaleron processes of larger masses $M_{\rm sph}$ at hadron colliders. 

Historically, the issue of instanton-induced decays
 was first noticed by Bjorken \cite{Bjorken:2000ni}. He pointed out  
that  $\eta_c$ decays have  3 large 3-body modes, several percents each
of the total width: $$\eta_c \rightarrow KK\pi;\,\,\, \pi\pi\eta;\,\,\, \pi\pi\eta'$$ Note that there is
no $\pi\pi\pi$ decay mode, or other decay modes we may think of without strangeness:  the
't Hooft vertex $must$ have all light quark flavors in it, including  $\bar{s}s$.

More generally, charmonium decays are especially nice since one can use the well known $J/\psi$
decays as a $^{\prime\prime}$control group$^{\prime\prime}$, since its three-gluon annihilation mode  does not go to operators 
that couple naturally to 
instantons. Indeed, the branchings of the 2- and 3-body decays of $J/\psi$ are much smaller, and
 the average multiplicity of its decays is much larger than 3. That is why the few-percent branching
of these 3-meson $\eta_c$ decays is remarkable by itself.

The actual calculations of $ \pi\pi, KK,\eta\eta$ decays  of scalar 
and  $KK\pi, \eta\pi\pi, \eta'\pi\pi $ decays of $\eta_c$
were made by Zetocha and Schafer \cite{Zetocha:2002as}.
 Their results contain rather high power of the instanton radius and therefore strongly depend
on its value. So the authors used the inverted logic, evaluating from each data point the
corresponding value of the mean instanton size $\bar\rho$.
The results 
reproduced the decay widths reasonably well.
Furthermore, these calculations provided about the most accurate
evaluation of the average instanton size available,  in the range of $\bar\rho=0.29-0.30\, fm$,
common to all decay channels.

Let us start with the flavor composition. The specific form of the effective operator between light quarks induced by instantons has been derived by 't Hooft
$$ (\bar u_R u_L)  (\bar d_R d_L)  (\bar s_R s_L) + (L\leftrightarrow R)$$
with the coefficients following from the LSZ reduction of the zero modes.
For the
the lightest quark clusters, of mass $M\sim 3\, {\rm GeV}$,
we expect the production of 6 quarks in combination $uu dd ss $.
Before discussing the decay modes, we need to look closer at $N_f=3$.

\subsection{The simplified forms of the effective Lagrangian } \label{sec_Hooft}

One issue is that in~\cite{Zetocha:2002as} the emitted mesons are treated in  the  ``local approximation",
with the vertex Lagrangian directly projected to meson ``decay constants" (which are the values of the wave functions at the origin). In reality, mesons fly with a momentum of about $1\, {\rm GeV}$, 
for which the projections from the initial quasi-local state to the final state  (the form factors) are not exactly one. This issue
of ``projection"  (which we have already addressed above) 
gets more important as one considers sphalerons with larger masses.

Another (more technical) issue is the 
correct inclusion of all diagrams. 
Whatever
form chosen for the 6-q Lagrangian,
there are $3!=6$ channels (various connections of quarks and antiquarks), and they all needs to be 
included. While it is possible to go from one form to 
to another with Fierz transformation of operators,
we were not sure all 6 connections were included on equal footing, and decided to repeat the calculation.

Another reason for redoing it is that  the
 3-flavor version of the Lagrangian  used in~\cite{Zetocha:2002as} (their (26))
is very complicated, containing
expressions with three color matrices and 
structure constants  of the $SU(3)$ group
 $$f^{abc}\lambda^a\lambda^b\lambda^c, \,\,\,\,d^{abc}\lambda^a\lambda^b\lambda^c $$
As we will show, these can be eliminated in favor of
a simpler form, which we use. (In  \ref{app_thooft} we will
also show one more generic form of this Lagrangian
based on the so called Weingarten coefficients of the
product of unitary matrices averaged over the groups.)

We start by explaining the two main technical complications   in this
problem. A single instanton (or sphaleron) is constructed in a color $SU(2)$ setting,
with the gauge fields $A^a_\mu T^a_{ij}$ with color indices $i,j=1,2$. The quark zero modes
$\psi_\alpha^j$  in the chiral representation ($\gamma_5$ is diagonal) carry a  matching spin index $\alpha=1,2$ and 
color index $j=1,2$, coupled to each other in a ``hedgehog" way. 

However,  the gauge group in  QCD is $SU(3)$ and in it, the instanton   appears
rotated from its standard form (just described) to some arbitrary
 2d plane  $\in SU(3)$  by a 
unitary $3\times 3$ matrix, e.g. $\psi^a_\alpha \rightarrow U^{a}_{i}\psi^{i}_\alpha$. The
't Hooft effective Lagrangian contains $2N_f=6$ (or 8 or 10) quark zero modes,
and therefore the 6-th (or  8-th or 10-th) power of this matrix. Although the
explicit expressions for ``isotropic" averaging of these powers of $U$ have been known for some time
(see     e.g. appendix of \cite{hep-ph/9510326}) these formulae are rather complicated, and contain convolutions with the structure constants $f^{abc}, d^{abc}$
of $SU(3)$. Schafer and Zetocha~\cite{Zetocha:2002as}  used such form, see eqn(26) of their paper.

Another technical issue is as follows. Multifermion operators can be identically 
represented in many different forms, since one can color-couple different 
$\bar q q$ pairs. (Or even $qq$ pairs as is convenient for color superconductivity and baryon production.)   Those ``Fierzing" transformations are jus identities, but
with flavor, color and Dirac indices involved, they can create a multitude of operators. 
Yet, whatever form of the vertex operator is used, one still need to 
include it in all possible channels. For example, for the operator  structure $\bar q \bar q \bar q q q q$ projected onto 3 mesons, there are obviously $3!=6$ ways to
relate it to three mesonic $\bar q q$ wave functions.
For 8-q operators there are $4!$, etc. 

Here and in Appendix~\ref{app_fierz} we show how one can simplify the operator 
structure, by keeping its flavor determinantal form intact, which has no
structure constants as quoted in~\cite{Nowak:1996aj},  and used for heavy-light  multi-quark systems~\cite{Chernyshev:1995gj}. We then explicitly do all possible convolutions with 
all mesonic wave functions in Mathematica, avoiding Fierzing altogether.   More specifically~\cite{Nowak:1996aj} (See eq. 2.56)

%

\begin{align}
\label{DET1}
{\cal V}^{L+R}_{qqq}=&\frac {\kappa}{N_c(N_c^2-1)}
\bigg(\frac{2N_c+1}{2(N_c+2)}\,{\rm det}(UDS)\nonumber\\
&+\frac 1{8(N_c+1)}\,
\bigg({\rm det}(U_{\mu\nu}D_{\mu\nu}S)+{\rm det}(U_{\mu\nu}DS_{\mu\nu})+{\rm det}(UD_{\mu\nu}S_{\mu\nu})\bigg)\bigg) +(L\leftrightarrow R)
\end{align}
with a strength

\be
\label{DENS}
\kappa=\frac{{n_{I+\bar I}}}2\bigg(\frac{4\pi^2\rho^3}{M\rho}\bigg)^3
\ee
and the short hand notations   ($Q\equiv U,D,S$)

\be
\label{DET2}
Q=\overline q_Rq_L\qquad Q_{\mu\nu} =\overline q_R\sigma_{\mu\nu}q_L\qquad Q^a=\overline q_R\sigma^aq_L
\ee
with $\sigma_{\mu\nu}=\frac 12[\gamma_\mu, \gamma_\nu]$. Note that the spin contribution is sizably
suppressed by $1/8N_c$ in the large $N_c$ limit,
when compared to the scalar contribution.
It is clear from our Fierzing arguments  in Appendix~\ref{app_fierz} and symmetry,  that  only two determinantal 
invariants  will survive after Fierzing as in  (\ref{DET1}), with only the structures $UDS$ and $U_{\mu\nu}D_{\mu\nu}S$ and their permutations allowed.
The only non-trivial results are their weight coefficients. This observation  holds 
for  4, 8 and 10 quark vertices as well, assuming they allow for zero modes. 
 For instance, for 4-quark vertices the general structure is 

\bea
\label{DET1-4}
{\cal V}^{L+R}_{qq}=\,{\kappa}_2\,A_{2N}\,
\bigg({\rm det}(UD)+B_{2N}\,{\rm det}(U_{\mu\nu}D_{\mu\nu})\bigg)+(L\leftrightarrow R)
\eea
which is readily  checked to hold with 

\bea
A_{2N}=\frac{(2N_c-1)}{2N_c(N_c^2-1)}\qquad B_{2N}=\frac 1{4(2N_c-1)}
\eea
Note that at large $N_c$, the suppression of the spin contribution  is still exactly $1/8N_c$, with  $A_{\#q}\approx 1/N_c^\#$.
Remarkably,  both the value of the $A,B$ coefficients and their determinantal structures are fixed uniquely in this limit by symmetry and scaling.
For completeness, the  8-quark vertices are  of the form 

\bea
\label{DET1-8}
{\cal V}^{L+R}_{qqqq}={\kappa}_4\,A_{4N}&&
\bigg({\rm det}(UDSC)+B_{4N}\,\bigg({\rm det}(U_{\mu\nu}D_{\mu\nu}SC)+\,{\rm perm.}\bigg)\nonumber\\
&&+C_{4N}\bigg({\rm det}(U_{\mu\nu}D_{\mu\nu}S_{\alpha\beta}C_{\alpha\beta})+\,{\rm perm.}\bigg)\bigg)
+(L\leftrightarrow R)
\eea
although for the heavy charm the use of the $L,R$ zero modes may not be justified.

In the Weyl basis 
$\sigma_{\mu\nu}\rightarrow  i\eta^a_{\mu\nu}\sigma^a$ with the $^\prime$t Hooft symbol satisfying $\eta^a_{\mu\nu}\eta^b_{\mu\nu}=4\delta^{ab}$, and (\ref{DET1})
can be simplified

\begin{align}
\label{DET3}
{\cal V}^{L+R}_{qqq}=&\frac {\kappa}{N_c(N_c^2-1)}
\bigg(\frac{2N_c+1}{2(N_c+2)}\,{\rm det}(UDS)\nonumber\\
&-\frac 1{2(N_c+1)}\,
\bigg({\rm det}(U^aD^aS)+{\rm det}(U^aDS^a)+{\rm det}(UD^aS^a)\bigg)\bigg)+(L\leftrightarrow R)
\end{align}
${\cal V}_{qqq}$ is only active in flavor singlet 6-quark states. The flavor determinantal interactions can be made more explicit by using the permutation 
operators in flavor space as the symmetric group $S_3$ of permutations is composed of $3!$ terms only, 3 cyclic-permutations with positive signature and 3 pair-permutations with negative signature.
Clearly, the 3-body instanton induced interaction (\ref{DET3}) does not vanish only in flavor singlet $uds$ states (repulsive).   
Its 2-body reduction is attractive in states with  a pair of antisymmetric 
flavor diquarks $ud$, $us$, $ds$ (attractive). A more  explicit writing of (\ref{DET3}) suitable for numerical analysis in terms of explicit $3\times 3$ flavor determinants is

\begin{align} 
\label{DET5}
&{\cal V}^{L+R}_{qqq}=\frac {\kappa}{N_c(N_c^2-1)}\bigg[
\bigg(\frac{2N_c+1}{2(N_c+2)}\bigg)
\begin{Vmatrix}
\overline u_Ru_L& \overline u_Rd_L & \overline u_Rs_L\\
\overline d_Ru_L & \overline d_Rd_L & \overline d_Rs_L\\
\overline s_Ru_L & \overline s_Rd_L & \overline s_Rs_L\\
\end{Vmatrix}
\nonumber\\
&-\frac 1{2(N_c+1)}\sum_{a=1}^3
\bigg(
\begin{Vmatrix}
\overline u_R\sigma^au_L& \overline u_R\sigma^ad_L & \overline u_Rs_L\\
\overline d_R\sigma^au_L & \overline d_R\sigma^ad_L & \overline d_Rs_L\\
\overline s_Ru_L & \overline s_Rd_L & \overline s_Rs_L\\
\end{Vmatrix}
+\begin{Vmatrix}
\overline u_R\sigma^au_L& \overline u_Rd_L & \overline u_R\sigma^as_L\\
\overline d_Ru_L & \overline d_Rd_L & \overline d_Rs_L\\
\overline s_R\sigma^au_L & \overline s_Rd_L & \overline s_R\sigma^as_L\\
\end{Vmatrix}
+\begin{Vmatrix}
\overline u_Ru_L& \overline u_Rd_L & \overline u_Rs_L\\
\overline d_Ru_L & \overline d_R\sigma^ad_L & \overline d_R\sigma^as_L\\
\overline s_Ru_L & \overline s_R\sigma^ad_L & \overline s_R\sigma^as_L\\
\end{Vmatrix}\bigg)\bigg]
\nonumber\\&+(L\leftrightarrow R)
\end{align}


\subsection{Mesonic decays of sphalerons}
Convoluting the vertex function in  the  form (\ref{DET5})
with various mesons wave functions
in all possible $3!=6$ ways
 we obtain the matrix elements for a number of 3-meson decay channels, listed in the Table. The meson definitions and couplings are defined in Appendix A.

\begin{table}[h!]
\begin{tabular}{|c|c|c|c|c|}
	\hline
	&  & PDG2020 & input \cite{Zetocha:2002as}  & |M|  \\
	\hline
$K\bar K \pi$	& $K^+K^-\pi^0$ &  $7.3\pm 0.4$ (all 4)  & $5.5\pm 1.7$ (all 4) &   5.07 $K_K^2 K_\pi$ \\
	\hline
	& $K^+ \bar K^0\pi^-$ &  &  &   7.27 $K_K^2 K_\pi$\\
	\hline
	&  $K^0 \bar K^0\pi^0$  &  &  &   5.07$K_K^2 K_\pi$ \\
	\hline
	&  $K^-  K^0\pi^+$  &  &  &   7.27$K_K^2 K_\pi$ \\
	\hline
$\pi\pi\eta$	& $\pi^+ \pi^-\eta$ &  $1.7\pm 0.5$  & $ 4.9\pm 1.8$  (both)&   4.92 $K_\pi^2 K_\eta^s $ \\
	\hline
	& $\pi^0 \pi^0\eta$   &  &  &   2.46$K_\pi^2 K_{\eta'}^s $\\
	\hline
$\pi\pi\eta'$	& $\pi^+ \pi^-\eta'$  &   $4.1\pm 1.7$ (both)& $4.1\pm 1.7$ (both) &   5.20 $K_\pi^2 K_{\eta'}^s$ \\
	\hline
	& $\pi^0 \pi^0\eta$  &  &  &   2.60$K_\pi^2 K_{\eta'}^s $\\
	\hline
$\bar K K \eta$	& $K^+K^-\eta$ & $1.36\pm 0.15$ (both)&  &   3.68 $K_K^2 F_\eta^q $\\
	\hline
 	& $K^0 \bar K^0\eta$ &  &  &   3.68 $K_K^2 F_\eta^q $\\
 $\bar K K \eta'$	& $ K^+K^- \eta'$ &  &  &   3.53  $K_K^2 F_{\eta'}^q $\\
	\hline
	& $ K^0 \bar K^0 \eta'$ &  &  &    3.53 $K_K^2 F_{\eta'}^q$ \\
	\hline
$\eta \eta \eta$	&  &  &  &   1.32$(K_\eta^q)^2 K_\eta^s $\\
\hline	
\end{tabular}
\caption{The first column gives the generic names of the decay channels of $\eta_c$,  while the second column records the specific channels. The third column contains the corresponding branching ratio (percents) according to the Particle Data Table 2020. For comparison, we show in the fourth column the corresponding numbers used in~\cite{Zetocha:2002as}. The last column gives the decay matrix elements. The meson-specific constants (wave function at the origin) are defined in Appendix A. 
}
\end{table}

Our first comment on the table is that in that two decades from the work in~\cite{Zetocha:2002as},  some of the 
experimental branching ratios have improved their accuracy, while some are  substantially modified.
This needs to be kept in mind when  comparing the  predictions to experiment.

Of course, we can construct many ratios out of the Table. Here, we will mention two in particular

\begin{eqnarray}
	{\Gamma(K \bar K\pi) \over \Gamma(\eta\pi\pi)}&=&{\sum | M|^2\over \sum | M|^2 }{0.111\over 0.135 }\approx 10\,\,\,\,\,\,\, \bigg({\rm exp}: {7.3 \pm 0.4 \over 2.55\pm 0.75 }\bigg)\\
	{\Gamma(\eta\pi\pi) \over \Gamma(\eta'\pi\pi)}&=&{\sum | M|^2\over \sum | M|^2 }{0.135\over 0.0893 } \approx 0.9\,\,\, \bigg({\rm exp}: {  2.55\pm 0.75  \over 4.1\pm 1.7 } \bigg) \nonumber
\end{eqnarray}
where the last ratio corresponds to the three-body phase space.  The ratios are only in qualitative agreement with the 
reported measurements. Clearly more studies are needed.

\subsection{Baryonic decays of sphalerons }

 Another fascinating issue, not so far discussed in the literature, is 
 whether the 6-fermion effective Lagrangian can be used to produce a {\em baryon-antibaryon} pair, rather than 3 mesons.
 
At first sight one finds a problem which looks quite severe. The operator  involves the quark set $u,d,s$,
 and the natural baryons to look at are $\Lambda$ or
 $\Sigma^0$ baryons. As for any baryon, their color wave function is antisymmetric in color, 
 $\epsilon_{c_1c_2c_3}$.  The flavor determinant is also  antisymmetric in flavor $\epsilon_{f_1f_2f_3}$. Fermi statistics then 
 require the remaining part of the wave function to  be
 antisymmetric. In the lowest shell of the quark model, 
 this remaining part is made of three quark spins.
It is not possible to create $\epsilon_{s_1s_2s_3}$ spin wave functions,  since the spin indices are $1,2$ only. 
Indeed it is well known, that the lowest baryonic octet does not have the ninth state, a singlet, unlike the mesons. 

This notwithstanding, let us also approach the issue phenomenologically. we assume, following Bjorken, that $\eta_c$ decays possess
topological paths, while $J/\psi$ decays do not. So, what do we see in the baryonic sector?

Both $\eta_c$ and  $J/\psi$ have observed decays into 
$\bar{\Lambda}\Lambda$, with similar branching $\sim 10^{-3}$. Yet their absolute widths differ by roughly a factor of 200, in favor of the former case. What speaks against the topological mechanism is the fact that in both cases the channels $\bar p p $ and $\Sigma^+\Sigma^-$ have similar yields to $\bar{\Lambda}\Lambda$, although they cannot follow from our 6-fermion effective vertex. 

In summary, topological production of $\bar{\Lambda}\Lambda$ does not appear to take place
in $\eta_c$ decays.

Yet, if the sphaleron mass is larger than 3 GeV,
one may think of production of two higher mass baryon resonances. Among those there are known $SU(3)$ singlets, which are anti-symmetric in flavor. The lowest of them 
are 
$$
 \Lambda(1405)\,\,:\,\,J^P=1/2^-, \qquad  \Lambda(1520)\,\,:\,\,J^P=3/2^- $$
Their negative parity is explained in the quark model by a unit of 
orbital momentum $L=1$. Unfortunately, this negative parity also protect them from any mixing with the usual $\Lambda$. 
As they are well known resonances, their back-to-back production should be quite noticeable.  Their widths are $\Gamma=50$ and $ 157 \, {\rm MeV}$, respectively. They
both have very characteristic decays, 
$$ \Lambda(1405)\rightarrow \Lambda \gamma, \,\,\, 
\Lambda(1520) \rightarrow  \Sigma \pi $$
with branching ratios of the order of 50\% for both.

\subsection{Chirality correlation in baryonic decays}
The $\Lambda \bar\Lambda$ channel is the most interesting for two reasons:

\indent {(i) Zero isospin means that $(ud)$ diquark has spin zero, and therefore the whole spin of $\Lambda$ is
carried by its strange quark;}\\
\indent {(ii) weak decays of $\Lambda$ hyperon allows to observe its polarization. Indeed, in the decay
$\Lambda \rightarrow p \pi^-$ of polarized hyperon, the direction of the proton is mostly along the initial polarization  direction.}

As a measure of $s$ quark chirality one can use
\begin{equation} \xi_{\Lambda}\equiv {\rm cos}\big( \theta(\vec p_{\Lambda} \vec p_p) \big)
\end{equation}
and calculate the distributions  $P(\chi_{ \Lambda \bar \Lambda})$ over the {\em relative chiralities}, the product of  
\begin{equation} \chi_{ \Lambda \bar \Lambda}\equiv    \xi_{\Lambda} \times \xi_{\bar \Lambda} 
\end{equation}

Ordinary perturbative diagrams with one or two gluons (or photons) leading to the production of a strange quark
pair are ``vector-like", meaning that  the
chiralities are the same, $$(\bar s_L s_L)+
 (\bar s_R s_R)$$
This means either both $\xi$ are positive, or both negative, leading to $ \chi_{ \Lambda \bar \Lambda}$ positive.

On the other hand, the
instanton/sphaleron-induced vertex is, 
$non-diagonal$ in chirality
$$ (\bar s_L s_R)+  (\bar s_R s_L) $$
 The produced 
$\Lambda$ and $ \bar\Lambda$ should therefore  have the $opposite$ chiralities, and $ \chi_{ \Lambda \bar \Lambda}$ is negative.
We are not aware of such study even in any inclusive 
reactions, in which a pair of Lambda hyperon decays
are identified with some reasonable statistics. 

As we discussed above, the exclusive production of $\bar \Lambda \Lambda$ from the  t' Hooft-like Lagrangian is not possible. Yet strong chirality correlations in question would perhaps persist in channels with other 
associate hadrons. For example,  in the production of 
 $\bar \Lambda(1405) \Lambda(1405)$ with their subsequent radiative decays into $\Lambda+\gamma$, there should remain rather
 significant correlation of polarizations.

If observed,
it would be an excellent indication of the  topological origin of the vertex, pointing to an  explicit
violation of the $U_A(1)$ symmetry.

\section{Sphaleron decays at medium masses:\\ \qquad $M=10-20\, {\rm GeV}$ and 10-fermion operators}

\subsection{Charm pairs and decays with 8-fermion operators}

The field magnitude at the center of the instantons
is comparable to $m_c^2$. Although charm is not
usually treated as a  light flavor, it must have a certain
coupling to instantons. The  fact of large instanton-induced decays of $\eta_c$  confirms this idea. 

The 8-flavor operators have also a flavor-asymmetric structure  $$V \sim (\bar u u) (\bar d d)(\bar s s)(\bar c c)$$
with a typical determinantal structure .
There are $4!=24$ 4-meson channels one should think of: for example those can be
the same 3-meson channels discussed above, $\pi\pi \eta, \pi\pi \eta', KK\pi$ with an $added$ $\eta_c$.
The other quite distinct channels are e.g. $KKDD$ $without$ any pions. Note also that 4 pseudoscalar
mesons without orbital momentum correspond to operator $G^2$, not the pseudoscalar one.

\subsection{Multi-glueball production}
We already mentioned several times, that the total
sphaleron mass is ${\cal O}(3\pi^2/g^2\rho)$ while the energy of
produced gluons are only $O(1/\rho$) . The gauge quanta multiplicity is therefore $\sim 1/g^2\sim {\cal O}(100)$ in the electroweak theory and  ${\cal O}(10)$ in QCD. 

Therefore, in the ``medium sphaleron mass" range indicated, the energy per gluon is in the range $E_g=1-2\, {\rm GeV}$. We know, from lattice studies 
and models, that under such conditions gluons are
not yet ``partons" with independent life. Instead,
they are paired into the lowest glueball states
with masses 
\begin{equation}
	M_{0^+}\approx 1.6\, {\rm GeV}\,\qquad 	M_{2^+}\approx 2.2\, {\rm GeV} 
\end{equation}
  So, we propose that in this mass range a signficant part of the energy will go into several lowest glueballs. All of them are now reasonably well
identified with scalar and tensor resonances.

\subsection{Semiclassical production in constant electric field and in sphaleron explosion}
\label{sec_c_and_b}

Another new interesting option is the production
of $\bar{b} b$ pair  by sphaleron decays.

Before we provide some estimates, let us  recall 
 how the related classic problem {\em of the
	 pair production of $e^+e^-$ in a constant electric field} can be described semiclassically. It is widely known as the Schwinger process,
as he solved it in detail in 1950. However, we will discuss neither the Schwinger paper, nor
even the earlier Heisenberg-Euler paper, but the much earlier semiclassical work \cite{Sauter:1931zz}
from 1931 (well before anyone else).  

The EOM of a charge relativistically moving in constant electric field is a classic
problem which everybody had encounter in E/M classes. Writing it in the usual Minkowski form
\begin{equation} {dp\over dt}={d \over dt}\bigg( {v \over \sqrt{1-v^2} } \bigg)={eE\over m}\equiv a \end{equation}
yields the hyperbolic solution

\begin{equation} v_M(t)={a t \over \sqrt{1+a^2 t^2}},\,\,\,\, x_M(t)={1\over a}\big( \sqrt{1-a^2t^2}-1\big) \end{equation}
for a particle that starts at rest with a
nonrelativistic acceleration, $x_m=at^2/2$.  At large times,  the motion turns   ultrarelativistic with 
$v_M \rightarrow 1$.

The analytical continuation of the the trajectory to Euclidean time $\tau=it$ yields 

\begin{equation}v_E={a\tau \over \sqrt{1-a^2\tau^2}},\,\,\,
	 x_E(\tau)={1\over a} \big( \sqrt{1+a^2\tau^2}-1 \big) 
\end{equation}
At time $\tau=0$ the particle is at the origin
with zero velocity, $x_E=0,v_E=0$.
For  $-1/a<\tau<1/a$ it describes an Euclidean path in shape of  a  semicircle. In  Euclidean space, the world-time is no different from
the other coordinates, and the electric field $G_{01}$ is not different from the magnetic field, so in the 0-1 plane
the paths are circles, like they are in all other planes.


To understand the physical meaning of the semicircle (blue dashed line) in the Dirac sea
interpretation
 one needs to split it into two halves.
The path
 describes tunneling through the ``mass gap",
 between energies $-M$ and $M$ 
  in the spectrum of states.
In Minkowski space   there
are no states
between $E=-\sqrt{p^2+m^2}$ and $E=\sqrt{p^2+m^2}$ with real momentum $p$,
   In the Euclidean
  world however, 
the momentum is imaginary! 

Let us note that the rapidity $y_M={\rm arctanh}(v_M)$
in Minkowski space, becomes in Euclidean space  a rotation angle 
\begin{equation} y_E={\rm arctan}(v_E) ={\rm arctan}\bigg({a\tau \over \sqrt{1-a^2\tau^2}}\bigg)\end{equation} 
and at $\tau=\pm 1/a$ it is $\pm \pi/2$. The action
$ S=\int (-m ds-eEx dt) $ when evaluated with the Euclidean path gives $S_E=\pi m^2/2eE$. The semiclassical probability (square of the amplitude) of
the pair production is then
\begin{equation} P\sim e^{-2S_E}\sim e^{- {\pi m^2}/{eE}} \end{equation}

	 
	 When discussing the heavy quark production we will assume that the back reaction of the produced quark on the gauge field can be neglected. 
	 This implies that the total sphaleron mass is
	 much larger than that of the produced pair
	 \begin{equation}
	 	{3\pi^2 \over g^2(\rho) \rho} \gg 2M
	 \end{equation}
	 For $\rho \sim 1/3 \,{\rm fm}$, corresponding to the
	  maximum in the  instanton size distribution,
	 the l.h.s. is about $M_{\rm sph}\approx 3\,{\rm  GeV}$.
	 Therefore, the condition is satisfied for the  strange quark pair $\bar s s$ but not for the charm quark pair $\bar{c} c$.
	 
\begin{figure}
	 	\centering
	 	\includegraphics[width=0.7\linewidth]{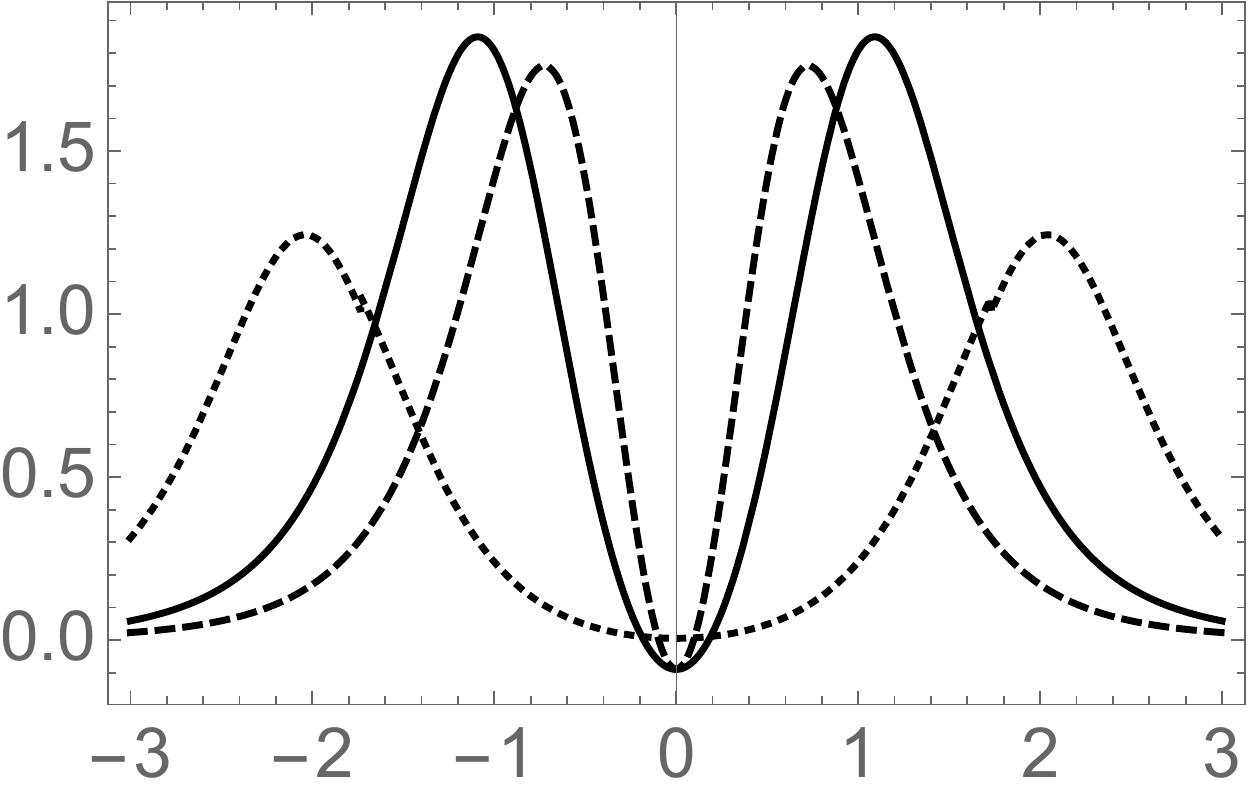}
	 	\caption{The snapshots of the electric field component 
	 		$E^3_3(r,t)$ in units of $1/g\rho^2$,	as a function of $x^3/\rho$, for
	 		times $t/\rho=0.5$ (dashed),  $t/\rho=1$ (solid) and $t/\rho=2$ (dotted) curves.
	 	}
	 	\label{fig:snapshotsofe}
	 \end{figure}

	 The electric field during the sphaleron explosion
	 $E^m_a(\vec x,t)$  follows from the expressions for the field potential (\ref{eqn_sph_field}). The component with $a=3$ corresponds to the diagonal
	 Pauli generator $\tau^3$, so that quark color remains unchanged. The maximal magnitude of 
	 the field corresponds to the $x^3$ or $z$ direction,  $E^3_3$. The formula is a bit long
	 to give here, but its behavior 
	 is shown in Fig.~\ref{fig:snapshotsofe}
	 
	 From our discussion of the semiclassical quark pair production in a $ homogeneus $ electric field it follows that the field must produce
	 an impulsion $$\Delta p= \int E dt \sim M $$ 
	 to reach, in Euclidean notations, the rotation angle $y_E=\pi/2$ needed to make it real.
	 For an estimate, one can take $E\approx 1.5/\rho^2$ and $\delta t\approx \rho$
	 ( following  Fig.~\ref{fig:snapshotsofe}),
	 from which the estimated impulsion  is
	 $$ \Delta p=E\delta t\approx {1.5/ \rho}
	 $$ 
	 This gives an estimated limit for  the mass $M$
	 of a quark which is $likely$ to be produced
	 \begin{equation}
	 	M <  1.5 / \rho
	 \end{equation}
	 Using $1/\rho=0.6\, GeV$, one finds the r.h.s. 
	 to be $\approx 1\, GeV$. This implies that
	 strange quarks, with $M_s\sim 0.14\, GeV$ can be produced, but not the charmed ones, 
	 with $M_c\sim 1.5\, GeV$. To satisfy this estimate, one would need to decrease $\rho$
	 by about a factor 2. To produce a $b$ quark, with $M_b\approx 5 \, GeV$, one would need to decrease $\rho$ by a factor 6 or so. 
	 
	 In order to get quantitative
	 semiclassical description on has to do the following: (i) convert the expressions for the field to Euclidean time;
	 and (ii) solve the relativistic classical EOM
	 $$ M{d u^\mu\over ds}=F^{\mu\nu} u_\nu
	 $$
	 where $u^\mu=dx^\mu/ds$ and $s$ is the proper time, $ds^2=dt^2-d\vec{x}^2$. Comparable and rather complex electric and magnetic fields
	 make the paths quite complicated. On top of that,
	 the result depends  on the starting location of the particle. So, at this time, we have no results on such a calculation to report.

\begin{figure}[h!]
	\centering
	\includegraphics[width=0.7\linewidth]{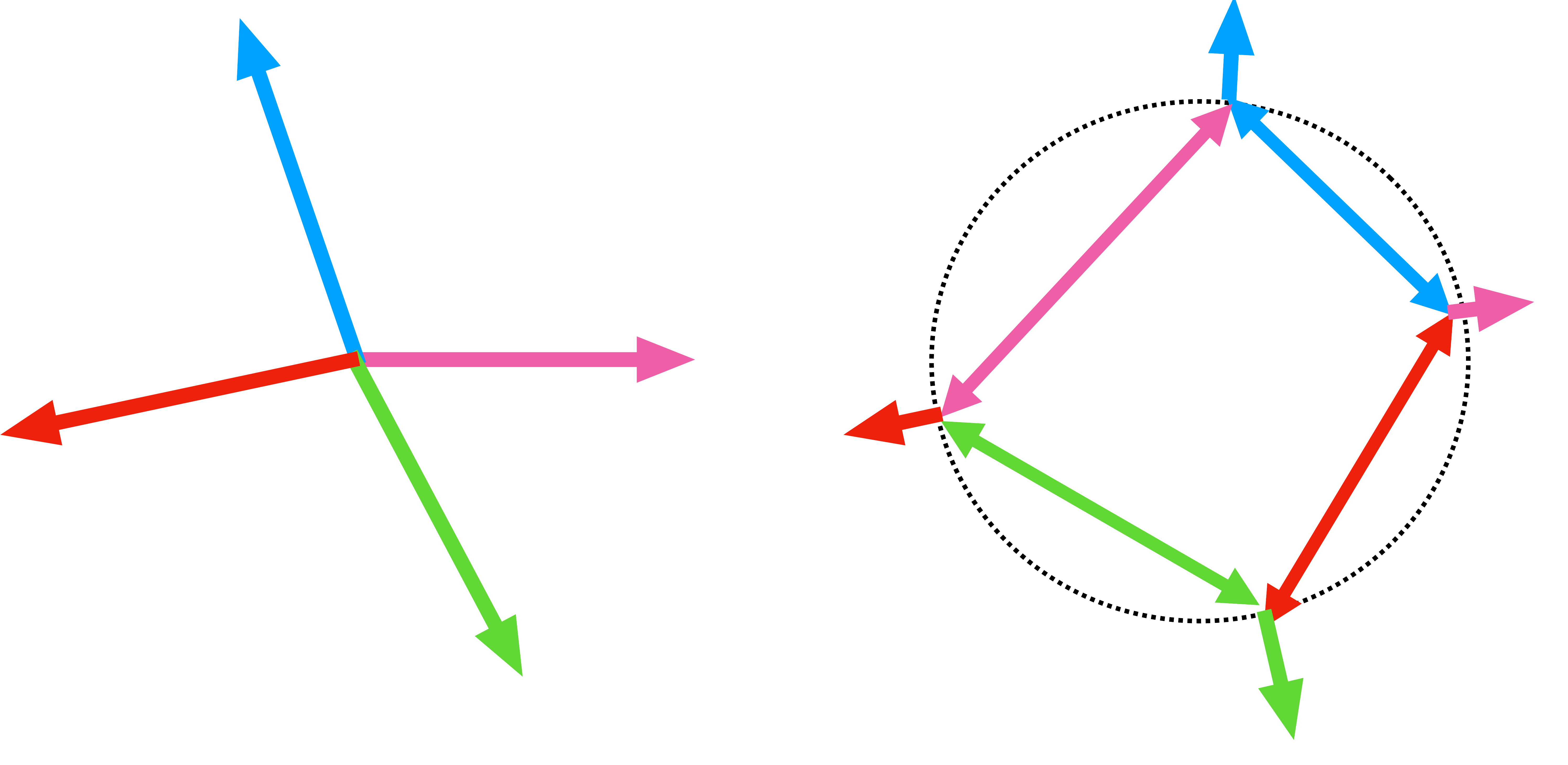}
	\caption{Schematic picture illustrating the 
		configuration of QCD strings in four-jet events. The left corresponds to the usual case, when jets originate from collisions at the origin in the transverse plane. The right corresponds to an exploding sphaleron in which the strings are
		not connected to the origin,  but are close to the expanding shell (dotted circle).}
	\label{fig:jetsstrings}
\end{figure}

\section{Sphaleron decays at large masses, $M \sim 100\, GeV$}

These masses fall in the range discussed in the theoretical literature, and mostly searched at HERA and LHC, as they
promise to have ${\cal O}(10)$ well recognized gluonic and quark jets. Background evaluation for such events
were attempted, using current event generators tuned to multi-jet events.  

Two comments are in order:

(i) In this mass range,  it is  possible to use double  diffractive events
at LHC, which are expected to reduce backgrounds substantially in comparison to ambient $pp$ collisions.

(ii) In the usual parton-parton collisions, the fragmentation function of ${\cal O}(10\, GeV)$ gluons is
 essentially the product of string breaking. One end of these strings is  on the leading gluon, the other ends
 at throughgoing partons, or the origin in the transverse plane, see Fig.~\ref{fig:jetsstrings} left.
But sphaleron decay leaves out
 the interior of the exploding shell ``empty" (more precisely, with a pure gauge configuration). As these
 gluons become physical and separated from each other, the strings would  go $between$ them close to the shell, rather than extending to he origin, 
 see Fig.~\ref{fig:jetsstrings} right. This suggests the appearance of some unusual gluon jet fragmentation functions, with a significant depletion  at small $p_t$.

\section{Magnetic field and Chiral Magnetic Effect}
In the presence of an electromagnetic field  $\vec b_{EM}$ and chiral disbalance of fermions, there exists the so called Chiral Magnetic Effect (CME), an electric current
along  $\vec b_{EM}$ is created, for a review see~\cite{Kharzeev:2013ffa}. Experiments aimed at observing it in heavy ion collisions at RHIC have been made, 
lately utilizing beams of nuclei with the same atomic weight but different charges, but the analysis  is not conclusive. The CME was
observed in semi-metals \cite{Kharzeev:2013ffa}. 

An interesting manifestation of the phenomenon must also be present in the instanton-sphaleron production processes, which we propose in this paper.

The first part of it, related with instantons,
has already been studied. 
While calculating nucleon form factors
using instantons~\cite{Faccioli:2003yy}, Faccioli observed that the  instanton zero mode acquires an electric dipole along the magnetic field. (Of course, adding the antiinstanton
contribution cancels it, unless the  topological $\theta$ angle is nonzero).   Lattice simulations in a magnetic field carried in~\cite{0812.1740} have confirmed reality of this effect.

The extensive analytic study of the electric dipole moment of a zero mode is due to Basar, Dunne and  Kharzeev \cite{1112.0532}. Let us just recall  a qualitative explanation of the mechanism from their work.  The Dirac equation in the instanton background conserves total $$\vec J=\vec L+\vec S+\vec T$$ the sum of orbital momentum, spin and (SU(2)) isospin. 
As noticed by t'Hooft, the isospin-orbit $(\vec T\cdot \vec L)$ term forces 
orbital momentum $not$ to be a good quantum number.
Therefore, it is possible to  mix opposite parity states\footnote{This discussion is in the instanton background, which is CP non-invariant. The QCD
vacuum and other states are CP conserving.
}. The external electromagnetic field $\vec b$ couples  spin  to angular momentum,
and induces such mixing. The zero fermion and antifermion modes are deformed in different
directions, creating a nonzero electric dipole
along the $\vec b$ field. (Antiinstantons of course would do the opposite: we discuss local, not global CP violation.)
No observable consequences of this effect have so far (to our knowledge) been proposed.

Now, consider e.g. very peripheral heavy ion collisions, in which Pomeron-Pomeron (or $\gamma$-Pomeron or $\gamma \gamma$) collision
initiate instanton-sphaleron production process. As usual, the large electric charges of the ions 
generate a strong magnetic field at the production point. The exploding instanton and sphaleron
 gauge field remains spherical, but, since the
quark zero mode is
deformed, the exploding  electric  field will 
carry them differently in direction of the magnetic   field, as compared to two other directions, producing anisotropy of the quark
momenta.

The effect of a magnetic field can be seen by
comparing $pp$ and  $PbPb$ peripheral collisions. The
orientation of the collision plane (to which
$\vec b$ is normal) can be deduced from the direction of two forward-moving protons (or ion remnants). 

The sign of the effect for a quark depends on its electric charge, and the light quarks in question, $u,d,s$ have charges $+2/3,-1/3,-1/3$. So, for the mesons $\pi^+$ versus $\pi^-$ (kaons, etc)   the effects are added, resulting in a charge-dependent deformation of the distribution.
The instanton and antiinstanton events still
produce opposite signs of the effect. So 
one may construct a meson $correlation$ observable inside an event ($all$ $\pi^+$ have
one preferred direction, opposite to $\pi^-$). 

Another (statistically challenging) option is
to select events with $\Lambda$ (or  $\bar\Lambda$) decays, which tells us about the strange quark chirality.
 Statistically selecting ``more-likely-instanton" or ``more-likely-antiinstanton" events, one can
 perhaps   see an electric dipole in
the event directly.   

Concluding this discussion, let us again emphasize the following. The instanton-sphaleron
process is the ``hydrogen atom" of topological
effects. The chiral correlations produced in it is maximally possible, as all quarks and antiquarks have fixed chiralities.  (For comparison, in heavy ion collisions we think
the axial charge of the fireball is between $\pm 20$ or so, on top of thousands  of unpolarized ones. While the proposed measurements may
appear quite exotic, the cross sections are
relatively large, allowing LHC with the right trigger to produce
billions of diffractive events.

\section{The ``entrance factor"}
So far our discussion was mostly semiclassical, we considered (Euclidean) tunneling paths and (Minkowski) sphaleron explosions. Then we projected the outgoing quanta
to some final hadronic states.

Now we focus on the very initial stage of the collision. Here we first have to 
make hard choices, selecting theoretical tools
for its description.  Here are these choices:
colliding protons can be seen as colliding (i) partons (gluons or quarks); (ii)
color dipoles; (iii) glueballs; or (iv) Pomerons. Let us briefly discuss them subsequently.

At high mass scale, e.g. the sphaleon mass squared $$s'=M^2_{\rm sph} \gg 100\, {\rm GeV}^2$$ the parton-parton collision is the natural perturbative description, developed since the
1970's in the context of jets or heavy quanta production. It was applied by Ringwald and Schrempp
for DIS at HERA~\cite{Ringwald:1998ek}, and now by Khoze et al.~\cite{1911.09726}
to LHC setting, incorporating
Mueller perturbative correction~\cite{Mueller:1991fa}.
 For  comparison of these predictions with
 HERA data see~\cite{1603.05567}, and with LHC~\cite{1805.06013}. At high mass regime  the theoretical predictions 
 have reasonable uncertainties, such as in the 
 semiclassical prefactor, but the main  
 difficulty in it is large background induced by ambient events. The main observable is a spherical multi-jet events.
 
\begin{figure}
 	\centering
 	\includegraphics[width=0.5\linewidth]{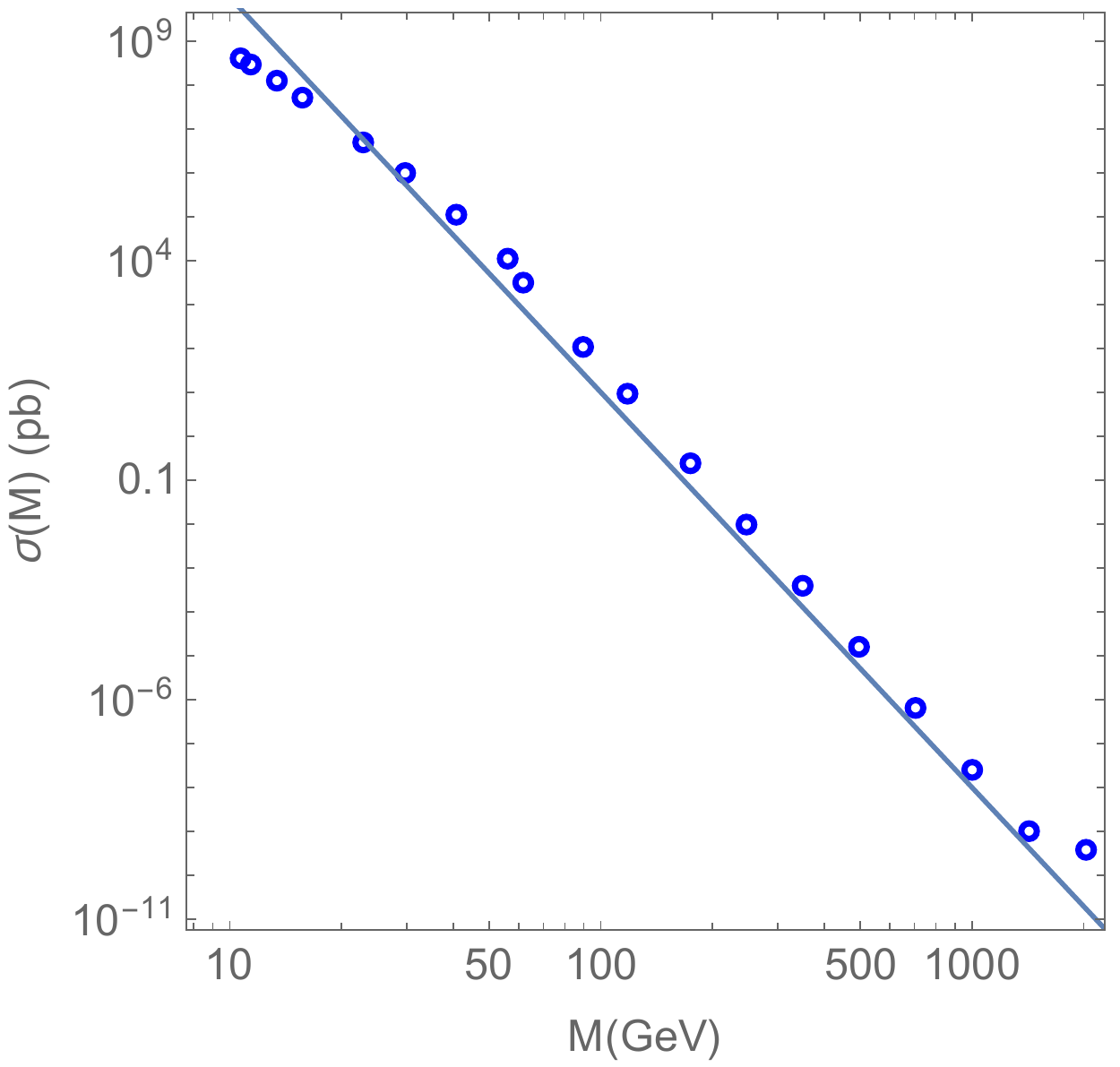}
 	\caption{Cross section $\sigma(M)$  in pb of the process gluon-gluon-instanton/sphaleron, versus the sphaleron mass $M (GeV)$.The points are explained in the text, the line is the $1/M^{9}$ background dependence shown for comparison.}
 	\label{fig:sigma-khoze}
 \end{figure}

  To illustrate the situation, we show the gluon-gluon   cross section $\sigma(\sqrt{s'})$ of the instanton-sphaleron process
 in~Fig.\ref{fig:sigma-khoze}.
   The points are from Table 1  by Khoze et al.~\cite{1911.09726}.
 The instanton size in the range of the plot changes from $1/\rho=1\, {\rm GeV}$ to $75 \, {\rm GeV}$, and the number of outgoing gluons at the high end reaches about a dozen.

 The
 line, shown for comparison, is $1/M^{9}$.  
 Recall that the
 one-loop coefficient of the QCD beta function is $b=(11/3)N_c-(2/3)N_f$, or  9 if  the number of light flavors $N_f=3$. The  effective action is twice that of the instanton $2S_{inst}$ minus its depletion 
for sphaleron production $-\Delta S$. The  cross section should be, by dimension,
 $$\sigma \sim {1 \over M^2} \bigg({\Lambda_{QCD}\over M}\bigg)^{b(2-\Delta S/S_0)}$$
One can see that the original estimates 
 \begin{equation}{ \Delta S\over S_0}\approx 1 \label{eqn_DeltaS}
 \end{equation}
are  supported by actual multidimensional integration. 

The calculation along a path including both the Euclidean and Minkowskian times
has been preformed in~\cite{Bezrukov:2003qm}, in the electroweak setting. Their result (solid line
in Fig.6 of that paper)  shows that the action is reduced from $\approx 12.5$ at the zero sphaleron mass to about $\approx 7$ at the large  sphaleron mass. It also supports 
the estimate (\ref{eqn_DeltaS}). 
 
 The backgrounds come from multiple QCD reactions,
 which have cross sections $$\sigma_{background}\sim {\alpha_s^2 \over M^2}$$
It is therefore clear that the task of separating the signal from the background becomes much harder as
the cluster mass $M$ grows. 
 
As the momentum transfer scale decreases,
one may naturally think of coherent
contribution of two (or more) partons.
Color neutrality can be implemented starting from color dipoles.
Unlike partons, the colliding color dipoles 
have a natural scale given by their sizes $a_1,a_2$. Correlating dipoles with instantons
via Wilson loops has been done by us ~\cite{Shuryak:2000df}. The cross sections obtained can be directly tested in double-inelastic electron-positron (or $\gamma^*\gamma^*$) collisions. Unfortunately,
a description of a proton in terms of color dipoles is not yet (to our knowledge)  developed.  

At small momentum transfer scale -- in diffractive processes we will focus on --
the colliding objects are described in terms
of (reggeized) hadron exchanges, 
especially by tensor glueballs
or (their extension to the whole Regge trajectory), the Pomerons. In a number of 
relatively recent works it has been shown that Pomeron exchanges require effective description in terms of symmetric tensor. 
These facts, together with overall development of holographic QCD models, had strengthen
the Pomeron-graviton connection.

 \section{Sphaleron production in Pomeron-Pomeron collisions} \label{sec_sph_in_PP}

We start with some general remarks related to experimental setting.
 
 Double-Pomeron production processes at LHC
are only at their initial stage. A  general but brief 
discussion of the existing detectors  was made in section~\ref{sec_PP}. Let us only add that apparently
in the standard high luminocity  LHC runs,  one cannot access
masses less than say hundreds of ${\rm GeV}$, but can go to any masses provided the dedicated low-luminocity runs are
performed.

As we will detail below, Pomeron-Pomeron vertices are
coupled to two operators,  scalar $G^2$ and pseudoscalar $G\tilde G$. The former is maximal when
electric field directions are parallel, in the second orthogonal. Momentum kicks to two protons are
directed with electric field. Therefore, locaing
roman pots in different azimuth, one can in principle
``tag" and separate these two contributions.

Returning now to current experiments at LHC, we note that exclusive channel
 $\pi^+\pi^-$ has been already studied by CMS~\cite{1706.08310}, for invariant masses till $M<2\, {\rm GeV}$.
 It shows a $\rho$-meson peak, and contains hints at some other resonances. 
 
 Heavier clusters unfortunately were only studied by the  old UA8
collaboration~\cite{Brandt:2002qr}   with  $p\bar{p}$ collisions at the SPS.
 The production cross section in Pomeron-Pomeron collisions at the peak $M\sim 3\, {\rm GeV}$
 is rather large $\sigma_{PP}\sim 4 \, {\rm mb}$.

But, as usual, what really matters is the Signal/Background (S/B) ratio. In this respect 
our only normalization process is the  $\eta_c$ decay.
Assuming standard $\bar c c\rightarrow gg $ annihilation,
into $G_{\mu\nu}\tilde G_{\mu\nu}$ pseudoscalar 
operator, one realizes that the gluon can either enter
the instanton/sphaleron tunneling, or produce quark pairs perturbatively. Most likely, at fixed $M_{gg}$ 
{\em it does not matter whether $gg$ came from $\bar c c$ annihilation or Pomeron-Pomeron collisions.}

In total, the three prominent 3-meson channels ascribed to the former have  certain branching ratio, from which one can conclude that
$$ \bigg({ Signal \over Background}\bigg)(M\approx 3\, {\rm GeV}) > 0.15 $$  
One can think of many other modes, e.g. with glueballs,
scalar or pseudoscalar, but at a mass of only $M_{\eta_c}\approx 3\, {\rm GeV}$ those would be suppressed by phase space~\footnote{
For a more accurate estimate, one needs to figure out
which other modes come from the signal. Unfortunately 
there are many modes and their branching ratios are at the level less than a percent, so it would be hard if not
impossible to get an accurate number.}. Perhaps it increases the signal by a factor  of 2 or something like that.
Alternatively, {\bf PP} collisions can be approximated as $gg+gg$ collisions, or operators with
 $two$ stress tensors. This perhaps increases the background, say also by a factor of 2, so that these effects cancel out in the S/B ratio.
 
  What is however even more important,
   if one looks in the ``instanton-induced" channels  there would hardly be any background. It follows from comparison of the $\eta_c$ decays to 
   $KK\pi,\pi\pi\eta$ yields with pure pion final states.

\begin{figure}
	\centering
	\includegraphics[width=0.45\linewidth]{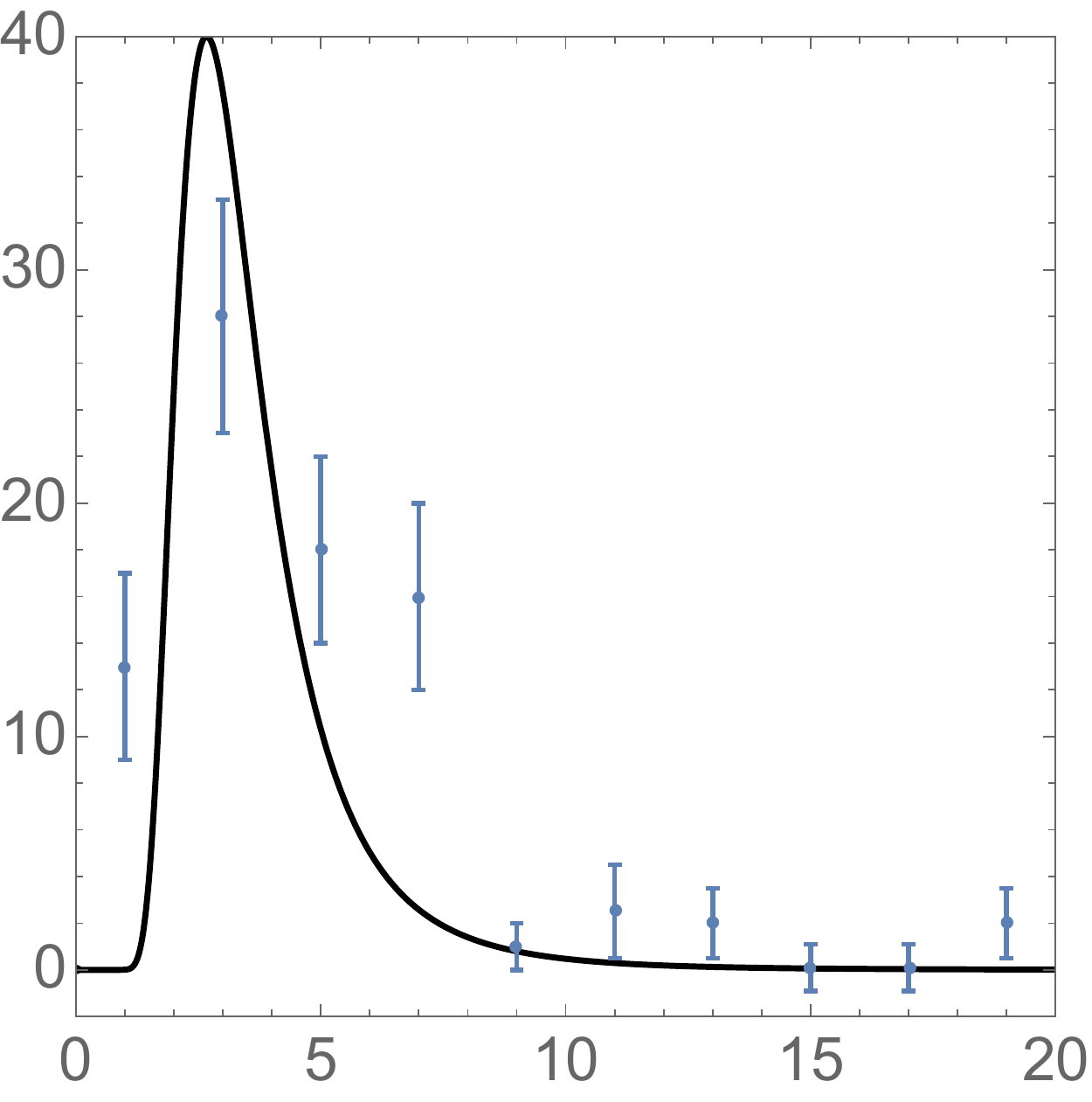}
		\includegraphics[width=0.45\linewidth]{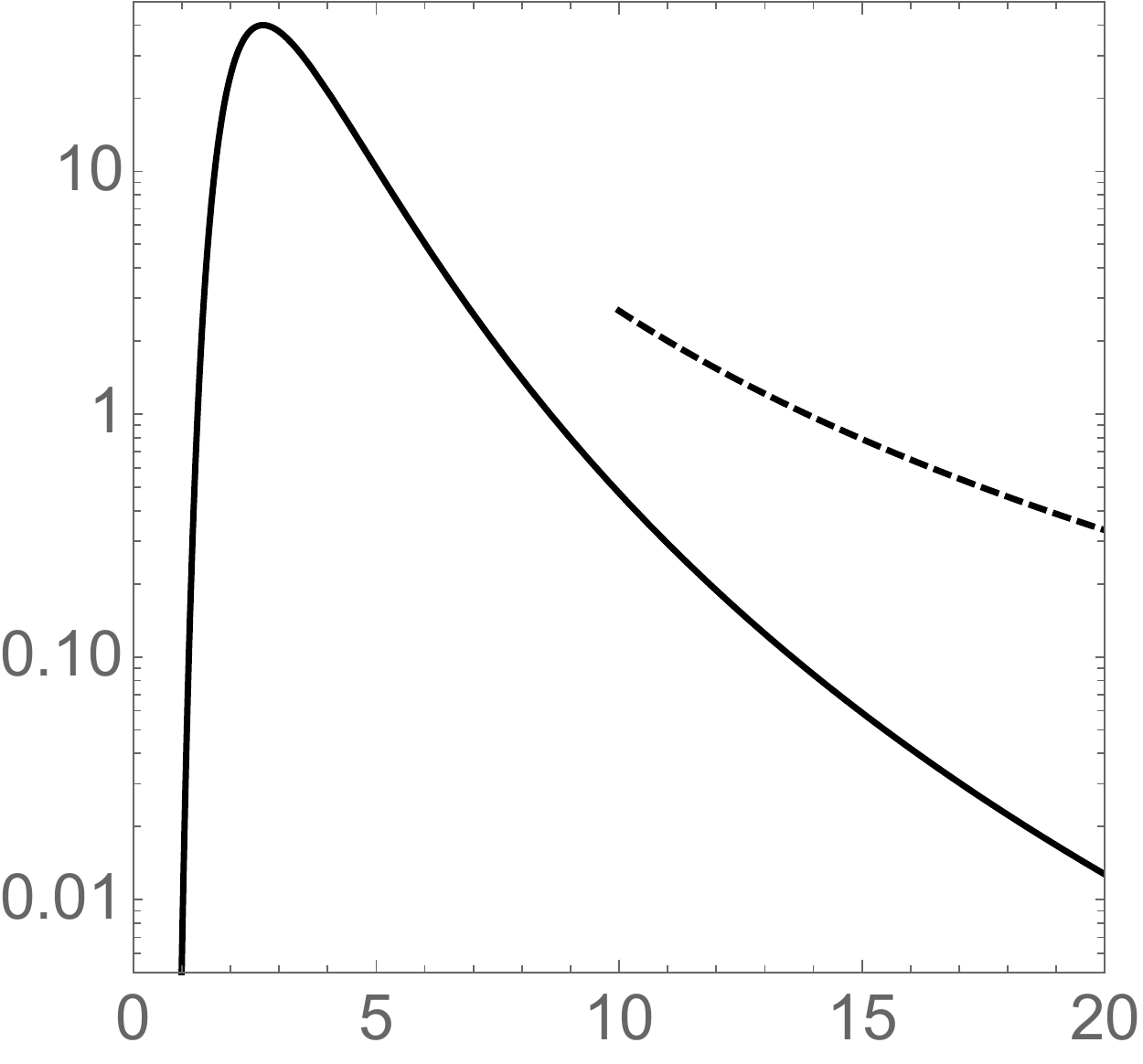}
	\caption{Left plot: Semiclassical distribution over the  cluster mass $M\, ({\rm GeV})$, compared to the data points from the UA8 experiment; Right plot is the logarithmic representation of the same curve (solid), now compared with the dashed line representing the perturbative background.     }
	\label{fig:clusters1}
\end{figure}

   Now, we address  the sphaleron mass distribution in {\bf PP}
   collisions. We already discussed in~\ref{sec_size_distribution} the 
instanton size distribution

\begin{equation}
	{dn \over d\rho}\sim\bigg({1\over \rho^5}\bigg) (\rho\Lambda)^{b} {\rm exp}\big(-2\pi \sigma\rho^2\big)
\end{equation}
with  $b=(11/3)N_c-(2/3)N_f$ is the first coefficient of the beta function,  and $\sigma\approx (0.42\, {\rm GeV})^2$  the string tension. 
When combined with the (pure gauge) expression of the
sphaleron mass $M_{sph}=3\pi^2/g^2\rho$, it yields the distribution over this mass. In Fig.~\ref{fig:clusters1} we compare it with the data
points from the UA8 experiment~\cite{Brandt:2002qr} (left plot), and with the  background (right plot) expected to 
scale as $d\sigma/dM\sim 1/M^3$ and normalized to the high-mass UA8 data.

As one can see, the vacuum distribution in instanton sizes and the experimental cluster mass
distribution are in qualitative agreement in shape, and in predicted position of a peak. 
This supports our suggestion that those clusters may in fact be QCD sphalerons.

Let us make few other points: 
\\(i)  the instanton-sphaleron production 
seems to be dominant at $M\sim {\rm few-GeV}$;\\
(ii) the cross section  to   $M\sim 20 \,{\rm GeV}$ 
decreases by about three orders of magnitude.
With the efficiency of  the forward trigger of UA8, about 1300 events were detected
at $the Sp\bar pS$. The luminocity of LHC is more than enough to 
cover these extra three orders of magnitude;
(iii) where the signal-to-background ratio is
about $1:20$. The background -- mostly two-jet
events -- is very anisotropic, so perhaps the identification of an isotropic  signal can still be possible,
even at $M\sim 20\, {\rm GeV}$.;\\
(iv) we  recall  that the clusters with $M\sim 20 \,{\rm GeV}$  are the ones in which  10 quark operators (with pairs of  $\bar b b$ quarks) can be produced.

\section{Sphaleron-gluon-quark vertex}
To make explicit the sphaleron-quark vertex, we note that
the instanton density in (\ref{DENS}) follows from the averaging of the gluon field in the instanton vacuum

\be
\label{IBARI}
n_{I+\bar I}=n_I+n_{\bar I}= \frac{\alpha_s}{8\pi}\langle {G^2}\rangle \qquad n_{I-\bar I}=n_I-n_{\bar I}= \frac{\alpha_s}{8\pi}\langle{G\tilde G}\rangle\rightarrow 0
\ee
where the second averaging is zero  for $\theta=0$. The normalizations in (\ref{IBARI}) are fixed by the QCD scale and axial anomalies
in the instanton vacuum, respectively.
Note that in (\ref{IBARI}) $G^2$ and $G\tilde G$ count the number of instantons
and anti-instantons. 

The single instanton-six-quark vertex and anti-instanton-six-quark vertex follow by omitting the vacuum averaging in (\ref{IBARI}),
and recalling that the left-vertex  is induced by an instanton, and the right vertex  by an  anti-instanton. More specifically, from (\ref{DET1}) and (\ref{IBARI})
we obtain

\bea
\label{GSIX}
{\cal L}_{Gqqq}=\bigg[\frac {\alpha_s}{16\pi}\bigg(G^2+G\tilde G\bigg)\bigg]\bigg(\frac{4\pi^2\rho^3}{M\rho}\bigg)^3\bigg[\frac{{\cal V}^L_{qqq}}\kappa\bigg]+
\bigg[\frac {\alpha_s}{16\pi}\bigg(G^2-G\tilde G\bigg)\bigg]\bigg(\frac{4\pi^2\rho^3}{M\rho}\bigg)^3\bigg[\frac{{\cal V}^R_{qqq}}\kappa\bigg]
\eea

The sphaleron produced in the diffractive process is half a tunneling process with half the topological charge, and not self-dual. 
At the turning point, the sphaleron drags six quark zero modes out of the QCD vacuum  with at $t=0$ the vertex mediated by 
$G_S^2=2B^2$ since  $G\tilde G_S=4E\cdot B=0$

\bea
\label{SSIX}
{\cal L}_{Sqqq}(t=0)=&&\bigg(\bigg[\frac {\alpha_s}{16\pi}\,G_S^2\bigg]\bigg(\frac{4\pi^2\rho^3}{M\rho}\bigg)^3\bigg[\frac{{\cal V}^L_{qqq}+{\cal V}^R_{qqq}}\kappa\bigg]\bigg)_{t=0}\nonumber\\
+&&\bigg(\bigg[\frac {\alpha_s}{16\pi}\,\dot{K}_{0S}\bigg]\bigg(\frac{4\pi^2\rho^3}{M\rho}\bigg)^3\bigg[\frac{{\cal V}^L_{qqq}-{\cal V}^R_{qqq}}\kappa\bigg]\bigg)_{t=0}
\eea
and by the rate of the Chern-Simons charge density $\dot{K}_{0S}$  following from $\partial^\mu K_\mu=G\tilde G$.

\section{Witten amplitude  for  diffractive production $pp\rightarrow ppX$}


QCD is difficult to track in the infrared since its fundamental quark and gluon constituents repackage in confined hadrons. The ensuing
hadronic dynamics is strongly coupled. Holographic QCD is a proposal guided by  the AdS/CFT or gauge/gravity duality discovered in
string theory. The holographic duality or principle states that boundary operators in the gauge theory can be mapped to a higher 
dimensional string theory in bulk in a curved anti-deSitter space. The original correspondence holds for type IIB superstring theory
in $AdS_5\times S_5$, but is commonly assumed to hold for a string theory in a general background.
The srting theory is in general difficult to solve, but in the double limit of a large number of colors and strong gauge coupling 
$\lambda=g^2N_c$, it reduces to a weakly coupled supergravity in the classical limit, with a weak string coupling $g_s=g^2/4\pi$. 
The gauge invariant operators at the boundary are mapped onto supergravity  fields
in bulk with pertinent anomalous dimensions.

\begin{figure}
	\centering
	\includegraphics[width=0.6\linewidth]{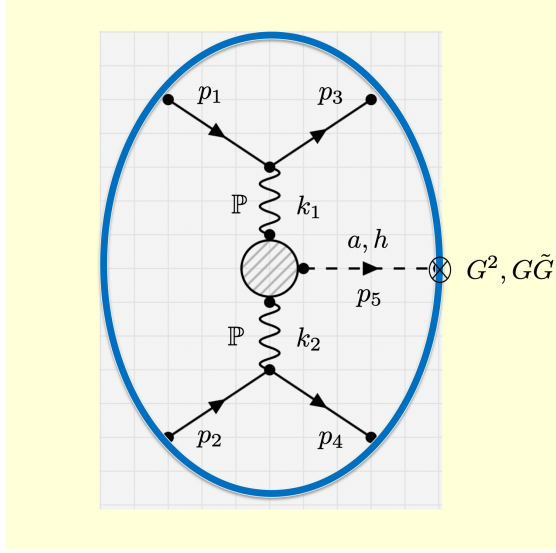}
	\caption{Witten diagram for diffractive $pp\rightarrow ppX$ with $X= (a,f)\equiv (G\tilde G,G^2)$  for the sphaleron at the boundary}
	\label{fig:pathconste}
\end{figure}

The  n-point functions at the boundary of $AdS_5$ follow from variation of the
on-shell supergravity action in bulk with respect to the boundary values. The results are  tree-level Feynman graphs with fixed end-points on the
boundary also known as Witten diagrams. The Witten diagram for the diffractive process of interest in this work $pp\rightarrow ppX$
is illustrated in Fig.~\ref{fig:pathconste}. The insertions on the $AdS_5$ boundary  refer to 4 
nucleon operator insertions ${\cal O}_{pi=1,...,4}$ and a glueball insertion ${\cal O}_{X=G^2,G\tilde G}$. ${\cal O}_p$ sources a Dirac fermion
in bulk with anomalous dimension $M=\tau-3/2$ ($\tau$ refers to the twist, with typically $\tau=3$ to reproduce the hard scattering rules),
 and the ${\cal O}_X$ sources a dilaton or axion in bulk with anomalous dimension
$\Delta_X=4$. In the Regge kinematics, the exchange in bulk is mediated by two closed string producing a dilaton $h$ or an axion $a$  which is equivalent
to  Pomeron-Pomeron fusion into $h,a$   (${\bf P}{\bf P}\rightarrow h, a$).

The Holographic construction provides a first principle description of the Pomeron (${\bf P}$) as a dual to a close string exchange~\cite{Rho:1999jm,Basar:2012ra}
or  Reggeized graviton in bulk~\cite{Polchinski:2002jw,Brower:2006ea}. 
In this section we will use this framework to extract the differential cross section for the reaction
$pp\rightarrow ppX$ through ${\bf PP}$  fusion as discussed in~\cite{Anderson:2014jia} for pseudoscalar emission, and in~\cite{Brower:2014ena}
for Higgs production.   After a brief description for the pertinent kinematics for this
process in the Reggeized limit, we will summarize the main formula for the production of $G^2$ and $G\tilde G$ glueballs
with most of the details given in \ref{app_PP1}-\ref{app_PP2}-\ref{app_PP3}.

\subsection{Kinematics for the Reggeized limit}

For the process $p(p_1)+p(p_2)\rightarrow p(p_3)+p(p_4)+X(p_5)$ we will set the incoming protons back-to-back 

\be
p_1=(E, 0,0,p)\qquad p_2=(E,0,0, -p)\qquad p_{i=3,4,5}=(E_i, q_{\perp i}, p_{iz})
\ee
with  longitudinal fractions $p_{3z}=x_1p$ and $p_{4z}=-x_2p$ for the outgoing protons.
The transverse momenta in the azimuthal plane are  2-vectors

\be
q_{\perp 3}=(q_3\,{\rm cos}\,\theta_3, q_3\,{\rm sin}\,\theta_3)\qquad q_{\perp 4}=(q_4\,{\rm cos}\,\theta_4, q_4\, {\rm sin}\,\theta_4)
\ee
The  Mandelstam kinematics  suggests  five invariants

\bea
s=(p_1+p_2)^2\qquad t_1=(p_1-p_3)^2\qquad t_2=(p_2-p_4)^2\qquad
s_1=(p_3+p_5)^2\qquad s_2=(p_4+p_5)^2\nonumber\\
\eea
which are reduced to four by azimuthal symmetry.
We will choose the four invariants as  $t_1, t_2$, the relative azimuthal angle $\theta_{34}=\theta_4-\theta_3$, and the relative
 momentum fraction $x_F=x_1-x_2$. Following~\cite{Anderson:2014jia}, 
the Reggeized limit is characterized by large $\sqrt{s}$ and small scattering or azimuthal angles, 

\be
s\gg s_1, s_2\gg -t_1,- t_2, m_N^2\qquad{\rm and\,\,fixed\,}\qquad \mu=\frac{s_1s_2}s
\ee
with the simplified kinematics

\be
\label{KIN}
s_{1,2}\approx \sqrt{\mu s}\qquad {t_{1,2}}\approx -q^2_{\perp 3,4} \qquad
\mu\approx p_5^2+t_1+t_2+2\sqrt{t_1t_2}\,{\rm cos}\,\theta_{34}
\ee


In this limit and using  (\ref{KIN}), the differential cross section for $pp\rightarrow pp0\mp$  is dominated by $x_F\approx 0$~\cite{Anderson:2014jia}, and reads

\be
\label{DIFFGENERAL}
\frac{d\sigma_\mp}{d\theta_{34} dt_1 dt_2}\approx \frac 1{16\pi^4s^2}\,{\rm ln}\bigg(\frac s\mu\bigg)\,\frac 14 \sum_{\rm spin}
\bigg|{\cal A}_{pp\rightarrow pp 0\mp}\bigg|^2
\ee
for  the production amplitudes (\ref{WPPFULL0-}) and (\ref{WPPFULL0+}) respectively. We now make explicit the production amplitude
for scalar and pseudoscalar production.

\subsection{Differential cross section: $G\tilde G$}

The main elements entering the Witten amplitude in Fig~\ref{fig:pathconste}  for  $pp\rightarrow pp0-$ through ${\bf PP}$ fusion  
are detailed in the Appendices~\ref{app_PP1}-\ref{app_PP2}-\ref{app_PP3}, and specialized to the soft wall model in a slab of AdS$_5$. 
Other models in AdS$_5$ yield similar results. The ensuing squared spin-averaged amplitude entering (\ref{DIFFGENERAL}) is

\bea
\label{DIFF1}
&&\frac 14 \sum_{\rm spins}\bigg|{\cal A}_{pp\rightarrow pp 0-}\bigg|^2=\frac{16}4\,A^2(t_1)\,A^2(t_2)\,\nonumber\\
&&\times\bigg[\bigg(\frac \lambda{\pi^2} +1\bigg)^2\bigg(\frac{\sqrt\lambda}{2\pi}\bigg(\gamma+\frac \pi 2\bigg)\bigg)^4
\frac{e^{-\sqrt\lambda (\gamma+\pi/2)^2(1/{\rm ln}(s_1/\tilde\kappa_N^2)+1/{\rm ln}(s_2/\tilde\kappa_N^2))}}{({\rm ln}(s_1/\tilde\kappa_N^2){\rm ln}(s_2/\tilde\kappa_N^2))^3}
\,\bigg(\frac{s_1}{{\tilde\kappa_N}^2}\bigg)^{2(\alpha_{\mathbb P}(t_1)-2)}\bigg(\frac{s_2}{{\tilde\kappa_N}^2}\bigg)^{2(\alpha_{\mathbb P}(t_2)-2)}\bigg]\nonumber\\
&&\times\bigg[ \bigg(B_1^-(t_1,t_2)\eta_{\beta\bar\beta}+B_2^-(t_1,t_2)k_{2\beta}k_{1\bar\beta}\bigg)\,\epsilon_{\alpha\bar\alpha\gamma\delta}k_1^\gamma k_2^\delta\bigg]\nonumber\\
&&\times\bigg[ \bigg(B_1^-(t_1,t_2)\eta_{\underline \beta\bar{\underline \beta}}
+B_2^-(t_1,t_2)k_{2\underline \beta}k_{1\bar{\underline \beta}}\bigg)\,\epsilon_{\underline \alpha\bar{\underline \alpha}\underline \gamma\underline \delta}
k_1^{\underline \gamma} k_2^{\underline \delta}\bigg]\nonumber\\
&&\times\bigg[p^\alpha p^\beta p^{\underline\alpha}p^{\underline\beta}
+\frac 1{16}\bigg(t_1\eta_{\alpha\underline\alpha}p_\beta p_{\underline\beta}+t_1\eta_{\alpha\underline\beta}p_\beta p_{\underline\alpha} +t_1\eta_{\beta\underline\alpha} p_\alpha p_{\underline\beta}
+(t_1\eta_{\beta\underline\beta}-k_{1\beta}k_{1\underline\beta})p_\alpha p_{\underline\alpha}\bigg)\bigg]\nonumber\\
&&\times\bigg[\underline p^{\bar \alpha} \underline p^{\bar \beta} \underline p^{\underline {\bar\alpha}}\underline p^{\bar{\underline \beta}}
+\frac 1{16}\bigg(t_2\eta_{\bar \alpha\bar{\underline \alpha}}p_{\bar \beta} p_{\bar{\underline \beta}}
+t_2\eta_{\bar  \alpha\bar{\underline \beta}}p_{\bar  \beta} p_{\bar{\underline \alpha}} +t_2\eta_{\bar  \beta\bar{\underline \alpha}} p_{\bar  \alpha} p_{\bar{\underline \beta}}
+(t_2\eta_{\bar \beta\bar{\underline \beta}}-k_{1\bar  \beta}k_{1\bar{\underline \beta}})p_{\bar  \alpha} p_{\bar{\underline \alpha}}\bigg)\bigg]
\eea
with the vertex

\bea
2B_1^-(t_1,t_2)-\mu B_2^-(t_1,t_2)=4\bigg({\bf C}^-_1-{\bf C}^-_2\sqrt{t_1t_2}\,{\rm cos}\,\theta_{34}\bigg)
\eea
and ${\bf C}^-_{1,2}$ given in (\ref{CS4}-\ref{XCS5}) for the soft-wall model.
Here, the Pomeron trajectory is the Reggeized graviton trajectory 

\bea
\alpha_{\mathbb P}(t)=\alpha_{\mathbb P}(0)+\frac{\alpha^\prime}2t\equiv 2-\frac 2{\sqrt\lambda}+\frac{\alpha^\prime}2t
\eea
after restoring the t-dependence, with $\alpha^\prime =l_s^2$  (the squared string length). For simplicity we have set $\lambda\rightarrow\infty$ in the A-form factors in (\ref{DIFF1}),  i.e.  $A(j_0\rightarrow  2,k)\equiv A(t)$, although this can be relaxed. A similar observation was made originally in~\cite{Shuryak:2003xz} (instantons) and more recently
 in~\cite{Anderson:2014jia} (holography) for the double diffractive production of $\eta, \eta^\prime$ .

 We recall that for  the nucleon as a Dirac fermion, 
the energy momentum tensor is characterized by three invariant form factors

\be
\label{EMT2}
\left<p_2|T^{\mu\nu}(0)|p_1\right>=\overline{u}(p_2)\left(
A(k)\gamma^{(\mu}p^{\nu)}+B(k)\frac{ip^{(\mu}\sigma^{\nu)\alpha}k_\alpha}{2m_N}+C(k)\frac{k^\mu k^\nu-\eta^{\mu\nu}k^2}{m_N}\right)u(p_1)\,,
\ee
with $p=(p_1+p_2)/2$ and $k=p_2-p_1$,  and  the normalization $\left<p|T^{\mu}_\mu|p\right>=2A(0)m_N^2$. In Appendix~\ref{app_PP1},
$A(t)$ is given explicitly for the soft wall model, which is found to be well parametrized by the  dipole form~\cite{Mamo:2019mka}

\be
\label{AKKX}
A(k)=\frac {A(0)}{\bigg(1+\frac {K^2}{m_A^2}\bigg)^2}=\frac {A(0)}{\bigg(1-\frac {t}{m_A^2}\bigg)^2}\equiv A(t)
\ee
with  $m_{A}=1.124$ GeV. It compares well  with the reported lattice value $m_{A,\,\rm lattice}=1.13$ GeV by the MIT group~\cite{Shanahan:2018pib}, for
$A(0)=0.58$.  The gravitational form factor $A(k)$ is saturated by the $2^{++}$
 glueball trajectory without any quark mixing, essentially a quenched result.  The holographic C-term $C(k)=-4A(k)$ is also in good agreement
 with the lattice results. The holographic B-term is found to vanish which is consistent with the lattice findings.	 Note that (\ref{AKKX})  corresponds 
 to a squared gravitational radius of the nucleon as a bulk Dirac fermion 
 
 \bea
 \left<r^2_G\right>=6\bigg(\frac{d{\rm log}A^(t)}{dt}\bigg)_{t=0}=(0.61\,{\rm fm})^2
 \eea
which is smaller than the nucleon squared charge radius $\left<r^2_C\right>=(0.831\,{\rm fm})^2$~\cite{Xiong:2019umf}.

In the Reggeized limit, the dominant contributions in the last two brackets stem from
the first two terms with $p^2, \underline{p}^2\sim s$ as the lead  gravitational coupling to the energy-momentum tensor, with the result

\bea
\label{DIFF2}
&&\frac 14 \sum_{\rm spins}\bigg|{\cal A}_{pp\rightarrow pp 0-}\bigg|^2\approx \frac 14 \,A^2(t_1)\,A^2(t_2)\,\nonumber\\
&&\times\bigg[\bigg(\frac \lambda{\pi^2} +1\bigg)^2\bigg(\frac{\sqrt\lambda}{2\pi}\bigg(\gamma+\frac \pi 2\bigg)\bigg)^4
\frac{e^{-\sqrt\lambda (\gamma+\pi/2)^2(1/{\rm ln}(s_1/\tilde\kappa_N^2)+1/{\rm ln}(s_2/\tilde\kappa_N^2))}}{({\rm ln}(s_1/\tilde\kappa_N^2){\rm ln}(s_2/\tilde\kappa_N^2))^3}
\,\bigg(\frac{s_1}{{\tilde\kappa_N}^2}\bigg)^{2(\alpha_{\mathbb P}(t_1)-2)}\bigg(\frac{s_2}{{\tilde\kappa_N}^2}\bigg)^{2(\alpha_{\mathbb P}(t_2)-2)}\bigg]\nonumber\\
&&\times s^4t_1t_2\bigg(2B_1^-(t_1,t_2)-\mu B_2^-(t_1,t_2)\bigg)^2\,{\rm sin}^2\,\theta_{34}
\eea
The corresponding differential cross section is

\bea
\label{DIFF3}
&&\frac{d\sigma_-}{d\theta_{34} dt_1 dt_2}\approx \frac 1{(8\pi)^4}\,A^2(t_1)\,A^2(t_2)\,t_1t_2\bigg(2B_1^-(t_1,t_2)-\mu B_2^-(t_1,t_2)\bigg)^2\,{\rm sin}^2\,\theta_{34}\nonumber\\
&&\times\bigg[\bigg(\frac \lambda{\pi^2} +1\bigg)^2\bigg(\frac{\sqrt\lambda}{2\pi}\bigg(\gamma+\frac \pi 2\bigg)\bigg)^4
\frac{e^{-\sqrt\lambda (\gamma+\pi/2)^2(1/{\rm ln}(s_1/\tilde\kappa_N^2)+1/{\rm ln}(s_2/\tilde\kappa_N^2))}}{({\rm ln}(s_1/\tilde\kappa_N^2){\rm ln}(s_2/\tilde\kappa_N^2))^3}\nonumber\\
&&\times\bigg[
\,\bigg(\frac{s_1}{{\tilde\kappa_N}^2}\bigg)^{2(\alpha_{\mathbb P}(t_1)-2)}\bigg(\frac{s_2}{{\tilde\kappa_N}^2}\bigg)^{2(\alpha_{\mathbb P}(t_2)-2)}\bigg]
\bigg[s^2{\rm ln}\bigg(\frac s\mu\bigg)\bigg]
\eea
For  $s_{1,2}\sim \sqrt{\mu s}$  and fixed $\mu$ in the Reggeized limit, (\ref{DIFF3}) asymptotes

\be
{\rm ln}\,s\,\bigg[\bigg(\frac{e^{-\#\sqrt\lambda/{\rm ln}\,s}}{({\rm ln}\,s)^{\frac 32}}\bigg)^2\,s^{1+\frac 12 (\alpha_{\mathbb P}(t_1)+\alpha_{\mathbb P}(t_2)-4)}\bigg]^2
\ee
with the first logarithm accounting for the $x_F\sim 0$ phase space enhancement~\cite{Anderson:2014jia}, and the remainder for the 2-Pomeron fusion exchange with its holographic diffusive signature in $5-2=3$ transverse directions. 
The rise in $s$ can be tamed through the Eikonalization of the exchange.

Finally, we note the transverse plane dependence on ${\rm sin}^2\theta_{34}$ of the differential cross section (also through $\mu$ in (\ref{KIN})) 
as inherited from the abnormal parity nature of the Chern-Simons vertex (\ref{CS}) at the origin of the pseudo-scalar emission in bulk. The diffractive emission of the pseudo-scalar glueball
is strongly anisotropic in the transverse plane.

\subsection{Differential cross section: $G^2$}

A similar reasoning applies for 
the spin-averaged squared amplitude for $pp\rightarrow pp0+$ which reads

\bea
\label{DIFFX1}
&&\frac 14 \sum_{\rm spins}\bigg|{\cal A}_{pp\rightarrow pp 0+}\bigg|^2=\frac{16}4\,A^2(t_1)\,A^2(t_2)\,\nonumber\\
&&\times\bigg[\bigg(\frac \lambda{\pi^2} +1\bigg)^2\bigg(\frac{\sqrt\lambda}{2\pi}\bigg(\gamma+\frac \pi 2\bigg)\bigg)^4
\frac{e^{-\sqrt\lambda (\gamma+\pi/2)^2(1/{\rm ln}(s_1/\tilde\kappa_N^2)+1/{\rm ln}(s_2/\tilde\kappa_N^2))}}{({\rm ln}(s_1/\tilde\kappa_N^2){\rm ln}(s_2/\tilde\kappa_N^2))^3}
\,\bigg(\frac{s_1}{{\tilde\kappa_N}^2}\bigg)^{2(\alpha_{\mathbb P}(t_1)-2)}\bigg(\frac{s_2}{{\tilde\kappa_N}^2}\bigg)^{2(\alpha_{\mathbb P}(t_2)-2)}\bigg]\nonumber\\
&&\times\bigg[ B^+(t_1,t_2,p_5^2)\eta_{\alpha\bar\alpha}\eta_{\beta\bar\beta}\bigg]\times
\bigg[ B^+(t_1,t_2,p_5^2)\eta_{\underline\alpha\underline{\bar\alpha}}\eta_{\underline \beta\bar{\underline \beta}}\bigg]\nonumber\\
&&\times\bigg[p^\alpha p^\beta p^{\underline\alpha}p^{\underline\beta}
+\frac 1{16}\bigg(t_1\eta_{\alpha\underline\alpha}p_\beta p_{\underline\beta}+t_1\eta_{\alpha\underline\beta}p_\beta p_{\underline\alpha} +t_1\eta_{\beta\underline\alpha} p_\alpha p_{\underline\beta}
+(t_1\eta_{\beta\underline\beta}-k_{1\beta}k_{1\underline\beta})p_\alpha p_{\underline\alpha}\bigg)\bigg]\nonumber\\
&&\times\bigg[\underline p^{\bar \alpha} \underline p^{\bar \beta} \underline p^{\underline {\bar\alpha}}\underline p^{\bar{\underline \beta}}
+\frac 1{16}\bigg(t_2\eta_{\bar \alpha\bar{\underline \alpha}}p_{\bar \beta} p_{\bar{\underline \beta}}
+t_2\eta_{\bar  \alpha\bar{\underline \beta}}p_{\bar  \beta} p_{\bar{\underline \alpha}} +t_2\eta_{\bar  \beta\bar{\underline \alpha}} p_{\bar  \alpha} p_{\bar{\underline \beta}}
+(t_2\eta_{\bar \beta\bar{\underline \beta}}-k_{1\bar  \beta}k_{1\bar{\underline \beta}})p_{\bar  \alpha} p_{\bar{\underline \alpha}}\bigg)\bigg]
\eea
Again, in the Reggeized limit, the dominant contributions in the last two brackets stem from
the first two terms with $p^2, \underline{p}^2\sim s$, with the result

\bea
\label{DIFFX3}
&&\frac{d\sigma_+}{d\theta_{34} dt_1 dt_2}\approx \frac 1{(8\pi)^4}\,A^2(t_1)\,A^2(t_2)\,B^{+ 2}(t_1,t_2,p_5^2)\nonumber\\
&&\times\bigg[\bigg(\frac \lambda{\pi^2}  +1\bigg)^2\bigg(\frac{\sqrt\lambda}{2\pi}\bigg(\gamma+\frac \pi 2\bigg)\bigg)^4
\frac{e^{-\sqrt\lambda (\gamma+\pi/2)^2(1/{\rm ln}(s_1/\tilde\kappa_N^2)+1/{\rm ln}(s_2/\tilde\kappa_N^2))}}{({\rm ln}(s_1/\tilde\kappa_N^2){\rm ln}(s_2/\tilde\kappa_N^2))^3}\bigg]\nonumber\\
&&\times\bigg[\bigg(\frac{s_1}{{\tilde\kappa_N}^2}\bigg)^{2(\alpha_{\mathbb P}(t_1)-2)}\bigg(\frac{s_2}{{\tilde\kappa_N}^2}\bigg)^{2(\alpha_{\mathbb P}(t_2)-2)}\bigg]
\bigg[s^4{\rm ln}\bigg(\frac s\mu\bigg)\bigg]\nonumber\\
\eea
with the scalar vertex $B^+(t_1,t_2,p_5^2)$ 

\be
\label{CUBIC4}
B^+(t_1, t_2, p_5^2)=\bigg({\bf C}_1^+(p_5^2+(t_1+t_2))-2{\bf C}_2^+-{\bf C}_3^+p_5^2\bigg)
\ee
and $p_5^2=m^2_{0^+}$ on mass-shell.
${\bf C}_{1,2,3}^+$ are given in (\ref{CUBICX2}-\ref{XZ2}) for the soft-wall model. Modulo $\mu$, the emission of the scalar glueball is isotropic in the transverse plane, in contrast to
the pseudo-scalar glueball emission.

\section{Summary}
The subject of this paper -- instanton-sphaleron processes, both in QCD and electroweak setting -- were at the forefront of the theoretical discussions in the late 1980s. The optimal tunneling path was suggested by Yung \cite{Yung:1987zp} and Verbaarschot
\cite{Verbaarschot:1991sq}. The idea to use it 
 to evaluate cross sections was suggested by
Ringwald and Khoze \cite{Khoze:1991sa},  as well as Verbaarschot and Shuryak \cite{Shuryak:1991pn}. A more detailed description of the topological landscape 
and the realization that this path
leads to the production of sphalerons came  in 2002, from  Ref.~\cite{Ostrovsky:2002cg}.
The analytic form of a (pure gauge) sphaleron classical explosion was also obtained in that work, with  a complementary 
derivation by us in~\cite{Shuryak:2002qz}, clarifying the production of fermions. So, from the theoretical
perspective, one might think the issue was completely clarified many years ago.

Ringwald and 
Schrempp \cite{Ringwald:1998ek} put a lot of efforts to identify  the  instanton-induced reactions in deep-inelastic scattering at HERA. The large $Q^2$ scale involved,
pushes the instantons to be
very small, causing the  corresponding cross sections to be also small. However,  the real problem turned out to
be the separation of  the  ``instanton events" from the multi-jet background. 
The more recent version for the LHC, by Khoze et al \cite{1911.09726}, relates the instanton size to  the  fraction
of the collision energy, again with a cross section that is strongly decreasing with the
invariant mass $M=\sqrt{s'}$ of the produced cluster. This mechanism was recently revisited using
single-diffractive medium cluster production and a more favorable choice of kinematics,  but  with stringent cuts~\cite{Khoze:2021jkd}.

While we do not disagree with these approaches, we still think one needs a different experimental strategy. One better starts with smaller
clusters, with invariant masses of several ${\rm GeV}$, which offer clearly
identifiable  mesonic and perhaps baryonic final states, which can be
related to existing theoretical predictions. The background issue would be
helped if instead of min.bias $pp$ collisions one would use the double diffraction, or {\bf PP} collisions.
Last but not least, using gluon-rich but colorless fusing Pomerons, rather than fusing gluonic partons, 
eliminates the need for color flux tubes and cleans up the stage.

Can it all be done in practice, and if so, {\em why now}?
It looks like LHC has reached certain stage of multiple searches
at which all which remains to do is to increase statistics.  The 
search for instanton/sphaleron clusters may not be ``beyond the standard model"
but is still worth trying. 

The
double diffractive direction we advocate is in especially nice position. The
first roman pot (RP) detectors  are  installed and running. ATLAS AFP is tuned to a
huge mass $M\sim 1\, {\rm TeV}$, while CMS-TOTEM has  carried studies of  exclusive reactions
at $M\sim 1\, {\rm GeV}$. The  in between region we suggest to explore is still basically {\em ``terra incognita"}.   It remained untouched for decades, since the pioneering
$UA8$ experiment.

Last but not least, the QCD sphalerons and their explosions are close relatives of the electroweak  sphalerons.
We do not know how one may produce those in the lab, but we do expect them
to be very important at the cosmological electroweak transition~\cite{Kharzeev:2019rsy}.  


\vskip 1cm
{\bf Acknowledgements} 
This work has been triggered by the CERN workshop " Topological Effects in the Standard Model: Instantons, Sphalerons and Beyond at LHC", Dec-2020.
We 
thank its organizers, and especially Matthias Schott,
for organizing it and for reviving this field in general.
This work is supported by the Office of Science, U.S. Department of Energy under Contract No. DE-FG-88ER40388.

\appendix

\section{Some meson couplings used}\label{seq_couplings}

For completeness,  we give  here some standard PCAC formulae and their numerical values, 
\begin{equation}
	K_\pi\equiv \langle 0 |\bar d\gamma^5 u | \pi^+\rangle= i{m_\pi^2 f_\pi \over m_u+m_d }\approx i (500 \, MeV)^2
\end{equation}
\begin{equation}
	K_K\equiv \langle 0 |\bar s\gamma^5 u | K^+\rangle= i{m_K^2 f_K \over m_u+m_s }\approx i (523\, MeV)^2
\end{equation}
The $\eta,\eta'$ couplings depend on  the mixing angle. We  will use the same  values adopted  in~\cite{Zetocha:2002as} 

\begin{eqnarray}
	K^q_\eta &\equiv&
	\langle 0 |\bar u \gamma^5 u |\eta \rangle= \langle 0 |\bar d \gamma^5 d |\eta \rangle=-i (358\, MeV)^2  \\
	K^q_{\eta'}&\equiv&
\langle 0 |\bar u \gamma^5 u |\eta' \rangle= \langle 0 |\bar d \gamma^5 d |\eta' \rangle=-i (320\, MeV)^2  \nonumber \\
K^s_\eta &\equiv & \langle 0 |\bar s \gamma^5 s |\eta \rangle  =i (435\, MeV)^2
    \nonumber \\
    K^s_{\eta'}&\equiv & \langle 0 |\bar s \gamma^5 s |\eta' \rangle  =i (481\, MeV)^2
\nonumber 
\end{eqnarray}	
	The other issue is to define
	the  $\bar s s$ part of $\eta,\eta'$   responsible  for the observed $\eta\pi\pi$ and $\eta'\pi\pi$
	decay rates.  The $\eta-\eta' $ mixing has been well researched.
	In terms of singlet and octet states
	\begin{eqnarray}
		\eta_8 =&{1\over \sqrt{6}}(\bar u u + \bar d d -2 \bar s s)\\ \nonumber
		\eta_1 =&{1\over \sqrt{3}}(\bar u u + \bar d d + \bar s s)
	\end{eqnarray} 
	the standard definition of the mixing is
	\begin{eqnarray}
		\eta' &=& {\rm sin}(\theta_p) \eta_8 + {\rm cos}(\theta_p)\eta_1 \\ \nonumber
		\eta &=& {\rm cos}(\theta_p)\eta_8 - {\rm sin}(\theta_p)\eta_1
	\end{eqnarray} 
	Using the mixing angle $\theta_p\approx -14.6^o$, 
	one finds that the ratio of the $\bar s s$ probabilities (squares of amplitudes) is
	\begin{equation}
		{ss_{\eta'} \over {ss}_{\eta}} =\bigg({0.764 \over 0.644}\bigg)^2=1.40
	\end{equation}

\section{Instanton-induced $uuddss$ interactions}\label{int_instanton}~\label{app_fierz}
In this Appendix  we outline the general structure of the induced $^\prime$t Hoot  vertex for multi-quark states.
In particular, we show the explicit steps that lead to the 6-quark vertex both in the color bleached and bleached form.
its generalization to 8- and 10-quark states follows. 
An alternative but tedious formula involving color averaging over the color moduli with also be discussed which can be 
generalized to even higher multi-quark states.

\subsection{Details of the color-spin reduction}~\label{app_thooft}

The momentum-space kernel for the LSZ reduced zero-mode  propagator in the instanton background reads

\bea
\label{V1}
L\equiv \bigg[\frac{(p^2\varphi^\prime (p))_0^2}{m\rho}\bigg]\bigg[\frac 18\,U(\delta_{\mu\nu}+i{\overline\eta}_{\mu\nu}^a\tau^a)U^\dagger\times \gamma_\mu\gamma_\nu\bigg]
\frac {1-\gamma_5}2
\eea
in the zero momentum limit $(p^2\varphi^{\prime 2}(p)_0\equiv \sqrt{C}$. 
The 6-quark  $uuddss$ vertex stemming from (\ref{V1}) follows from  averaging over the color matrices $U$
the product 

\bea
\label{V2}
\int dU\, [L]^3&=&
\bigg[\frac 18\,U(\delta_{\alpha\beta}+i{\overline\eta}_{\alpha\beta}^a\tau^a)U^\dagger\times \gamma_\alpha\gamma_\beta\bigg]\nonumber\\
&\times&\bigg[\frac 18\,U(\delta_{\mu\nu}+i{\overline\eta}_{\mu\nu}^b\tau^b)U^\dagger\times \gamma_\mu\gamma_\nu\bigg]\times
\bigg[ \frac 18\,U(\delta_{\lambda\tau}+i{\overline\eta}_{\lambda\tau}^c\tau^c)U^\dagger\times \gamma_\lambda\gamma_\tau\bigg]
\eea
using the identity~\cite{Chernyshev:1995gj}

\bea
\label{V3}
&&\int dU\, [U^a_iU^{\dagger j}_b]^3=\bigg[\frac 1{N_c}\delta^{a_1}_{b_1}\delta_{i_1}^{j_1}\,{\bf 1}_1\bigg]
\bigg[\frac 1{N_c}\delta^{a_2}_{b_2}\delta_{i_2}^{j_2}\,{\bf 1}_2\bigg]\bigg[\frac 1{N_c}\delta^{a_3}_{b_3}\delta_{i_3}^{j_3}\,{\bf 1}_3\bigg]\nonumber\\
&&+\bigg(\bigg[\frac 1{N_c}\delta^{a_1}_{b_1}\delta^{j_1}_{i_1}\,{\bf 1}_1\bigg]
\bigg[\frac 1{4(N_c^2-1)}[\lambda_2^A]^{a_2}_{b_2}[\lambda_3^A]^{a_3}_{b_3}[\lambda_2^B]^{j_2}_{i_2}[\lambda_3^B]^{j_3}_{i_3}\bigg]+2\,{\rm  perm.}\bigg)\nonumber\\
&&+\bigg(\frac{N_c}{8(N_c^2-1)(N_c^2-4)}\bigg[d^{ABC}([\lambda_1^A]^{a_1}_{b_1}[\lambda_2^B]^{a_2}_{b_2})[\lambda_3^C]^{a_3}_{b_3}\bigg]
\bigg[d^{IJK}([\lambda_1^I]^{j_1}_{i_1}[\lambda_2^J]^{j_2}_{i_2})[\lambda_3^K]^{j_3}_{i_3}\bigg)\bigg]\nonumber\\
&&+\bigg(\frac{1}{8N_c(N_c^2-1)}\bigg[f^{ABC}([\lambda_1^A]^{a_1}_{b_1}[\lambda_2^B]^{a_2}_{b_2})[\lambda_3^C]^{a_3}_{b_3}\bigg]
\bigg[f^{IJK}([\lambda_1^I]^{j_1}_{i_1}[\lambda_2^J]^{j_2}_{i_2})[\lambda_3^K]^{j_3}_{i_3}\bigg]
\bigg)
\eea
which follows from the projection onto the color singlet channel.
The result after some algebra is

\bea
\label{V4}
\bigg[\frac{C}{m\rho N_c}\bigg]^3
&\bigg(&UDS+\frac{N_c(N_c-2)}{4(N_c^2-1)}\bigg[U^AD^AS+U^ADS^A+UD^AS^A\bigg]\nonumber\\
&&-\frac{N_c^2}{16(N_c^2-1)}\bigg[U_{\mu\nu}^AD_{\mu\nu}^AS+U_{\mu\nu}^ADS_{\mu\nu}^A+UD_{\mu\nu}^AS_{\mu\nu}^A\bigg]\bigg)\nonumber\\
&&-\frac{N_c^3(N_c-2)}{96(N_c^2-1)(N_c^2-4)}d^{ABC}\bigg[U_{\mu\nu}^AD_{\mu\nu}^BS^C+U_{\mu\nu}^AD^BS_{\mu\nu}^C+U^AD_{\mu\nu}^BS_{\mu\nu}^C\bigg]\nonumber\\
&&+\frac{N_c^2(N_c-2)}{8(N_c^2-1)(N_c+2)}d^{ABC}U^AD^BS^C-\frac{N_c^2}{32(N_c^2-1)}if^{ABC}U_{\mu\nu}^AD_{\nu\rho}^BS_{\rho\mu}^C\bigg)
\eea
where we have defined

\bea
\label{V5}
&&U=\overline u_Ru_L\qquad U^A=\overline u_R\lambda^A u_L\qquad U_{\mu\nu}^A=\overline u_R\lambda^A \gamma_\mu\gamma_\nu u_L\nonumber\\
&&D=\overline d_Rd_L\qquad D^A=\overline d_R\lambda^A d_L\qquad D_{\mu\nu}^A=\overline d_R\lambda^A \gamma_\mu\gamma_\nu d_L\nonumber\\
&&S=\overline s_Rs_L\qquad S^A=\overline s_R\lambda^A s_L\qquad S_{\mu\nu}^A=\overline s_R\lambda^A \gamma_\mu\gamma_\nu s_L
\eea
Throughout in the notation $\gamma_\mu\gamma_\nu$ the condition $\mu<\nu$ is subsumed.
(\ref{V3}) was originally derived in~\cite{Nowak:1988bh} for $N_c=3$.

(\ref{V4}) can be considerably reduced by successive Fierzing to bleach the color and  bring it to a determinantal form typical of instanton induced interactions
that are manifestly $SU(N_f)$ symmetric but $U(1_f)$ violating. Since the procedure is considerably lengthy, we will show how the Fierzing 
works for the typical blocks   in (\ref{V4}).


Consider the  typical  colored scalar block in (\ref{V4})

\bea
\label{V6}
U^AD^A\equiv&& \overline u_R\lambda^Au_L\overline d_R\lambda^A d_L\nonumber\\
=&&\overline u_{Ra}u_{Lb}\overline d_{Rc} d_{Ld}\,[\lambda^A]^{ab}[\lambda^A]^{cd}
=2\overline u_{Ra}u_{Lb}\overline d_{Rb} d_{La}-\frac 2{N_c}\overline u_Ru_L\overline d_Rd_L\nonumber\\
=&&-\frac 2{N_c}\overline u_Ru_L\overline d_Rd_L-\overline u_Rd_L\overline d_Ru_L+\frac 14 \overline u_R\gamma_\mu\gamma_\nu d_L\overline d_R\gamma_\mu\gamma_\nu u_L
\eea
where we made use of the color identity  in the second line

\bea
\label{V7}
[\lambda^A]^{ab}[\lambda^A]^{cd}=2\delta^{ad}\delta^{bc}-\frac 2{N_c}\delta^{ab}\delta^{cd}
\eea
and the Fierz re-arrangement in the third line

\bea
\overline u_{Ra}u_{Lb}\overline d_{Rb} d_{La}
=-\frac 12 \overline u_Rd_L\overline d_Ru_L+\frac 18 \overline u_R \gamma_\mu\gamma_\nu d_L \overline d_R\gamma_\mu\gamma_\nu  u_L
\eea
The same procedure carries  for the colored  tensor block with the result

\bea
\label{V8}
U_{\mu\nu}^AD_{\mu\nu}^A\equiv&& \overline u_R\lambda^A \gamma_\mu\gamma_\nu u_L\overline d_R\lambda^A \gamma_\mu\gamma_\nu d_L\nonumber\\
=&&-\frac 2{N_c}\overline u_R\gamma_\mu\gamma_\nu u_L\overline d_R \gamma_\mu\gamma_\nu d_L+12\overline u_Rd_L\overline d_Ru_L
+\overline u_R\gamma_\mu\gamma_\nu d_L\overline d_R\gamma_\mu\gamma_\nu u_L
\eea
using the Fierz re-arrangement

\bea
\label{V9}
\overline u_{Ra}\gamma_\mu\gamma_\nu u_{Lb}\overline d_{Rb}\gamma_\mu\gamma_\nu  d_{La}
=6  \overline u_R d_L\overline  d_Ru_L+\frac 12  \overline u_R \gamma_\mu\gamma_\nu d_L \overline d_R\gamma_\mu\gamma_\nu  u_L
\eea

The bleaching of color from the contributions involving the structure factors $f^{ABC}$ and $d^{ABC}$ can be simplified by noting the 
respective identities

\bea
\label{V10}
if^{ABC}[\lambda^A]^{ab}[\lambda^B]^{cd}[\lambda^C]^{ef}=2\bigg(\delta^{ad}\delta^{cf}\delta^{eb}-\delta^{af}\delta^{cb}\delta^{ed}\bigg)
\eea
and

\bea
\label{V11}
d^{ABC}[\lambda^A]^{ab}[\lambda^B]^{cd}[\lambda^C]^{ef}=&&
2\bigg(\delta^{ad}\delta^{cf}\delta^{eb}+\delta^{af}\delta^{cb}\delta^{ed}\bigg)-\frac{4}{N_c^2}\delta^{ab}\delta^{cd}\delta^{ef}\nonumber\\
-&&\frac 2{N_c}\bigg(\delta^{ab}[\lambda^A]^{cd}[\lambda^A]^{ef}+\delta^{cd}[\lambda^A]^{ab}[\lambda^A]^{ef}+\delta^{ef}[\lambda^A]^{ab}[\lambda^A]^{cd}\bigg)
\eea
(\ref{V10}-\ref{V11}) can be established by Fierzing in color and using the identities

\bea
&&\frac 18 d^{ABC}[\lambda^A]^a_b[\lambda^B]^b_c[\lambda^C]^c_d=\frac{N_c^2-4}{2N_c}C_F\,\delta^a_d\nonumber\\
&&\frac 18 f^{ABC}[\lambda^A]^a_b[\lambda^B]^b_c[\lambda^C]^c_d=\frac i2 N_c C_F\, \delta^a_d
\eea
with $C_F=(N_c^2-1)/2N_c$ the Casimir in the fundamental representation.
(\ref{V11}) can be totally  color reduced using (\ref{V7}).  Inserting (\ref{V6}-\ref{V8}-\ref{V10}-\ref{V11}) in (\ref{V4}) yield (\ref{DET1}) after
lengthy algebraic re-arrangements. 
Note that in (\ref{DET1}) we are still using the notation $Q_{\mu\nu}$ but with $\sigma_{\mu\nu}$ instead 
of $\gamma_\mu\gamma_\nu$ with $\mu<\nu$ subsumed.

\subsection{Alternative form}

An alternative form for the 6-quark $^\prime$t Hooft vertex, can be reached by reconsidering the full LSZ reduced amplitude
in Weyl notations

\bea
\label{HOOFT1}
\bigg[\frac{n_I}2\bigg]_\rho\,&&\int \frac{d^4k_1}{(2\pi)^4}\frac{d^4k_2}{(2\pi)^4} \frac{d^4p_1}{(2\pi)^4}\frac{d^4p_2}{(2\pi)^4} \frac{d^4q_1}{(2\pi)^4}\frac{d^4q_2}{(2\pi)^4}
\bigg[(2\pi)^4\delta^4(k_1+p_1+q_1-k_2-p_2-q_2)\bigg]\nonumber\\
 &&\times\bigg[\bigg<
u_R^\dagger(k_2)k_2\bigg[\sqrt{2}\varphi^\prime(k_2)\hat k_2\epsilon U\bigg]\frac 1{m}
\bigg[\sqrt{2}\varphi^\prime(k_1) U^\dagger\epsilon\hat k_1\bigg]k_1 u_L(k_1)\nonumber\\
&& \qquad\times d_R^\dagger(p_2)p_2\bigg[\sqrt{2}\varphi^\prime(p_2)\hat p_2\epsilon U\bigg]\frac 1{m}
\bigg[\sqrt{2}\varphi^\prime(p_1) U^\dagger\epsilon\hat p_1\bigg]p_1 d_L(p_1)\nonumber\\
&&\qquad\times s_R^\dagger(q_2)q_2\bigg[\sqrt{2}\varphi^\prime(q_2)\hat q_2\epsilon U\bigg]\frac 1{m}
\bigg[\sqrt{2}\varphi^\prime(q_1) U^\dagger\epsilon\hat q_1\bigg]q_1 s_L(q_1)\bigg>_{U}\bigg]+{L\leftrightarrow R}
\eea
after averaging over the instanton Z-position in 4-Euclidean space, without taking the zero momentum limit as in (\ref{V1}). The left zero mode in the instanton background
in momentum space is

\bea
\psi^\alpha_{iI}(p)=\sqrt{2}\varphi^\prime(p)(\hat p\epsilon U)^\alpha_i
\qquad\qquad
\varphi^\prime(p)=\pi\rho^2\bigg(I_0K_0-I_1K_1\bigg)^\prime(z=\rho p/2)
\eea
in terms of which (\ref{HOOFT1}) reads

\bea
\bigg[\frac{n_I}2\bigg]_\rho\,&&\int \frac{d^4k_1}{(2\pi)^4}\frac{d^4k_2}{(2\pi)^4} \frac{d^4p_1}{(2\pi)^4}\frac{d^4p_2}{(2\pi)^4} \frac{d^4q_1}{(2\pi)^4}\frac{d^4q_2}{(2\pi)^4}
\bigg[(2\pi)^4\delta^4(k_1+p_1+q_1-k_2-p_2-q_2)\bigg]\nonumber\\
 &&\times
 \bigg[\frac 1{m^3} (2k_1k_2\varphi^\prime(k_1)\varphi^\prime(k_2))(2p_1p_2\varphi^\prime(p_1)\varphi^\prime(p_2))(2q_1q_2\varphi^\prime(q_1)\varphi^\prime(q_2))\bigg]\nonumber\\
&&\times
 \bigg<
\bigg[u_R^\dagger(k_2)\epsilon U\bigg]\bigg[ U^\dagger\epsilon u_L(k_1)\bigg]\,
\bigg[d_R^\dagger(p_2)\epsilon U\bigg]\bigg[ U^\dagger\epsilon d_L(p_1)\bigg]\,
\bigg[s_R^\dagger(q_2)\epsilon U\bigg]\bigg[ U^\dagger\epsilon s_L(q_1)
\bigg>_U\bigg]
+{L\leftrightarrow R}\nonumber\\
\eea
The color bracket is explicitly

\bea
&&\bigg<U^{a_1}_{c_1}U^{\dagger b_1}_{d_1}
U^{a_2}_{c_2}U^{\dagger b_2}_{d_2}
U^{a_3}_{c_3}U^{\dagger b_3}_{d_3}\bigg>_U
\bigg[\epsilon^{i_1c_1}\epsilon_{j_1b_1}\,\epsilon^{i_2c_2}\epsilon_{j_2b_2}\,\epsilon^{i_3c_3}\epsilon_{j_3b_3}\bigg]
\nonumber\\
\times&&\bigg[ u^\dagger_{Ra_1i_1}(k_2)u_L^{d_1j_1}(k_1)\,d^\dagger_{Ra_2i_2}(p_2)d_L^{d_2j_2}(p_1)\,s^\dagger_{Ra_3i_3}(q_2)s_L^{d_3j_3}(q_1)\bigg]
\eea
The color averaging over the unitary matrices can be written in terms of the Weingarten coefficients

\bea
&&\bigg<U^{a_1}_{c_1}U^{\dagger b_1}_{d_1}
U^{a_2}_{c_2}U^{\dagger b_2}_{d_2}
U^{a_3}_{c_3}U^{\dagger b_3}_{d_3}\bigg>_U=\nonumber\\
&&\frac{(N_c^2-2)}{N_c(N_c^2-1)(N_c^2-4)}\sum_{n=1}^{3!}\delta^{a_1a_2a_3}_{(d_1d_2d_3)_n}\delta^{c_1c_2c_3}_{(b_1b_2b_3)_n}\nonumber\\
-&&\frac{1}{(N_c^2-1)(N_c^2-4)}\sum_{n=1}^{3!}\delta^{a_1a_2a_3}_{(d_1d_2d_3)_n}
\bigg(\delta^{c_1c_2c_3}_{(b_2b_1b_3)_n}+\delta^{c_1c_2c_3}_{(b_1b_3b_2)_n}+\delta^{c_1c_2c_3}_{(b_3b_2b_1)_n}\bigg)\nonumber\\
+&&\frac{2}{N_c(N_c^2-1)(N_c^2-4)}\sum_{n=1}^{3!}\delta^{a_1a_2a_3}_{(d_1d_2d_3)_n}
\bigg(\delta^{c_1c_2c_3}_{(b_3b_1b_2)_n}+\delta^{c_1c_2c_3}_{(b_2b_3b_1)_n}\bigg)
\eea
The short-hand notation 

\bea
\delta^{a_1a_2a_3}_{(d_1d_2d_3)_n}\equiv \sum^{3!}_{n=1}\bigg(\delta^{a_1}_{d_1}\delta^{a_2}_{d_2}\delta^{a_3}_{d_3}+{\rm perm.}\bigg)
\eea 
refers to the product of three kroneckers in the sum over the n! permutations of the permutation group $S_3$.
For completeness, we note  the analogous relations for lower color averagings  for two flavors

\bea
&&\left<U^{a_1}_{c_1}U^{\dagger b_1}_{d_1}U^{a_2}_{c_2}U^{\dagger b_2}_{d_2}\right>_U=\nonumber\\
&&\frac 1{N_c^2-1}\sum_{n=1}^{2!}\delta^{a_1a_2}_{(d_1d_2)_n}\delta^{c_1c_2}_{(b_1b_2)_n}
-\frac 1{N_c(N_c^2-1)}\sum_{n=1}^{2!}\delta^{a_1a_2}_{(d_1d_2)_n}\delta^{c_1c_2}_{(b_2b_1)_n}
\eea
and the well known one for one flavor

\bea
\left<U^{a}_{c}U^{\dagger b}_{d}\right>_U=\frac 1{N_c}\delta^a_d\delta_b^c
\eea

\section{Exploding quark modes in Sphaleron background}

In so far, we have extracted the sphaleron-six-quark vertex using the $O(4)$ exact zero modes in a self-dual instanton (anti-instanton) background. 
While at $t=0$ this is the case, as $t>0$ the zero modes become real and turn to $O(3)$ exploding quark modes as we discussed in~\cite{Shuryak:2002qz}.
We now briefly review how these modes are constructed and then substitute them in (\ref{SSIX}) to generate ${\cal L}(t>0)$  to describe 
the final state explosive vertex in the diffractive $pp\rightarrow ppX$ process,

\subsection{O(4) and O(3) fermionic zero mode}

A Dirac fermion in a general gauge field solves the equation

\be
\label{OZ1}
\left(\partial_\mu-iA_\mu\right)\gamma_\mu\,{\psi}=0
\ee
We will use the chiral basis  with spin matrices $\bar\sigma_{Ms}=(1, -i\vec\sigma_s)$
 and

\begin{eqnarray}
\label{OZ2}
\gamma_5=
\left(
\begin{array}{cc}
  1 &   0  \\
 0  &   -1
\end{array}
\right)
\qquad
\gamma_\mu=
\left(
\begin{array}{cc}
  0 &   \sigma_{M s}   \\
  \bar\sigma_{Ms}  &   0
\end{array}
\right)
\end{eqnarray}
If we recall  that the t$^{\prime}$ Hooft symbol satisfies the color identity
$\bar\sigma_{M c} \sigma_{N c}=\sigma_{ac} \overline{\eta}_{a MN}$
with  $\sigma_{Mc}=(1, -i\vec \sigma_c)$,
then the positive chirality mode $\psi_+$ solves  (\ref{OZ1})

\be
\label{OZ4}
\left(\bar\sigma_{M s}\partial_M+\frac 12 \bar\sigma_{N s}\bar\sigma_{N c}\,\sigma_{M
c}\partial_M\,F\right)\,{\psi}_+ =0
\ee
with the spin and color matrices commuting. $F(\xi(y))$ 
for an $O(4)$ symmetric gauge field is given in (\ref{FO4}).
Note that while writing  (\ref{OZ4}) we have added a U(1) part to the gauge field
for notational simplicity, and will remove it in the final answer by inspection.  A formal
solution to  (\ref{OZ4}) is $(\psi_+)^a_\mu=\varphi \epsilon_\mu^a$
wich is a singlet (hedgehog)  in color-spin space 

\be
\label{OZ5X1}
\sigma_{Ms}\chi_Q=\bar\sigma_{Mc}\epsilon\qquad \bar\sigma_{Ms}\epsilon=\sigma_{Mc}\epsilon
\ee
Inserting  (\ref{OZ5X1}) in (\ref{OZ4})  yields

\be
\label{OZ5X2}
\left(\bar\sigma_{M s}\partial_M+\frac 12 \bar\sigma_{N s}\sigma_{N s}\,\bar\sigma_{M
s}\partial_M\,F\right)\,\varphi\epsilon=0
\ee
To remove the redundant  U(1) contribution noted above we use

\be
\label{OZ5X3}
\bar\sigma_{N s}\sigma_{N s}\,\epsilon=
\left(1+(\vec\sigma_s)^2\right)\epsilon\rightarrow (\vec\sigma_s)^2\epsilon=3\epsilon
\ee
after which $\varphi$ is seen to solve $\varphi^\prime+\frac 32 F^\prime \varphi=0$, hence

\be
\label{OZ7}
({\psi}_+)^a_\mu(y)= {\bf C}\,e^{-\frac 32 F(\xi (y))}\epsilon^a_\mu
\ee
with

\be
\xi(y)=
\frac 12 {\rm ln}\left(\frac {(t-\rho)^2+r^2}{(t+\rho)^2+r^2}\right)
\ee
The overall normalization ${\bf C}$ is fixed by

\be
\label{O28}
{\bf C}=\left|\int_{T_{\frac 12}} d^4y\, e^{-3F(\xi(y))}\right|^{-\frac 12}
\ee
with $T_{\frac 12}$ the sphaleron tunneling time.
(\ref{O28}) is  agreement with the result in~\cite{Shuryak:2002qz}
(see Eq. 22  with the exponent 
$2\rightarrow \frac 32$ when (\ref{OZ5X3}) is enforced).

The chiral O(3) symmetric zero mode  follows by applying the off-center inversion  (\ref{OFFCENTER})
in Section~\ref{sec_explode}, 
onto  the O(4) symmetric zero mode in (\ref{OZ7}). The corresponding tansform is

\be
{\tilde \psi}_+(x) = \frac{\sigma^\dagger_\mu \,(y +a)_\mu}{1/(y+a)^2}
\,{ \psi}_+ (y)
\label{x1}
\ee
or more explicitly ($r=|\vec x|$)

\bea
({\tilde \psi}_+)^a_\mu (t,r) =
\frac{{8\bf\, C}\rho^6}{((t+\rho)^2+r^2)^2}\,e^{-\frac 32 F(\xi(y))}\,
\big[ \left((t+\rho) +i\vec{\sigma}\cdot\vec{x}\right)\,\epsilon\big]^a_\mu
\label{ZX19}
\eea
This result is in agreement with the one derived in~\cite{Shuryak:2002qz}
 prior to the analytical continuation to Minkowski space.

\section{Graviton-nucleon coupling}~\label{app_PP1}

In bulk, the graviton couples to the Dirac fermion through its  energy momentum tensor~\cite{Abidin:2009hr,Mamo:2019mka}
 
  \be
  \label{GTMUNU}
-\frac{\sqrt{2\kappa^2}}{2}\int d^5x\,\sqrt{g}\,h_{\mu\nu}T_F^{\mu\nu}=
-\frac{\sqrt{2\kappa^2}}{2}\int d^5x\,\sqrt{g}\,
\bigg(\frac{e^{-\phi}}{2g_5^2}\frac{i}{2}\,z\,\overline\Psi\gamma^\mu\overset{\leftrightarrow}{\partial^\nu}\Psi-\eta^{\mu\nu}\mathcal{L}_F\bigg)
 \ee
with typically for the Dirac fermions in the soft-wall model

 \bea
 \label{DLF}
\mathcal{L}_F=&& \frac{e^{-\phi(z)}}{2g_{5}^2}\,
 \bigg( \frac{i}{2} \bar{\Psi} e^N_A \Gamma^A\big(\overrightarrow{D}_N-\overleftarrow{D}_N\big)\Psi-(M+V(z))\bar{\Psi}\Psi\bigg)\,,
 \eea
with $M=\tau-3$ the anomalous dimension. The potential $V(z)=\tilde{\kappa}_{N}^2z^2$ is added to make the Dirac fermions massive in bulk.
Here $e^N_A=z \delta^N_A$ denotes the inverse vielbein, and   the covariant derivatives are defined as

\bea
\overrightarrow{D}_N=&&\overrightarrow{\partial}_N +\frac{1}{8}\omega_{NAB}[\Gamma^A,\Gamma^B]\nonumber\\
\overleftarrow{D}_N=&&\overleftarrow{\partial}_N +\frac{1}{8}\omega_{NAB}[\Gamma^A,\Gamma^B]
\eea
The components of the spin connection are $\omega_{\mu z\nu}=-\omega_{\mu\nu z}=\frac{1}{z}\eta_{\mu\nu}$, 
the Dirac gamma matrices  satisfy anti-commutation relation $\{\Gamma^A,\Gamma^B\}=2\eta^{AB}$  with the
explicit choice  $\Gamma^A=(\gamma^\mu,-i\gamma^5)$.

In the nucleon as a bulk Dirac fermion, the energy momentum tensor is characterized by the three invariant form factors in (\ref{EMT2}).
For the  soft wall model, we have explicitly~\cite{Abidin:2009hr,Mamo:2019mka}

\be\label{Aff}
A(k)=\frac{1}{2}\int dz\sqrt{g}\,e^{-\phi}z\,\big(\psi_R^2(z)+\psi_L^2(z)\big)\,\mathcal{H}(k,z)=-\frac{C(k)}{(\alpha z_0m_N/2)^2}
\ee
with ${\cal H}(K,z^\prime)$ the non-normalizable bulk-to-boundary graviton propagator~\cite{Hong:2004sa,Abidin:2008ku,BallonBayona:2007qr}

\bea
\label{BTOBH}
\mathcal{H}(K, z)=&&4z^{\prime 4}\Gamma(a_K+2)U\Big(a_K+2,3;2\tilde{\kappa}_N^2z^{ 2}\Big) =\Gamma(a_K+2)U\Big(a_K,-1;2\tilde{\kappa}_N^2z^{2}\Big)\nonumber\\
=&&\frac{\Gamma(a_K+2)}{\Gamma(a_K)} \int_{0}^{1}dx\,x^{a_K-1}(1-x){\rm exp}\Big(-\frac{x}{1-x}(2\tilde{\kappa}_N^2z^{ 2})\Big)
\eea
and  $a_K={K^2}/{8\tilde{\kappa}_N^2}\geq 0$. We  have used the transformation $U(m,n;y)=y^{1-n}U(1+m-n,2-n,y)$. 
(\ref{BTOBH}) satisfies the normalization condition ${\cal H}(0,z^\prime)={\cal H}(K,0)=1$.
The bulk Dirac fermions $\Psi(x,z)=\psi_{L,R}(z)e^{-ip\cdot x}u_{L,R}(p)$ are

\bea
\label{BDIRAC}
\psi_R(z)=\frac{\tilde{n}_R}{\tilde{\kappa}_N^{\tau-2}} z^{\frac{5}{2}}\xi^{\frac{\tau-2}{2}}L_0^{(\tau-2)}(\xi)
\qquad \psi_L(z)=\frac{\tilde{n}_L}{\tilde{\kappa}_N^{\tau-1}} z^{\frac{5}{2}}\xi^{\frac{\tau-1}{2}}L_0^{(\tau-1)}(\xi)\,
\eea
Here  $\xi=\tilde\kappa_N^2 z^2$, $\tau=7/2-1/2=3$ is the twist parameter, and 
$L_n^{(\alpha)}(\xi)$ are the generalized Laguerre polynomials, with 

\be
\tilde{n}_R=\tilde{n}_L \tilde{\kappa}_N^{-1}\sqrt{\tau-1}\qquad
\tilde{n}_L=\tilde{\kappa}_N^{\tau}\sqrt{{2}/{\Gamma(\tau)}}\nonumber\\
\ee
More specifically~\cite{Abidin:2009hr,Mamo:2019mka} ($k^2=-K^2=t\leq 0$)

\bea
\label{FFj2}
A(k)&=&A(0)\,(a_k+1)\bigg(-\left(1+a_k+2a_k^2\right)+2\left({a_k}+2{a^3_k}\right)\Phi(-1,1,a_k)\bigg) \nonumber\\
&=&A(0) \bigg((1-2a_k)(1+a_k^2)+a_k(1+a_k)(1+2a_k^2)\bigg(\psi\bigg(\frac{1+a_k}{2} \bigg)-\psi\bigg(\frac{a_k}{2}\bigg)\bigg)\bigg)
\eea
with $a_k={k^2}/{8\tilde\kappa_N^2}$, and $B(0)=0$. $\Phi(-1,1,a')$ refers to the LerchPhi function, and $\psi(x)$ refers to the digamma function 
or harmonic number $H_x=\psi(x)+\gamma$. 
The gravitational form factor $C(k)$ is proportional to $A(k)$ modulo a negative  overall constant
$-(\alpha z_0 m_N/2)^2<0$ which is left undetermined since $\alpha$ is arbitrary in the tensor
decomposition of the graviton~\cite{Mamo:2019mka}.
Note that ${\cal H}(k,0)={\cal H}(0,z)$ is arbitrary (1-point function),  so that 
$A(0)$ is not fixed in holography.

\section{Spin-j-nucleon coupling}~\label{app_PP2}

For diffractive scattering at large $\sqrt{s}$,  the spin-2 graviton coupling reggeizes to spin-j coupling.
The resummed spin-j couplings transmutes to the Pomeron coupling. The generic form of the spin-j
coupling is~\cite{Mamo:2019mka}

\bea
\label{VHPP}
\mathcal{V}^{\alpha\beta(TT)}_{h\bar\Psi\Psi}(j,p_1,p_2,k_z)=
-\frac{\sqrt{2\kappa^2}}{2}\int dz\,\sqrt{g}\,e^{-\phi}\,\mathcal{H}(j,K,z)\,
\bigg[z^{1+2(j-2)}\frac 12 \bar\Psi(p_2,z)\gamma^{[\alpha} p^{\beta]}\,\Psi(p_1,z)z^{-(j-2)}\bigg]\nonumber\\
\eea
where ${\cal H}(j,K,z)$ is the spin-j bulk to boundary space-like propagator. 
For the pure spin-2 graviton $h_\mu^\mu=0$ is traceless, so the coupling to ${\cal L}_F$ in (\ref{GTMUNU}) drops out.
The z-pre-factors in the internal braket in (\ref{VHPP}) reflect
on the spin-j  generalization of the energy-momentum tensor (higher twist vertices).
For the soft-wall model,  (\ref{VHPP})  takes the explicit form

\be
\mathcal{V}^{\alpha\beta(TT)}_{h\bar\Psi\Psi}(j,p_1,p_2,k_z)=
-\sqrt{2\kappa^2}\times g_5^2A(j,K)\times\frac 12 \bar u(p_2)\gamma^{[\alpha} p^{\beta]} u(p_1)
\ee
with

\bea
\label{Aj1}
A(j,K)=&&\frac 1{2g_5^2}\int dz\,\sqrt{g}\,e^{-\phi}\,\mathcal{H}(j,K,z)\bigg[z^{1+2(j-2)}\,\big(\psi_R^2(z)+\psi_L^2(z)\big)\,z^{-(j-2)}\bigg]\nonumber\\
=&&\frac{1}{2g_5^2}\frac{2^{2-\Delta(j)}\tilde{\kappa}_N^{j-2-\Delta(j)}}{\Gamma(\tilde{a}(j))}
\int_{0}^{1}dx\,{x^{\tilde{a}(j)-1}}{(1-x)^{-\tilde{b}(j)}}\big(I_z^R(x)+I_z^L(x)\big)\,,
\eea
where  the fermionic integrals are defined as ($\xi=\tilde\kappa_N^2 z^2$)

\be
\label{IFERMION}
I_z^{R/L}(x)=\int dz\,\sqrt{g}\,e^{-\phi}\,z^{1+2(j-2)}\,
\psi_{R/L}^2(z)\,\xi^{\frac{-(j-2)}{2}}\xi^{2-\frac{\Delta(j)}{2}}
{\rm exp}\Big(-\frac{2x\xi}{1-x}\Big)\,,
\ee
and  j-dependent parameters are

\bea
\label{PARAX}
\tilde a(j)=a_k+2-\frac 12 \Delta(j)\qquad &{\rm and}&\qquad \tilde b(j)=3-\Delta(j)
\nonumber\\
\Delta(j)=2+\sqrt{2\sqrt{\lambda}(j-j_0)}\qquad &{\rm and}&\qquad j_0=2-\frac{2}{\sqrt{\lambda}}
\eea
 Using the bulk Dirac wavefunctions (\ref{BDIRAC}) for the soft-wall model, we can further reduce the
fermionic integrals (\ref{IFERMION}) to have~\cite{Mamo:2019mka}


\bea
\label{Iz3}
&&I_z^{R}(x)=\frac{1}{2}\times\tilde{\kappa}_N^{-2(j-2)}\times\bigg(\frac{\tilde{n}_R}{\tilde{\kappa}_N^{\tau-1}}\bigg)^2\times\Gamma\bigg(\frac{j-2}{2}+\tau -\frac{\Delta(j)}{2}+1\bigg)\times\bigg(\frac{1+x}{1-x}\bigg)^{-\frac{j-2}{2}-\tau +\frac{\Delta(j)}{2}-1}\,,\nonumber\\
&&I_z^{L}(x)=\frac{1}{2}\times\tilde{\kappa}_N^{-2(j-2)}\times\bigg(\frac{\tilde{n}_L}{\tilde{\kappa}_N^{\tau}}\bigg)^2\times\Gamma\bigg(\frac{j-2}{2}+\tau -\frac{\Delta(j)}{2}+2\bigg)\times\bigg(\frac{1+x}{1-x}\bigg)^{-\frac{j-2}{2}-\tau +\frac{\Delta(j)}{2}-2}\,.\nonumber\\
\eea
Using (\ref{Iz3}) in (\ref{Aj1}), the spin-j form factor of the Dirac fermion reduces to

\bea
\label{Aj2}
A(j,K)=&&\frac{1}{4g_5^2}\frac{\tilde{\kappa}_N^{-(j-2)-\Delta(j)}}{\Gamma(\tilde{a}(j))}\int_{0}^{1}dx\,x^{\tilde{a}(j)-1}(1-x)^{-\tilde{b}(j)}\nonumber\\&&\times\bigg(\bigg(\frac{\tilde{n}_R}{\tilde{\kappa}_N^{\tau-1}}\bigg)^2\times\Gamma(c(j))\bigg(\frac{1+x}{1-x}\bigg)^{-c(j)}
+\bigg(\frac{\tilde{n}_L}{\tilde{\kappa}_N^{\tau}}\bigg)^2\times\Gamma(c(j)+1)\bigg(\frac{1+x}{1-x}\bigg)^{-(c(j)+1)}\bigg)\,,\nonumber\\
\eea
with $\Delta(j)$, $\tilde a(j), \tilde b(j)$ given in (\ref{PARAX}) and

\bea
c(j)=(\tau+1)+\frac{j-2}{2}-\frac{\Delta(j)}{2}
\eea
$A(j,K)$ is  the spin-j generalization of  the spin-2 graviton form factor (\ref{FFj2}).  The x-integral form (\ref{Aj2}) can be interpreted as 
the holographic {\it partonic} content of the Dirac fermion at strong coupling for spin-j, with $x$ playing the role of the Bjorken momentum fraction.

\section{Pomeron-Pomeron-glueball  bulk coupling}~\label{app_PP3}

The sphaleron at the boundary of the Witten vertex sources either $G^2$ or $G\tilde G$. 
$G^2$ is dual to the dilaton in bulk, while $G\tilde G$ is dual to the axion, both of which 
are described by massive glueballs of even and odd parity respectively.  $G^2$ yields a
direct scalar Pomeron-Pomeron-dilaton coupling times the pertinent bulk-to-boundary 
propagato. $G\tilde G$ yields an indirect pseudoscalar Pomeron-Pomeron-pseudoscalar 
coupling by mixing to the bulk singlet and Chern-Simons term,
times pertinent bulk-to-bulk and bulk-to-boundary propagators

\subsection{Bulk dilaton-graviton-graviton coupling}

The graviton  in bulk is dual to a a scalar glueball or $G^2$ on the boundary.  The effective action for the graviton is given by the standard Einstein-Hilbert action

\be
\label{EH}
\frac 1{2\tilde g_5^2}\int d^5x\sqrt{g}e^{-2\phi}\sqrt{R}
\ee
with $R$ the scalar curvature and $2\tilde{g}_5^2=2\kappa^2=16\pi G_N={8\pi^2}/{N_c^2}$.  To extract the bulk gravitational coupling, we  linearize the
metric around the flat metric $\eta_{\mu\nu}\rightarrow \eta_{\mu\nu}+\sqrt{2\kappa^2} h_{\mu\nu}$.  In 
de Donder gauge with $\partial_\alpha h^\alpha_\mu=\frac 12\partial_\mu f$ and $f=h_\mu^\mu$, the expansion of (\ref{EH})  yields the quadratic contributions

\be
(\sqrt{2\kappa^2})^2\bigg(\frac 1{8\tilde g_5^2}\,g^{MN}\eta^{\alpha\bar\alpha}\eta^{\beta\bar\beta}
\partial_M h_{\alpha\bar\alpha}\,\partial_N h_{\beta\bar\beta} -\frac 1{4\tilde g_5^2}\,g^{MN}\eta^{\alpha\bar\alpha}\eta^{\beta\bar\beta}
\partial_{M}h_{\alpha\beta}\partial_N h_{\bar\alpha\bar\beta}\bigg)
\ee
and the cubic contributions

\be
\label{CUBIC}
(\sqrt{2\kappa^2})^3
\bigg(-\frac 1{8\bar g_5^2}\,g^{MN}\eta^{\alpha\bar\alpha}\eta^{\beta\bar\beta}f\, \partial_M h_{\alpha\bar\beta}\,\partial_N h_{\bar\alpha\beta}
-\frac 1{4\bar g_5^2}\,g^{MN}\eta^{\alpha\bar\alpha}\eta^{\beta\bar\beta}\,\partial_M f\, h_{\alpha\bar\beta}\,\partial_N h_{\bar\alpha\beta}\bigg)
\ee
where only the couplings through the tracefull $f$ and traceless $h_{\alpha\beta}$ were retained.
Note that $f(z=0,x)$ couples to  $T^\mu_\mu(x)$ of the gauge theory on the boundary, hence the identification with the dual of $G^2$.
 
The induced interaction between the scalar and the graviton on the boundary, follows from the bulk decomposition  $h_{\mu\nu}(x, z)=h(z)h_{\mu\nu}(x)$ and $f(x, z)=s(z)f(x)$, 
in the form

\bea
\label{CUBICX1}
\int d^4x\,\eta^{\alpha\bar\alpha}\eta^{\beta\bar\beta}
\bigg(&&
{\bf C}^+_1\,f(x)\,\eta^{\mu\nu}\partial_\mu h_{\alpha\bar\beta}(x)\partial_\nu h_{\bar\alpha\beta} (x)\nonumber\\
+&&{\bf C}^+_2\,f(x)\,h_{\alpha\bar\beta}(x)h_{\bar\alpha\beta}(x)
+{\bf C}_3^+\,\eta^{\mu\nu}\partial_\mu f(x)h_{\alpha\bar\beta}(x)\partial_\nu h_{\bar\alpha\beta}(x)
\bigg)
\eea
with the induced bulk coefficients

\bea
\label{CUBICX2}
{\bf C}_1^+=&&
-\frac{(\sqrt{2\kappa^2})^3}{4\tilde{g}_5^2}\int dz\sqrt{g}e^{-\phi}\bigg(\frac {z^2}2s(z)h^2(z)\bigg)\nonumber\\
{\bf C}_2^+=&&
+\frac{(\sqrt{2\kappa^2})^3}{4\tilde{g}_5^2}\int dz\sqrt{g}e^{-\phi}\bigg( 
\frac {z^2}2s(z)h^{\prime 2}(z)+\frac{z^2}2s^\prime (z)(h^2)^\prime(z)\bigg)\nonumber\\
{\bf C}_3^+=&&
-\frac{(\sqrt{2\kappa^2})^3}{4\tilde{g}_5^2}\int dz\sqrt{g}e^{-\phi}\bigg( 
z^2s(z)h^2(z)\bigg)
\eea
with $2\tilde{g}^2_5=2\kappa^2=8\pi^2/N_c^2$. 
For the Witten diagram in Fig.~\ref{fig:pathconste}, we note that  for a massive $0^+$ production, the vertex is close to the boundary.
If we denote by $z_5=z\approx 0$ the position of the vertex in bulk, each of the two  bulk-to-bulk graviton propagators factors out

\bea
\label{XZ1}
G(z\approx 0, z^\prime)\approx \frac{z^4}4{\cal H}(K,z^\prime)\approx h(z){\cal H}(K,z^\prime)
\eea
with ${\cal H}(K,z^\prime)$ the non-normalizable bulk-to-boundary graviton propagator in (\ref{BTOBH}). Finally, we identify $s(z)$ with the normalizable
scalar $0^+$ glueball state in bulk.  So the fields in (\ref{CUBICX2}) read

\bea
\label{XZ2}
h(z)\approx \frac{z^4}4 \qquad s(z)\approx {\cal S}(M_{0^+},z)
\eea

\subsection{Bulk axion-graviton-graviton coupling}

The boundary pseudoscalar glueball or $G\tilde G$ is dual to the axion field $a$ in bulk. We now note that the
axion mixes with the flavor singlet U(1) gauge field $A_M$ in bulk and the latter interacts with the bulk gravitons through the
chiral-gravitational anomaly or Chern-Simons term~\cite{Anderson:2014jia}

\be
\label{CS}
\frac{N_c}{1536\pi^2}\int d^5x \sqrt{g}e^{-\phi}\,\epsilon^{MNPQR}g^{S\bar S}g^{T\bar T}\,{\rm Tr}A_M\,R_{NPST}R_{QR\bar T\bar S}
\ee
with $R_{NPST}$ the Riemann  tensor. The mixing between the axion  $a$ and the scalar part of the singlet U(1) gauge field $A_{M=z}$ that is consistent
with the QCD U$_A$(1) axial anomaly can be captured in (\ref{CS}) through the minimal gauge shift $A_z\rightarrow A_z+\partial_z a$, 


\be
\label{CS1}
\frac{N_fN_c}{1536\pi^2}\int d^5x \sqrt{g}e^{-\phi}\,\epsilon^{zNPQR}g^{S\bar S}g^{T\bar T} \,\partial_za\,R_{NPST}R_{QR\bar T\bar S}
\ee
The coupling of the axion to the graviton follows by expanding the metric $g_{MN}=\eta_{MN}+\delta_{\mu M}\delta_{\nu N}\sqrt{2\kappa^2} h_{\mu\nu}$,
with the result

\bea
\label{CS2}
\frac{N_fN_c\kappa^2}{764\pi^2}\int d^5x \sqrt{g}e^{-\phi}\,\epsilon^{\mu\nu\rho\sigma}\,\partial_za\,&&
\bigg(
\frac 92 \eta^{\alpha\beta}\eta^{\gamma\delta}\partial_\mu h_{\nu\alpha}\partial_\sigma h_{\rho\beta}
-6\eta^{\alpha\beta}\partial_z\partial_\mu h_{\nu\alpha}\partial_\sigma h_{\rho\beta}\nonumber\\
&&\,\,\,\,+2\eta^{\alpha\beta}\partial_z\partial_\mu h_{\nu\alpha}\partial_z\partial_\sigma h_{\rho\sigma}
+2\eta^{\alpha\beta}\eta^{\gamma\delta}\partial_\gamma\partial_\mu h_{\nu\alpha} 
\partial_\sigma(\partial_\delta h_{\rho\beta}-\partial_\beta h_{\rho\delta})
\bigg)\nonumber\\
\eea
Note that $a(z=0,x)$ couples to the topological charge density or $G\tilde G(x)$ of the gauge theory on the boundary. 

Using the bulk decomposition for the axion and the graviton

\be
a_z(x,z)=\theta(z) a(x)\qquad h_{\mu\nu}(x,z)=h(z)h_{\mu\nu}(x)
\ee
(\ref{CS2}) yields the boundary interaction

\bea
\label{CS3}
\int d^4x\,a(x)\,
\bigg({\bf C}^-_1\,
\eta^{\alpha\beta}\eta^{\gamma\delta}\partial_\mu h_{\nu\alpha}(x)\partial_\sigma h_{\rho\beta}(x)
+{\bf C}^-_2\,\eta^{\alpha\beta}\eta^{\gamma\delta}\partial_\gamma\partial_\mu h_{\nu\alpha} (x)
\partial_\sigma(\partial_\delta h_{\rho\beta}(x)-\partial_\beta h_{\rho\delta}(x))
\bigg)\nonumber\\
\eea
with the induced  coefficients

\bea
\label{CS4}
{\bf C}^-_1=&&\frac{N_fN_c\kappa^2}{764\pi^2}\int dz\sqrt{g}e^{-\phi}\,\theta^\prime(z)\,\bigg(\frac 92 h^2(z)-6hh^\prime(z) +2h^{\prime 2}(z)\bigg)\nonumber\\
{\bf C}^-_2=&&\frac{N_fN_c\kappa^2}{764\pi^2}\int dz\sqrt{g}e^{-\phi}\,\theta^\prime(z)\,h^2(z)
\eea
Again,  for the Witten diagram in Fig.~\ref{fig:pathconste}, we note that  for a massive $0^-$ production, the vertex is also close to the boundary.
The same reasoning as in (\ref{XZ1}-\ref{XZ2}) shows that the bulk fields in (\ref{CS4}) read

\bea
\label{XCS5}
h(z)\approx \frac{z^4}4 \qquad \theta(z)\approx {\cal P}(M_{0^-},z)
\eea
where the latter refers to the normalized pseudo-scalar gueball state in bulk.

\section{Diffractive amplitudes}

The use of Witten diagrams for DIS scattering in holographic QCD was pioneered by~\cite{Polchinski:2002jw,Polchinski:2002jw}. In particular,   the Pomeron
was identified in bulk with  a Reggeized graviton which transmutes to a string exchange. Numerous studies along these lines followed, which we cannot
fully and fairly  cover here. We only mention  the original suggestion made in~\cite{Rho:1999jm}  and based on a holographic string exchange which triggered 
all these studies, and which ultimatly  identifies the holographic Pomeron exchange with a stringy instanton exchange~\cite{Basar:2012jb}. In this Appendix, we use the Witten 
diagram approach as details how the Pomeron interacts in bulk, and follow the conventions detailed in~\cite{Mamo:2019mka}

\subsection{Amplitude for emitting a  pseudo-scalar}

For the Witten diagram with outgoing pseudo-scalar glueball $X=G\tilde G=0-$, the contribution is

\bea
\label{WPP1}
{\cal A}_{pp\rightarrow pp0-}(j_1,j_2, s, t)&&=[{\cal V}^{\alpha\beta(TT)}_{h\Psi\Psi}(j_1, p_1, p_3, k_1)]
\times[{\mathbb B}^-_{\alpha\beta\bar\alpha\bar\beta}(t_1, t_2, k_1, k_2)]\times [{\cal V}^{\bar\alpha\bar\beta(TT)}_{h\Psi\Psi}(j_2, p_2, p_4, k_2)]\nonumber\\
\eea
with $k_1=p_1-p_3$ and  $k_2=p_2-p_4$  and $t_{1,2}=-k_{1,2}^2$, for spin $j_{1,2}$ exchanges in the t-channel. The ${\mathbb P\mathbb P}0-$ vertex is

\bea
\label{WPP2}
{\mathbb B}^-_{\alpha\beta\bar\alpha\bar\beta}(t_1, t_2, k_1, k_2)=(B_1^-(t_1, t_2)\eta_{\beta\bar\beta}
+B_2^-(t_1, t_2)k_{2\beta}k_{1\bar\beta})\epsilon_{\alpha\bar\alpha\gamma\delta}k_1^\gamma k_2^\delta
\eea
(\ref{WPP2}) is consistent with charge conjugation and Lorentz symmetries, 
and follows from the Chern-Simons vertex (\ref{CS3}) after inserting the plane wave decomposition

\be
 a(x)=e^{-ip_5x}\qquad h_{\mu\nu}(x)=\epsilon_{\mu\nu}(k)e^{ik\cdot x}
\ee
with the explicit vertex factors

\bea
B_1^-(t_1, t_2)=2({\bf C}^-_1-{\bf C}^-_2 k_1\cdot k_2)\qquad B_2^-(t_1, t_2)=2{\bf C}^-_2
\eea





\subsection{Amplitude for emitting a scalar}

For the Witten diagram with outgoing scalar glueball $X=G^2=0+$, the contribution is

\bea
\label{WPP8}
{\cal A}_{pp\rightarrow pp0+}(j_1,j_2, s, t)=[{\cal V}^{\alpha\beta(TT)}_{h\Psi\Psi}(j_1, p_1, p_3, k_1)]
\times[{\mathbb B}^+_{\alpha\beta\bar\alpha\bar\beta}(t_1, t_2, k_1, k_2)]\times [{\cal V}^{\bar\alpha\bar\beta(TT)}_{h\Psi\Psi}(j_2, p_2, p_4, k_2)]\nonumber\\
\eea
 with the general scalar ${\mathbb P\mathbb P}0+$ vertex for the process $h(k_1)+h(k_2)\rightarrow f(p_5)$ 

\bea
\label{WPP9}
{\mathbb B}^+_{\alpha\beta\bar\alpha\bar\beta}(t_1, t_2, k_1, k_2)=B^+(t_1,t_2, p_5^2)\,\eta_{\alpha\bar\alpha}\eta_{\beta\bar\beta}
\eea
as quoted in  (\ref{CUBIC4}). Note that
(\ref{WPP9}) is consistent  with parity, charge conjugation and Lorentz symmetries.  More explicitly, 
for the  fusion process $h(k_1)+h(k_2)\rightarrow f(p_5)$ with

\be
\label{CUBICX3}
h_{\mu\nu}(x)= \epsilon_{\mu\nu}^{TT}(k)\,e^{+ik_{1,2}\cdot x}\qquad f(x)=e^{-ip_5\cdot x}
\ee
and  the transverse and traceless polarizations
$k^\mu\epsilon_{\mu\nu}^{TT}(k_{1,2})=0$ and $\eta^{\mu\nu}\epsilon_{\mu\nu}^{TT}(k)=0$,
(\ref{WPP9}) follows from the cubic graviton coupling (\ref{CUBIC}).

\subsection{Reggeization}

The Reggeized form of the amplitude (\ref{WPP1}) follows from the double Mellin transforms

\bea
\label{WPP4}
&&{\cal A}_{pp\rightarrow pp0-}(s_1, s_2, s, t)\nonumber\\
&&=\int_{\mathbb C_1}\frac{dj_1}{2\pi i}
\left(\frac{s_1^{j_1-2}+(-s_1)^{j_1-2}}{{\rm sin}\,\pi j_1}\right)
\int_{\mathbb C_2}\frac{dj_2}{2\pi i}
\left(\frac{s_2^{j_2-2}+(-s_2)^{j_2-2}}{{\rm sin}\,\pi j_2}\right)
{\cal A}_{pp\rightarrow pp0-}(j_1, j_2, s, t)
\eea
The contours ${\mathbb C_{1,2}}$  are at the rightmost of the branch-points of $A(j_{1,2}, k)$   as defined in (\ref{Aj2}). 
Each of the Mellin transform in (\ref{WPP4})  factorizes generically to 

\bea
\label{WPP5}
\int_{\mathbb C}\frac{dj}{2\pi i}
\left(\frac{s^{j-2}+(-s)^{j-2}}{{\rm sin}\,\pi j}\right)\,A(j, k)
\eea
The Pomeron amplitude follows by closing the contour to the left. The imaginary part follows from the discontinuity 
of the $\Gamma$-function in $A(j, k)$ with the result

\bea
\label{WPP6}
-\tilde{s}^{j_{0}}\int_{-\infty}^{j_0}\frac{dj}{\pi}
\left(\frac{1 +e^{-i\pi}}{{\rm sin}\,\pi j}\right)\tilde{s}^{j-j_{0}}\,\sin\left[\tilde{\xi}\sqrt{2\sqrt{\lambda}(j_0-j)}\right]
\frac{\kappa_N^{2j}}{s^2}\Gamma(\Delta(j)-2)A(j,k)\bigg\vert_{j\rightarrow j_0,\,\Delta(j)\rightarrow 2}
\eea
with $j_0=2-2/\sqrt{\lambda}$, and the t-dependence set to zero in the exponent. It will be restored in the final result by inspection.
We have set $\tilde{s}\equiv {s}/{\tilde{\kappa}_N^2}$, and defined $\tilde{\xi}\equiv\gamma+{\pi}/{2}$ with Euler constant $\gamma=0.55772$.
 In the high energy limit $\sqrt{\lambda}/\tilde{\tau}\rightarrow 0$
with $\tilde{\tau}\equiv\log\tilde{s}$, the j-integration yields

\bea
\label{WPP7}
e^{j_0\tilde{\tau}} \left[(\sqrt{\lambda}/\pi)+ i\right] ( \sqrt{\lambda}/ 2 \pi )^{1/2}\; \tilde{\xi}  \; \frac{e^ {-\sqrt\lambda  \tilde{\xi}^2 / 2\tilde{\tau}}}{\tilde{\tau}^{3/2}}\left(1 + {\cal O}\bigg(\frac{\sqrt{\lambda}}{\tilde{\tau}}\bigg) \right)
\times \frac{\kappa_N^{2j}}{s^2}\Gamma(\Delta(j)-2)A(j,k)\bigg\vert_{j\rightarrow j_0,\,\Delta(j)\rightarrow 2}
\eea

\subsection{Final amplitudes}

Combining the above results in (\ref{WPP1}) gives for the pseudo-scalar glueball emission amplitude

\bea
\label{WPPFULL0-}
 &&\left[ \left[(\sqrt{\lambda}/\pi)+ i\right] ( \sqrt{\lambda}/ 2 \pi )^{1/2}\; \tilde{\xi}  \; \frac{e^ {-\sqrt\lambda  \tilde{\xi}^2 / 2\tilde{\tau}_1}}{\tilde{\tau}_1^{3/2}}
\frac{A(j_0, k_1)}{\tilde s_1^{2-j_0}}\left[\frac 12\overline{u}(p_3)\gamma^{[\alpha} p^{\beta]} u(p_1)\right]\right]\nonumber\\
\times&&\bigg[{\mathbb B}^-_{\alpha\beta\bar\alpha\bar\beta}(t_1, t_2, k_1, k_2)\bigg]\nonumber\\
\times&&\left[ \left[(\sqrt{\lambda}/\pi)+ i\right] ( \sqrt{\lambda}/ 2 \pi )^{1/2}\; \tilde{\xi}  \; \frac{e^ {-\sqrt\lambda  \tilde{\xi}^2 / 2\tilde{\tau}_2}}{\tilde{\tau}_2^{3/2}}
\frac{A(j_0, k_2)}{\tilde s_2^{2-j_0}}\left[\frac 12 \overline{u}(p_4)\gamma^{[\bar \alpha} {\underline  p}^{\bar \beta]} u(p_2)\right]\right]
 \eea
with $p=(p_1+p_3)/2$ and $\underline p=(p_2+p_4)/2$.  Similarly, 
combining the above results in (\ref{WPP8}) yields the scalar glueball emission amplitude

\bea
\label{WPPFULL0+}
 &&\left[ \left[(\sqrt{\lambda}/\pi)+ i\right] ( \sqrt{\lambda}/ 2 \pi )^{1/2}\; \tilde{\xi}  \; \frac{e^ {-\sqrt\lambda  \tilde{\xi}^2 / 2\tilde{\tau}_1}}{\tilde{\tau}_1^{3/2}}
\frac{A(j_0, k_1)}{\tilde s_1^{2-j_0}}\left[\frac 12 \overline{u}(p_3)\gamma^{[\alpha} p^{\beta]} u(p_1)\right]\right]\nonumber\\
\times&&\bigg[{\mathbb B}^+_{\alpha\beta\bar\alpha\bar\beta}(t_1, t_2, k_1, k_2)\bigg]\nonumber\\
\times&&\left[ \left[(\sqrt{\lambda}/\pi)+ i\right] ( \sqrt{\lambda}/ 2 \pi )^{1/2}\; \tilde{\xi}  \; \frac{e^ {-\sqrt\lambda  \tilde{\xi}^2 / 2\tilde{\tau}_2}}{\tilde{\tau}_2^{3/2}}
\frac{A(j_0, k_2)}{\tilde s_2^{2-j_0}}\left[\frac 12\overline{u}(p_4)\gamma^{[\bar \alpha} {\underline p}^{\bar \beta]} u(p_2)\right]\right]
 \eea

\bibliographystyle{elsarticle-num}
	 \bibliography{exclusive-JAN6,inst_spha,allbib}
 \end{document}